\documentclass[12pt,preprint]{emulateapj}
\def\be{\begin{equation}} 
\def\ee{\end{equation}} 
\def\gsim{\lower.5ex\hbox{\gtsima}} 
\def\gtsima{$\; \buildrel > \over 
\sim \;$}  
\def\lcii{$L_{[{\rm CII}]_{158\mu m}}$}
\def\loiii{$L_{[{\rm OIII}]_{88\mu m}}$}
\def\cii{[CII]$_{158\mu m}$}
\def\oiii{[OIII]$_{88\mu m}$}

\begin{document}
\title{Reionization Era Bright Emission Line Survey: Selection and
  Characterization of Luminous Interstellar Medium Reservoirs in the
  $z>6.5$ Universe} 
\title{Reionization Era Bright Emission Line Survey: Selection and
  Characterization of Luminous Interstellar Medium Reservoirs in the
  $z>6.5$ Universe} \author{R.J. Bouwens$^{1}$, R. Smit$^{2}$,
  S. Schouws$^{1}$, M. Stefanon$^{1}$, R.A.A. Bowler$^{3}$,
  R. Endsley$^{4}$, V. Gonzalez$^{5,6}$, H. Inami$^{7}$,
  D. Stark$^{4}$, P. Oesch$^{8,9}$, J. Hodge$^{1}$, M. Aravena$^{10}$,
  E. da Cunha$^{11}$, P. Dayal$^{12}$, I. de Looze$^{13,14}$,
  A. Ferrara$^{15}$, Y. Fudamoto$^{8,16,17}$, L. Graziani$^{18,19}$,
  C. Li$^{20,21}$, T. Nanayakkara$^{22}$, A. Pallottini$^{15}$,
  R. Schneider$^{18,23}$, L. Sommovigo$^{15}$, M. Topping$^{4}$,
  P. van der Werf$^{1}$, H. Algera$^{7,24}$,
  L. Barrufet$^{8}$, A. Hygate$^{1}$,
  I. Labb{\'e}$^{22}$, D. Riechers$^{25}$, J. Witstok$^{26,27}$}
\altaffiltext{1}{Leiden Observatory, Leiden University, NL-2300 RA
  Leiden, Netherlands} \altaffiltext{2}{Astrophysics Research
  Institute, Liverpool John Moores University, 146 Brownlow Hill,
  Liverpool L3 5RF, United Kingdom} \altaffiltext{3}{Astrophysics, The
  Denys Wilkinson Building, University of Oxford, Keble Road, Oxford,
  OX1 3RH, United Kingdom} \altaffiltext{4}{Steward Observatory,
  University of Arizona, 933 N Cherry Ave, Tucson, AZ 85721, United
  States} \altaffiltext{5}{Departmento de Astronomia, Universidad de
  Chile, Casilla 36-D, Santiago 7591245, Chile}
\altaffiltext{6}{Centro de Astrofisica y Tecnologias Afines (CATA),
  Camino del Observatorio 1515, Las Condes, Santiago, 7591245, Chile}
\altaffiltext{7}{Hiroshima Astrophysical Science Center, Hiroshima
  University, 1-3-1 Kagamiyama, Higashi-Hiroshima, Hiroshima 739-8526,
  Japan} \altaffiltext{8}{Observatoire de Gen{\`e}ve, 1290 Versoix,
  Switzerland} \altaffiltext{9}{Cosmic Dawn Center (DAWN), Niels Bohr
  Institute, University of Copenhagen, Jagtvej 128, K\o benhavn N,
  DK-2200, Denmark} \altaffiltext{10}{Nucleo de Astronomia, Facultad
  de Ingenieria y Ciencias, Universidad Diego Portales, Av. Ejercito
  441, Santiago, Chile} \altaffiltext{11}{International Centre for
  Radio Astronomy Research, University of Western Australia, 35
  Stirling Hwy, Crawley,26WA 6009, Australia}
\altaffiltext{12}{Kapteyn Astronomical Institute, University of
  Groningen, PO Box 800, NL-9700 AV Groningen, the Netherlands}
\altaffiltext{13}{Sterrenkundig Observatorium, Ghent University,
  Krijgslaan 281 - S9, 9000 Gent, Belgium} \altaffiltext{14}{Dept. of
  Physics \& Astronomy, University College London, Gower Street,
  London WC1E 6BT, United Kingdom} \altaffiltext{15}{Scuola Normale
  Superiore, Piazza dei Cavalieri 7, 56126 Pisa, Italy}
\altaffiltext{16}{Research Institute for Science and Engineering,
  Waseda University, 3-4-1 Okubo, Shinjuku, Tokyo 169-8555, Japan}
\altaffiltext{17}{National Astronomical Observatory of Japan, 2-21-1,
  Osawa, Mitaka, Tokyo, Japan} \altaffiltext{18}{Dipartimento di
  Fisica, Sapienza, Universita di Roma, Piazzale Aldo Moro 5, I-00185
  Roma, Italy} \altaffiltext{19}{INAF/Osservatorio Astrofisico di
  Arcetri, Largo E. Femi 5, I-50125 Firenze, Italy}
\altaffiltext{20}{Department of Astronomy \& Astrophysics, The
  Pennsylvania State University, 525 Davey Lab, University Park, PA
  16802, USA} \altaffiltext{21}{Institute for Gravitation and the
  Cosmos, The Pennsylvania State University, University Park, PA
  16802, USA} \altaffiltext{22}{Centre for Astrophysics \&
  Supercomputing, Swinburne University of Technology, PO Box 218,
  Hawthorn, VIC 3112, Australia} \altaffiltext{23}{INAF/Osservatorio
  Astronomico di Roma, via Frascati 33, 00078 Monte Porzio Catone,
  Roma, Italy}
\altaffiltext{24}{National Astronomical Observatory of Japan, 2-21-1,
  Osawa, Mitaka, Tokyo, Japan}
\altaffiltext{25}{Cornell University, 220 Space
  Sciences Building, Ithaca, NY 14853, USA} \altaffiltext{26}{Kavli
  Institute for Cosmology, Cambridge, University of Cambridge,
  Madingley Road, Cambridge, CB3 0HA, United Kingdom}
\altaffiltext{27}{Cavendish Laboratory, University of Cambridge, 19 JJ Thomson Avenue, Cambridge CB3 0HE, UK}
\begin{abstract}
The Reionization Era Bright Emission Line Survey (REBELS) is a cycle-7
ALMA Large Program (LP) that is identifying and performing a first
characterization of many of the most luminous star-forming galaxies
known in the $z>6.5$ universe.  REBELS is providing this probe by
systematically scanning 40 of the brightest $UV$-selected galaxies
identified over a 7-deg$^2$ area for bright [CII]$_{158\mu m}$ and
[OIII]$_{88 \mu m}$ lines and dust-continuum emission.  Selection of
the 40 REBELS targets was done by combining our own and other
photometric selections, each of which is subject to extensive vetting
using three completely independent sets of photometry and
template-fitting codes.  Building on the observational strategy
deployed in two pilot programs, we are increasing the number of
massive interstellar medium (ISM) reservoirs known at $z>6.5$ by
$\sim$4-5$\times$ to $>$30.  In this manuscript, we motivate the
observational strategy deployed in the REBELS program and present
initial results.  Based on the first-year observations, 18 highly
significant $\geq 7\sigma$ [CII]$_{158\mu m}$$\,$lines have already
been discovered, the bulk of which (13/18) also show
$\geq$3.3$\sigma$ dust-continuum emission.  These newly
discovered lines more than triple the number of bright ISM-cooling
lines known in the $z>6.5$ universe, such that the number of
ALMA-derived redshifts at $z>6.5$ rival Ly$\alpha$ discoveries.  An
analysis of the completeness of our search results vs. star formation
rate (SFR) suggests an $\sim$79\% efficiency in scanning for
[CII]$_{158\mu m}$ when the SFR$_{UV+IR}$ is $>$28
M$_{\odot}$/yr.  These new LP results further demonstrate ALMA's
efficiency as a ``redshift machine,'' particularly in the epoch of
reionization.
\end{abstract}

\section{Introduction}

The first galaxies are thought to have started forming in the first
200-300 Myr of the Universe (Bromm \& Yoshida 2011; Wise et al.\ 2012)
and then rapidly built up their stellar mass.  With existing
telescopes, it has been possible to obtain a glimpse of galaxies
forming close to this time (e.g., Zheng et al.\ 2012; Ellis et
al.\ 2013; Oesch et al.\ 2014; Bouwens et al.\ 2015, 2016; Bowler et
al.\ 2015; Finkelstein et al.\ 2015; Stefanon et al.\ 2017a; Ono et
al.\ 2018; Dayal \& Ferrara 2018), with galaxies identified as far
back as $z\sim 11$ (e.g., Coe et al.\ 2013; Oesch et al.\ 2016),
$\sim$400 Myr after the Big Bang.

The discovery of impressively massive galaxies in the first 1-2
billion years of the universe, including an especially massive
$2\times10^{11}$ M$_{\odot}$ passive galaxy at z=3.717 (Glazebrook et
al.\ 2017), a $\sim$3$\times$10$^{11}$ M$_{\odot}$ star-forming galaxy
at $z=6.9$ (Strandet et al.\ 2017; Marrone et al.\ 2018), a $z=7.54$
quasar with a supermassive black hole of mass 8$\times$10$^{8}$
$M_{\odot}$ (Ba{\~n}ados et al.\ 2018), and even a $10^9$ M$_{\odot}$
galaxy at $z\sim 11$ (Oesch et al.\ 2016; Jiang et al.\ 2021), have
reignited long-standing questions about how rapidly galaxies could
have begun forming large numbers of stars. Besides these few
well-known examples of $z>3$ galaxies with particularly high masses
(and other work by e.g. D{\'\i}az-Santos et al.\ 2016 and Tanaka et
al.\ 2019), other evidence for significant early star-formation
include a few $z>6$ galaxies with apparently older stellar populations
(e.g., Hashimoto et al.\ 2018; Roberts-Borsani et al.\ 2020) and
evidence for lower-mass $z\sim 4$-8 galaxies with elevated
stellar-mass-to-halo-mass ratios (e.g., Behroozi et al.\ 2013).

Until just recently, the only physical information we had on the early
epochs of star formation for most massive galaxies is from sensitive
near-IR observations from {\it Hubble} Space Telescope ({\it HST}),
mid-IR observations from {\it Spitzer}, and some rest-UV spectroscopy
from the ground.  {\it HST} provides us with high spatial resolution
images revealing newly formed stars not obscured by dust, while {\it
  Spitzer} probes line emission and the stellar population age in
galaxies through features like the Balmer break. Quantifying the
masses in $z>6$ galaxies has proven to be particularly challenging due
to the limited information in {\it Spitzer}/IRAC imaging observations
and degeneracy between the impact of older stellar populations and
strong nebular emission lines (e.g., Schaerer \& de Barros 2009;
Raiter et al.\ 2010; Roberts-Borsani et al.\ 2020). Rest-$UV$
spectroscopy has resulted in the detection of Ly$\alpha$ and some high
ionization $UV$ lines in select $z>6$ galaxies (e.g., Zitrin et
al.\ 2015; Stark et al.\ 2017), suggestive of very young stellar
populations with moderately hard radiation fields (Stark et al.\ 2015;
Mainali et al.\ 2017; Schmidt et al.\ 2017; Mainali et al.\ 2018;
Hutchison et al.\ 2019). While such radiation fields are likely
important in driving the reionization of the Universe at $z>6$,
determining the typical stellar population properties of $z>6$
galaxies has remained very challenging.

Fortunately, thanks to the improving capabilities of the Atacama Large
Millimeter Array (ALMA), we can make significant progress
characterizing the physical properties of massive star-forming sources
in the early universe (e.g., Hodge \& da Cunha 2020). ALMA efficiently
allows for a probe of the redshift and dynamical state of galaxies in
the reionization epoch using bright interstellar medium (ISM)-cooling
lines like 157.74$\mu$m [CII] and 88.36$\mu$m [OIII] (e.g., Smit et
al.\ 2018; Hashimoto et al.\ 2019; Tamura et al.\ 2019; Jones et
al.\ 2021), while simultaneously probing the far-IR dust-continuum
radiation (e.g., Capak et al.\ 2015; Bowler et al.\ 2018; Fudamoto et
al.\ 2020; Schouws et al.\ 2021). This provides us with an essential
measure of the star formation in galaxies obscured by dust. The
effectiveness of ALMA in probing the physical properties of galaxies
has been demonstrated by a few particularly noteworthy observational
programs with ALMA, e.g., the ASPECS ALMA Large Program in cycle 4
probing the dust and gas in $z>0.5$ galaxies over the Hubble Ultra
Deep Field (e.g., Gonz{\'a}lez-L{\'o}pez et al. 2020; Aravena et
al.\ 2020; Bouwens et al.\ 2020; Decarli et al.\ 2020; Walter et
al.\ 2020) and the ALPINE ALMA Large Program in cycle 5 targeting
$>$110 $z=4$-6 sources (Le Fevre et al.\ 2020; B{\'e}thermin et
al.\ 2020; Faisst et al.\ 2020).  By contrast, the number of sources
in the reionization epoch which have been characterized in detail with
ALMA has been modest (e.g., Watson et al.\ 2015; Maiolino et
al.\ 2015; Inoue et al.\ 2016; Pentericci et al.\ 2016; Knudsen et
al.\ 2017; Matthee et al.\ 2017, 2019; Laporte et al.\ 2017, 2019;
Smit et al.\ 2018; Hashimoto et al.\ 2018, 2019; Tamura et al.\ 2019;
Bakx et al.\ 2020; Schouws et al.\ 2021).

Until recently, the ALMA view of galaxies in the $z>6$ universe has
been restricted to those galaxies which show prominent Ly$\alpha$
emission, whether these sources be QSOs (e.g., Venemans et al.\ 2019,
2020) or simply massive star-forming galaxies, e.g., B14-65666 at
$z=7.15$ (Bowler et al.\ 2018; Hashimoto et al.\ 2019) or CR7 at
$z=6.59$ (Matthee et al.\ 2017).  This is especially a concern for more
massive galaxies ($\gtrsim 10^{9}$ M$_{\odot}$) which frequently do
not show Ly$\alpha$ in emission at all (e.g., Stark et al.\ 2010;
Schenker et al.\ 2014; Pentericci et al.\ 2018; Jung et al.\ 2020;
Endsley et al.\ 2021b), giving us a potentially biased view of the
characteristics of luminous galaxies at early times.  Given such, it is
clearly desirable for us to have an alternate strategy of selecting
massive galaxies in the $z>6$ universe.

Recently, it has been shown that scanning $UV$-bright galaxies for
prominent ISM-cooling lines like \cii$\,$or \oiii$\,$can be a very
efficient way to identify luminous galaxies in the $z>6$ universe.
This strategy appears to be most efficient (1) when targets are
especially $UV$ bright and (2) when the redshift of targets is
particularly well constrained, e.g., from the position of the Lyman
break or the position of strong nebular emission lines within the {\it
  Spitzer}/IRAC bands (Smit et al.\ 2015).  Smit et al.\ (2018) used
just an hour of ALMA observations to identify what were then two of
the most luminous \cii$\,$lines known at $z>6.5$.  Employing a similar
spectral scan strategy to observe six more luminous $z\sim7$ galaxies
with well-constrained redshifts, Schouws et al.\ (2022) found bright
\cii$\,$lines in three more $z\sim7$ galaxies with an 11-hour program
(2018.1.00085.S).  The Smit et al.\ (2018) and Schouws et al.\ (2022)
program demonstrated how effective ALMA could be in scanning for
extremely bright \cii$\,$lines in $UV$-bright sources at $z>6.5$ and
effectively served as pilots for the ALMA Large Program we introduce
here.

In parallel with these \cii$\,$studies, there were simultaneous
efforts exploring the use of the \oiii$\,$line for spectral scans at
$z>8$, taking advantage of both the brightness of the line and its
greater accessibility in band 7 at $z>8$.  The first detection of the
\oiii$\,$line in the $z>7$ universe was by Inoue et al.\ (2016) at
$z=7.21$ in a Ly$\alpha$-emitting galaxy.  Hashimoto et al.\ (2018)
then demonstrated the first successful scan for \oiii at $z>8$,
securing a 7.4$\sigma$ detection of line in the $z=9.1096 \pm 0.0006$
source MACS1149-JD (Zheng et al.\ 2012).  Tamura et al.\ (2019)
similarly made use of ALMA to successfully scan for the \oiii$\,$line
in a magnified Lyman-break galaxy behind MACS0416, MACS0416\_Y1,
finding a redshift of 8.3118$\pm$0.0003 based on a 6.3$\sigma$
detection.

With multiple ALMA programs demonstrating the feasibility of using
spectral scans to search for especially luminous ISM reservoirs at
$z>6.5$, the necessary preparatory work had been done to consider
executing a much more significant survey program with ALMA for
luminous ISM reservoirs in the $z>6.5$ universe.  It was in this
environment that the Reionization Era Bright Emission Line Survey
(REBELS) program (2019.1.01634.L) was approved for execution in cycle
7 as an extragalactic ALMA Large Program.  The goal of the REBELS
program is to create the first significant sample of especially
luminous ISM reservoirs in the Reionization Epoch.  REBELS is doing so
by scanning for bright ISM cooling lines on a large sample of
$UV$-bright ($-23.0<M_{UV,AB}<-21.5$) galaxies at $z>6.5$, while
simultaneously probing the dust-continuum flux of sources.

In this paper, we summarize the observational strategy employed by the
REBELS ALMA Large Program and highlight some first results
demonstrating the effectiveness of this strategy.  This includes an
overview of the target selection and major scientific goals.  The
standard concordance cosmology $\Omega_0 = 0.3$, $\Omega_{\Lambda} =
0.7$, and $H_0 = 70\,\textrm{km/s/Mpc}$ is assumed for consistency
with previous studies.  SFR and stellar masses are quoted assuming a
Chabrier (2003) IMF.  All magnitudes are in the AB system (Oke \& Gunn
1983).

\begin{figure}
\epsscale{1.17}
\plotone{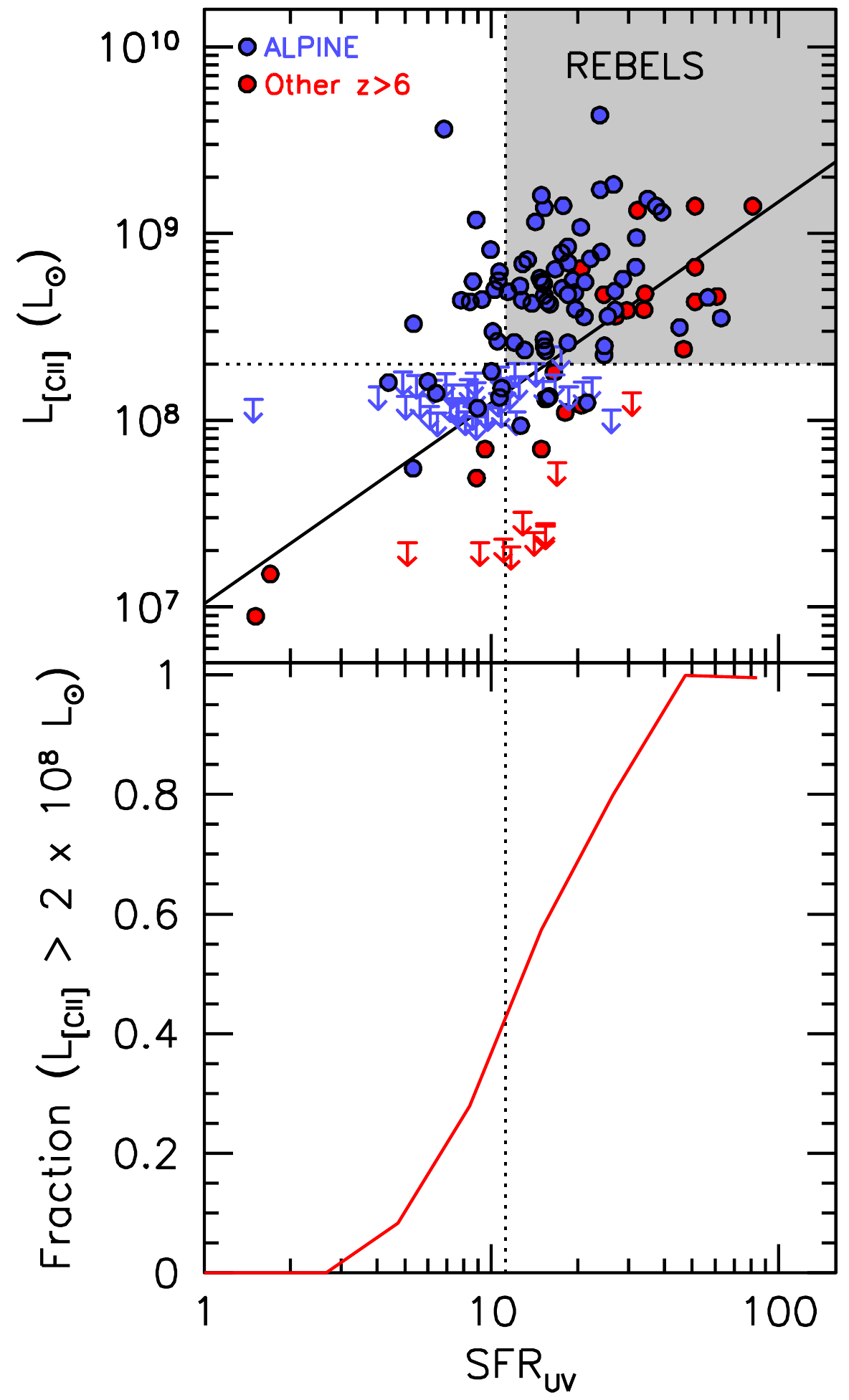}
\caption{(\textit{upper}) Observed luminosity of the \cii$\,$cooling
  line seen in galaxies at $z\sim6$-7 vs. the observed star formation
  rate of galaxies in the rest-UV.  The plotted results are drawn from
  the ALPINE (Le Fevre et al.\ 2020; B{\'e}thermin et al.\ 2020;
  Faisst et al.\ 2020) and Capak et al.\ (2015) $z=4$-6 samples
  (\textit{blue circles}) as well as the REBELS pilots (Smit et
  al.\ 2018; Schouws et al.\ 2022) and other $z>6$ sources
  (\textit{red circles}: Matthee et al.\ 2019). The black solid line
  shows the $z\sim0$ \cii-SFR$_{UV+IR}$ de Looze et al.\ (2014)
  relation.  Particularly noteworthy is the increase in the
  \cii$\,$luminosities of galaxies as rest-$UV$ SFRs increase from 5
  to 30 M$_{\odot}$/yr.  REBELS only includes sources with rest-$UV$
  SFRs in excess of 11 M$_{\odot}$/yr (\textit{dotted vertical
    line}).  The horizontal dotted line shows the approximate
  $5\sigma$ sensitivity limit adopted in the REBELS program.
  (\textit{lower}) The fraction of galaxies with \cii$\,$luminosities
  in excess of $2\times10^8$ $L_{\odot}$ (the typical
  sensitivity limit for observations taken as part of REBELS)
  vs. the unobscured SFR of galaxies at $z>6$.  A significant fraction
  of galaxies with rest-$UV$ SFRs in excess of 10 M$_{\odot}$/yr and
  especially 20 M$_{\odot}$/yr have \cii$\,$luminosities in excess of
  $2\times10^8$ $L_{\odot}$ (\S2.1).  It is these high SFR sources at
  $z>6.5$ that the REBELS large program particularly targets in
  scanning for bright ISM cooling lines. \label{fig:cii_vs_muv}}
\end{figure}

\section{$UV$-Bright Sample Selection}

In this section, we motivate our use of a $UV$-bright selection for
the REBELS ALMA LP, and then discuss procedurally how we identify the
most robust set of targets for the program.

\begin{figure*}
\epsscale{1.17}
\plotone{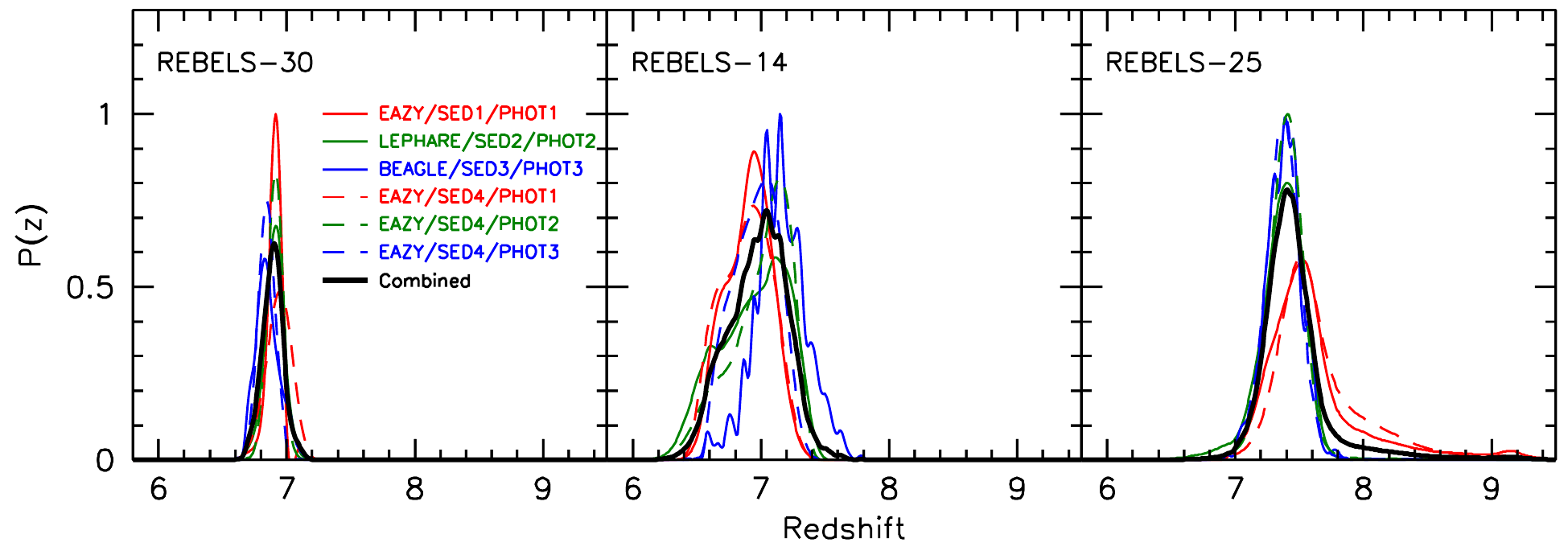}
\caption{Redshift likelihood distributions derived for three different
  sources in the REBELS program REBELS-30 (\textit{left}), REBELS-14
  (\textit{center}), and REBELS-25 (\textit{right}) by team members
  Rebecca Bowler, Ryan Endsley, and Mauro Stefanon (shown as the red,
  green, and blue lines, respectively) by applying the
  \textsc{lephare}, \textsc{beagle}, and \textsc{eazy} photometric
  redshift codes, respectively, to their own independent photometry
  PHOT1, PHOT2, and PHOT3, respectively.  Three additional redshift
  likelihood distributions have been derived on the basis of each set
  of photometry by using the \textsc{eazy} photometric redshift code
  and a separate set of SED templates SED4 are shown as the dashed
  red, green, and blue lines, respectively.  The thick black lines
  show the redshift likelihood distributions derived by averaging the
  results of the six separate likelihood distributions.
\label{fig:pz}}
\end{figure*}

Briefly, the REBELS LP selects 40 $UV$-bright galaxies at $z>6.5$ for
spectral scan observations, drawing from a 7 deg$^2$ area with the
deepest wide-area optical, near-IR, and Spitzer/IRAC imaging
observations and features the 2 deg$^2$ COSMOS/UltraVISTA field, the 5
deg$^2$ VIDEO/XMM-LSS + UKIDSS/UDS fields, and an 0.2 deg$^2$ area
composed of various HST legacy fields including CANDELS (Grogin et
al.\ 2011; Koekemoer et al.\ 2011), CLASH (Postman et al.\ 2012),
RELICS (Coe et al.\ 2019; Salmon et al.\ 2020), and various
BoRG/HIPPIES pure parallel fields (Trenti et al.\ 2011; Yan et
al.\ 2011; Bradley et al.\ 2012; Schmidt et al.\ 2014; Morishita et
al.\ 2020; Roberts-Borsani et al.\ 2021).  Sources are chosen both
because of their $UV$ luminosities (Figure~\ref{fig:cii_vs_muv}) and
because of the tight constraints we have on their redshifts from the
available imaging data (Figure~\ref{fig:pz}).

\begin{figure*}
\epsscale{1.17}
\plotone{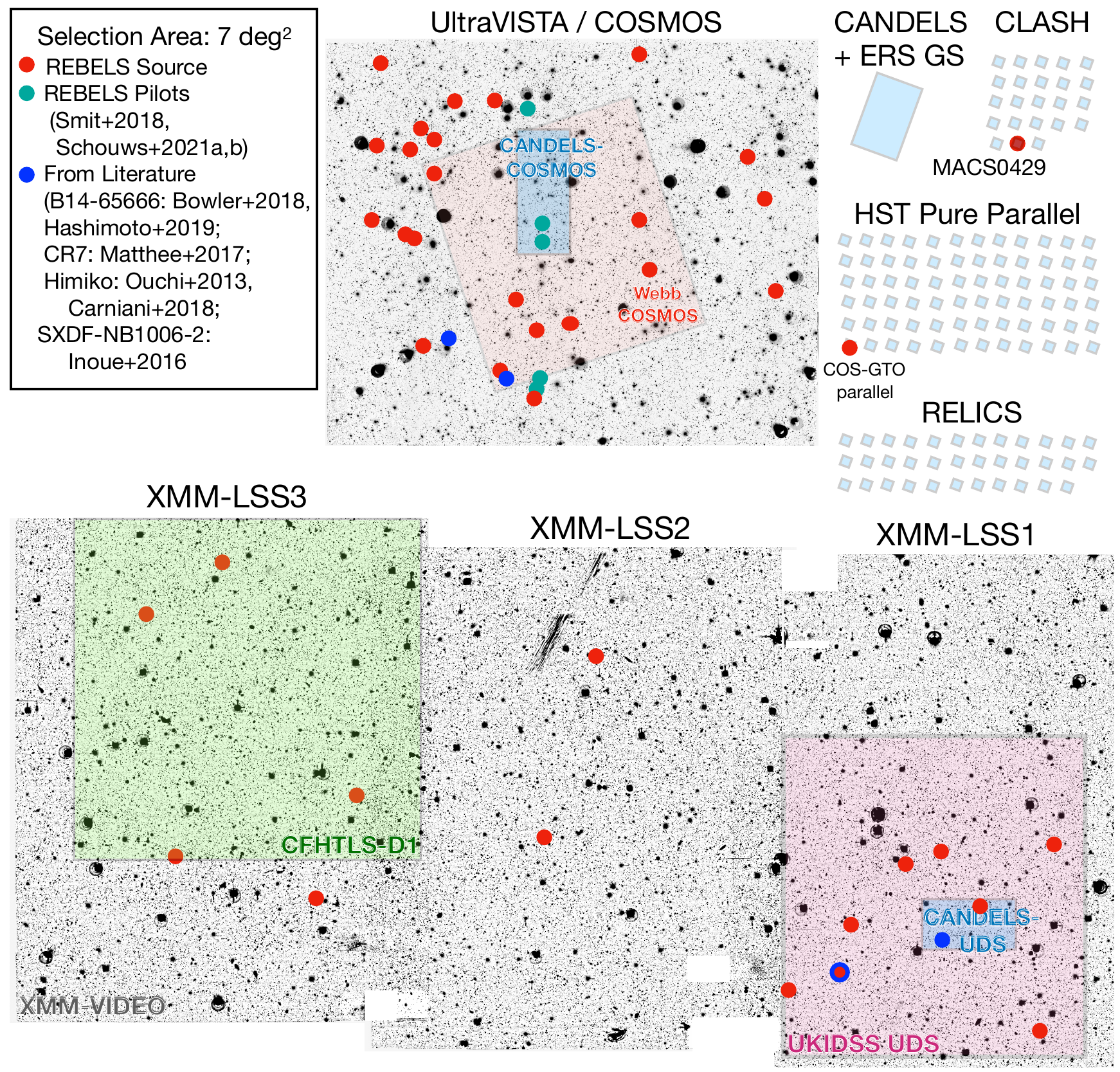}
\caption{The layout of the two wide-area ground-based fields + {\it
    HST} data sets used to identify $z=6.5$-9.5 targets for the REBELS
  ALMA Large Program.  The red circles indicate the position of REBELS
  targets, and the cyan circles indicate the position of targets in
  the REBELS pilot programs (Smit et al.\ 2018; Schouws et
  al.\ 2022).  The blue circles indicate the position of other
  sources in the literature that also meet the selection criteria for
  REBELS but already have substantial ALMA observations (Ouchi et
  al.\ 2013; Inoue et al.\ 2016; Matthee et al.\ 2017; Bowler et
  al.\ 2018; Hashimoto et al.\ 2019; Carniani et al.\ 2018).  Also
  shown is the array of individual HST fields considered in the
  selection of targets for REBELS, including CANDELS GOODS South
  (Grogin et al.\ 2011), CLASH (Postman et al.\ 2012), various pure
  parallel fields (e.g., Morishita et al.\ 2018; Bridge et al.\ 2019),
  and RELICS (Coe et al.\ 2019).  The position of the CANDELS COSMOS
  and CANDELS UDS fields within the COSMOS/UltraVISTA and UDS/XMM-LSS
  fields is indicated.  The planned layout of the Webb-COSMOS survey
  (Kartaltepe et al.\ 2021) is also included on this
  figure.  \label{fig:layout}}
\end{figure*}

Figure~\ref{fig:layout} illustrates the search fields used for
constructing the REBELS sample as well as the distribution of the
REBELS targets over the sky.  Figure~\ref{fig:twod} illustrates the
distribution of $UV$ luminosities and redshifts for targets in the
REBELS program and compares this distribution against those sources in
the REBELS pilot programs, in the ALPINE program, galaxies with
confirmed Ly$\alpha$ emission at $z>6.5$, and sources in various HST
legacy fields.  A complete list of the targets in the program is
provided in Table~\ref{tab:targlist}.

\subsection{Prevalence of Massive Star-Forming Galaxies Amongst $UV$-Bright Galaxies at $z>4$}

Over the last few years, it has become increasingly clear that many of
the brightest known ISM-cooling lines at $z\gtrsim 5.5$ are found in
especially $UV$-bright galaxies ($M_{UV,AB}$$\lesssim$$-$22 mag:
$\geq$2 $L_{UV}^*$).  Examples include the ten $z\sim5.5$ sources
targeted by Capak et al.\ (2015), two $z\sim6.1$ galaxies targeted by
Willott et al.\ (2015), CR7 by Matthee et al.\ (2017), B14-65666 by
Hashimoto et al.\ (2019), VR7 by Matthee et al.\ (2019), and the three
$z\sim6.1$-6.2 galaxies targeted by Harikane et al.\ (2020).  All show
bright ($\gtrsim$2 mJy$\,$km/s) \cii$\,$lines.  Meanwhile, follow-up
of $z\geq 6$ galaxies with redshifts from Ly$\alpha$ have frequently
revealed only moderate to low luminosity ($\lesssim$1 mJy$\,$km/s)
\cii$\,$lines (Ouchi et al.\ 2013; Ota et al.\ 2014; Pentericci et
al.\ 2016; Brada{\v{c}} et al.\ 2017; Carniani et al.\ 2018, 2020),
strongly suggesting that there may be an inverse correlation between
Ly$\alpha$ equivalent width (EW) and the mass of the ISM reservoirs
(Figure 18 from Harikane et al.\ 2020).  Given such, it seems the best
way to obtain a representative sample of massive galaxies at $z>6$ is
to identify galaxies on the basis of their $UV$ luminosity (rather
than Ly$\alpha$ emission alone).

To quantify the relationship between the $UV$ luminosity $M_{UV}$ of
galaxies and their \lcii, we present a compilation of some recent
results at $z>4$ (Capak et al.\ 2015; Smit et al.\ 2018; Matthee et
al.\ 2019; B{\'e}thermin et al.\ 2020; Schouws et al.\ 2022) in
Figure~\ref{fig:cii_vs_muv}.  A conversion factor of $7.1\times
10^{-29} L_{\nu}
[\textrm{M}_{\odot}/\textrm{yr}/(\textrm{ergs/s/Hz})]$ (M. Stefanon et
al.\ 2022, in prep) is used to transform the $UV$ luminosities of
galaxies into SFRs.  What is particularly striking here is the strong
correlation we observe between \lcii$\,$and the unobscured SFR$_{UV}$,
with a particularly clear correlation around 10 M$_{\odot}$/yr.  The
fraction of sources with \cii$\,$luminosities in excess of
$2\times10^8 L_{\odot}$, the typical sensitivity limit achieved
  in REBELS observations, increases from 35\% at 10 M$_{\odot}$/yr to
80\% at 20 M$_{\odot}$/yr.

Because of the strong correlation between \lcii$\,$and $M_{UV}$, one
can very efficiently derive redshifts for galaxies which are bright in
the $UV$ by scanning for the \cii$\,$ISM cooling line with ALMA, as
demonstrated in the introduction by both pilot programs to REBELS
(Smit et al.\ 2018; Schouws et al.\ 2022).

\subsection{Search Fields}

In constructing a sample of 40 $UV$-bright galaxies to follow up with
the REBELS LP, we made use of deep wide-area optical + near-IR
observations over the $\sim$2 deg$^2$ COSMOS/UltraVISTA field
(Scoville et al.\ 2007; McCracken et al.\ 2012), the $\sim$5 deg$^2$
UKIDSS/UDS + VIDEO/XMM-LSS fields (Lawrence et al.\ 2007), and a wide
range of {\it HST} search fields, including CANDELS, CLASH, and the
BoRG/HIPPIES pure parallel fields.

A significant fraction of the $UV$-bright targets for REBELS are drawn
from the $\sim$2 deg$^2$ COSMOS/UltraVISTA field.  In constructing our
source catalogs for this field, we made use of the very sensitive $Y$,
$J$, $H$, and $K_s$ near-IR observations (both data release 3 and 4)
from the UltraVISTA program (McCracken et al.\ 2012), the deep optical
Subaru Suprime-Cam $BgVriz$ observations from Taniguchi et
al.\ (2005), the deep optical $ugriyz$ observations from the CFHT Deep
Legacy survey (Erben et al.\ 2009; Hildebrandt et al.\ 2009), and the
$ugrizy$ Subaru Hyper Suprime-Cam (HSC) observations (Aihara et
al.\ 2018a, 2018b).  Use of the {\it Spitzer}/IRAC 3.6$\mu$m and
4.5$\mu$m observations from SCOSMOS (Sanders et al.\ 2007),
\textit{Spitzer} Large-Area Survey with HSC (SPLASH: Steinhardt et
al.\ 2014), SMUVS (Caputi et al.\ 2017; Ashby et al.\ 2018), COMPLETE
(Labb{\' e} et al.\ 2016), and COMPLETE2 (Stefanon et al.\ 2018)
programs were also made.

The bulk of the remaining targets are drawn from the $\sim$5 deg$^2$
UKIDSS/UDS and VIDEO/XMM-LSS area.  The UDS/XMM-LSS area includes
sensitive near-IR observations from both the UKIDSS/UDS (Lawrence et
al.\ 2007: $J$, $H$, and $K$ bands) and VISTA Deep Extragalactic
Observations (VIDEO) programs (Jarvis et al.\ 2013: $Y$, $J$, $H$, and
$K_s$), sensitive optical observations from the Subaru Hyper
Suprime-Cam deep and wide-area programs (Aihara et al.\ 2018a, 2018b),
and {\it Spitzer}/IRAC observations from the SPLASH and
\textit{Spitzer} Extragalactic Representative Volume Survey (SERVS:
Mauduit et al.\ 2012) programs.

We also considered including $UV$-bright $z>6.5$ galaxies from the
CANDELS, CLASH, and BoRG/HIPPIES fields as targets in the REBELS
program.  In total, the CANDELS, CLASH, and BoRG/HIPPIES fields that
we used for selecting targets covered an area of 0.2 deg$^2$
(excluding those HST CANDELS fields within COSMOS and UDS).

\subsection{Selection of the Targets for REBELS}

From each of these search fields, we considered a range of different
catalogs in arriving at our final target list.  Here we provide a
brief summary of each sample we consider:\\

\noindent \textit{Bowler et al.\ 2014, 2017 ($z\sim7$):} The Bowler et
al.\ (2014) $z\sim7$ selection was constructed based on deep $z$, $Y$,
$J$, $H$, and $K_s$ observations obtained over the COSMOS/UltraVISTA
and UKIDSS/UDS fields with Subaru Suprime-Cam + VISTA and Subaru,
VISTA, and UKIRT, respectively.  The deep $Y+J$ and $J$ band images
over UltraVISTA/COSMOS and the UKIDSS/UDS fields, respectively, were
used as the detection image in contructing source catalogs for the
search.  Then, after removing all sources detected at $2\sigma$ in the
optical imaging data, the \textsc{lephare} photometric redshift
software (Arnouts et al.\ 1999; Ilbert et al.\ 2006) was run, using
Bruzual \& Charlot (2003) models with a range of exponentially
declining star-formation histories, metallicities of 0.2 $Z_{\odot}$
and $Z_{\odot}$, a range of dust attenuation ($A_V < 4$), Ly$\alpha$
EWs to 240\AA, and Madau (1995) IGM absorption.  Then, Bowler et
al.\ (2014) compared their \textsc{lephare} fit results with similar
fits to stars from the SpecX library (Burgasser 2014) to identify and
exclude any possible low-mass stars from their selection.  Bowler et
al.\ (2017) refined the Bowler et al.\ (2014) selection, taking
advantage of subsequent WFC3/IR F140W imaging they obtained over the
fields with {\it HST}, identifying two additional and fainter $z\sim7$
galaxies in the neighborhood of their bright candidates and removing
three cross-talk artefacts from their catalogs (\S3.2 of Bowler et
al.\ 2017).  In total, 22 bright $z\sim7$ galaxies were identified as
part of the Bowler et al.\ (2014) and (2017) studies, with $M_{UV}$
luminosities ranging from $-20.7$ mag to $-23.2$ mag.\\

\noindent \textit{Stefanon et al.\ 2017b, 2019a ($z\sim8$-9):} The
selection of $z\sim8$-9 galaxies in Stefanon et al.\ (2017b, 2019a) was
performed by first creating a large parent catalog over the
COSMOS/UltraVISTA fields based on the $Y$, $J$, $H$, and $K_s$ images
and then applying an optical non-detection and two color Lyman-break
galaxy-like criteria.  Stefanon et al.\ (2019a) then made use of the
\textsc{mophongo} package (Labb{\' e} et al.\ 2006, 2013, 2015) to do
careful optical, near-IR, and Spitzer/IRAC photometry for each source
in the parent catalog, modeling and subtracting the flux from nearby
neighbors to improve the robustness of the flux measurements.
Candidate $z\sim8$-9 galaxies were then selected by running the
\textsc{eazy} photometric redshift code (Brammer et al.\ 2008) on the
optical+near-IR+{\it Spitzer}/IRAC photometry that had been derived.
18 $z\sim8$-9 candidate galaxies were identified by Stefanon et
al.\ (2017b, 2019a) over the $\sim$2 deg$^2$ COSMOS field.  Through
similar fits of the source photometry to the SpecX dwarf star spectral
libaries (Burgasser 2014), Stefanon et al.\ (2019a) explicitly verified
that all sources were better fit ($\Delta \chi^2 > 1$) by galaxy SED
templates than dwarf star templates.\\

\begin{deluxetable*}{ccccccc}
\tablewidth{0cm}
\tabletypesize{\footnotesize}
\tablecaption{Bright $z>6.5$ Candidate Galaxies Targeted by the REBELS Program\label{tab:targlist}}
\tablehead{
\colhead{REBELS ID} & \colhead{ALMA ID} & \colhead{R.A.} & \colhead{Dec} & \colhead{$z_{phot}$} & \colhead{$M_{UV}$} & \colhead{Ref\tablenotemark{$\dagger$}}}
\startdata
REBELS-01\tablenotemark{$\ddagger$}  & XMM1-Z-276466 & 02:16:25.09 & $-$04:57:38.5 & $7.31_{-0.10}^{+0.11}$ & $-22.9\pm0.1$ & \\
REBELS-02\tablenotemark{$\ddagger$}  & XMM1-35779 & 02:16:32.43 & $-$05:30:05.6 & $6.65_{-0.13}^{+0.18}$ & $-22.1\pm0.2$ & \\
REBELS-03\tablenotemark{$\ddagger$}  & XMM1-Z-1664 & 02:17:15.23 & $-$05:07:45.8 & $6.99_{-0.20}^{+0.24}$ & $-21.8\pm0.3$ & \\
REBELS-04  & XMM-J-355 & 02:17:42.46 & $-$04:58:57.4 & $8.57_{-0.09}^{+0.10}$\tablenotemark{b} & $-22.3\pm0.1$ &  [6]\\
REBELS-05  & XMM1-1591 & 02:18:11.51 & $-$05:00:59.3 & $6.68_{-0.17}^{+0.18}$ & $-21.6\pm0.2$ & [2] \\
REBELS-06\tablenotemark{$\ddagger$}  & XMM1-Z-151269 & 02:18:47.47 & $-$05:10:20.3 & $6.79_{-0.11}^{+0.13}$ & $-21.7\pm0.3$ & \\
REBELS-07\tablenotemark{$\ddagger,c$}  & XMM1-Z-1510 & 02:18:56.53 & $-$05:19:58.6 & $7.15_{-0.14}^{+0.20}$ & $-22.1\pm0.3$ & [3] \\
REBELS-08  & XMM1-67420 & 02:19:35.13 & $-$05:23:19.2 & $6.71_{-0.10}^{+0.13}$ & $-21.8\pm0.4$ &  [7]\\
REBELS-09\tablenotemark{$*$}  & XMM2-Z-1116453 & 02:21:54.15 & $-$04:24:12.3 & $7.58_{-0.13}^{+0.27}$ & $-23.0\pm0.3$ & [6] \\
REBELS-10\tablenotemark{$*,\ddagger$}  & XMM2-Z-564239 & 02:22:32.59 & $-$04:56:51.2 & $7.42_{-0.91}^{+0.23}$ & $-22.7\pm0.3$ & \\
REBELS-11\tablenotemark{$\ddagger$}  & XMM3-Y-217016 & 02:24:39.35 & $-$04:48:30.0 & $8.24_{-0.37}^{+0.65}$ & $-22.8\pm0.2$ & \\
REBELS-12\tablenotemark{$\ddagger$}  & XMM3-Z-110958 & 02:25:07.94 & $-$05:06:40.7 & $7.40_{-0.20}^{+0.15}$ & $-22.5\pm0.3$ & \\
REBELS-13\tablenotemark{$*$}  & XMM-J-6787 & 02:26:16.52 & $-$04:07:04.1 & $8.19_{-0.50}^{+0.84}$ & $-22.9\pm0.2$ &  [6]\\
REBELS-14\tablenotemark{$\ddagger$}  & XMM3-Z-432815 & 02:26:46.19 & $-$04:59:53.5 & $7.00_{-0.27}^{+0.20}$ & $-22.6\pm0.4$ & \\
REBELS-15\tablenotemark{$\ddagger$}  & XMM3-Z-1122596 & 02:27:13.11 & $-$04:17:59.2 & $6.78_{-0.09}^{+0.11}$ & $-22.6\pm0.3$ & \\
REBELS-16  & MACS0429-Z1 & 04:29:37.20 & $-$02:53:49.1 & $6.74_{-0.09}^{+0.09}$ & $-21.5\pm0.1$ &  [1]\\
REBELS-17\tablenotemark{$\ddagger$}  & UVISTA-Z-1373 & 09:57:36.99 & 02:05:11.3 & $6.66_{-0.22}^{+0.16}$ & $-21.7\pm0.2$ & \\
REBELS-18  & UVISTA-Y-001 & 09:57:47.90 & 02:20:43.7 & $8.20_{-0.37}^{+0.63}$ & $-22.5\pm0.1$ &  [4,6]\\
REBELS-19  & UVISTA-Y-879 & 09:57:54.69 & 02:27:54.9 & $7.52_{-0.21}^{+0.29}$ & $-21.6\pm0.2$ &  [6]\\
REBELS-20  & UVISTA-Z-734 & 09:59:15.88 & 02:07:31.9 & $7.07_{-0.08}^{+0.10}$ & $-21.8\pm0.1$ & [2] \\
REBELS-21  & UVISTA-Z-013 & 09:59:19.35 & 02:46:41.3 & $6.63_{-0.12}^{+0.09}$ & $-21.9\pm0.2$ &  [7,8]\\
REBELS-22  & UVISTA-Y-657 & 09:59:20.35 & 02:17:22.7 & $7.31_{-0.10}^{+0.11}$ & $-22.2\pm0.1$ &  [6]\\
REBELS-23  & UVISTA-Z-1410 & 10:00:04.36 & 01:58:35.5 & $6.68_{-0.09}^{+0.12}$ & $-21.6\pm0.5$ &  [7]\\
REBELS-24  & UVISTA-Y-005 & 10:00:31.89 & 01:57:50.2 & $8.35_{-0.51}^{+0.66}$ & $-22.0\pm0.2$ &  [4,6]\\
REBELS-25  & UVISTA-Y-003 & 10:00:32.32 & 01:44:31.3 & $7.40_{-0.19}^{+0.22}$ & $-21.7\pm0.2$ &  [4,6]\\
REBELS-26  & UVISTA-Z-011 & 10:00:42.12 & 02:01:57.1 & $6.64_{-0.13}^{+0.22}$ & $-21.8\pm0.1$ &  [2,7,8]\\
REBELS-27  & UVISTA-Y-004 & 10:00:58.49 & 01:49:56.0 & $7.40_{-0.14}^{+0.13}$ & $-22.0\pm0.2$ &  [4,6]\\
REBELS-28\tablenotemark{$\ddagger$}  & UVISTA-Z-1595 & 10:01:04.60 & 02:38:56.7 & $6.82_{-0.13}^{+0.13}$ & $-22.5\pm0.3$ & \\
REBELS-29  & UVISTA-Z-004 & 10:01:36.85 & 02:37:49.1 & $6.82_{-0.11}^{+0.13}$ & $-22.3\pm0.1$ &  [2,7,8]\\
REBELS-30  & UVISTA-Z-009 & 10:01:52.30 & 02:25:42.3 & $6.90_{-0.09}^{+0.08}$ & $-22.4\pm0.1$ &  [2,8]\\
REBELS-31  & UVISTA-Z-005 & 10:01:58.50 & 02:33:08.2 & $6.65_{-0.06}^{+0.10}$ & $-22.4\pm0.2$ &  [2,7,8]\\
REBELS-32  & UVISTA-Z-049 & 10:01:59.07 & 01:53:27.5 & $6.79_{-0.11}^{+0.17}$ & $-21.7\pm0.1$ &  [7,8]\\
REBELS-33  & UVISTA-Z-018 & 10:02:03.81 & 02:13:25.1 & $6.68_{-0.12}^{+0.12}$ & $-21.6\pm0.1$ & [2,8] \\
REBELS-34  & UVISTA-Z-002 & 10:02:06.47 & 02:13:24.2 & $6.75_{-0.07}^{+0.09}$ & $-22.5\pm0.1$ &  [2,7,8]\\
REBELS-35  & UVISTA-Z-003 & 10:02:06.70 & 02:34:21.4 & $6.98_{-0.10}^{+0.10}$ & $-22.5\pm0.1$ &  [8]\\
REBELS-36  & UVISTA-Y-002 & 10:02:12.56 & 02:30:45.7 & $7.88_{-0.20}^{+0.58}$ & $-22.2\pm0.2$ &  [4]\\
REBELS-37  & UVISTA-J-1212 & 10:02:31.81 & 02:31:17.1 & $7.75_{-0.17}^{+0.09}$\tablenotemark{b} & $-22.2\pm0.1$ &  [6]\\
REBELS-38  & UVISTA-Z-349 & 10:02:54.05 & 02:42:12.0 & $6.67_{-0.10}^{+0.16}$ & $-21.9\pm0.2$ &  [7]\\
REBELS-39  & UVISTA-Z-068\tablenotemark{a} & 10:03:05.25 & 02:18:42.7 & $6.76_{-0.05}^{+0.06}$ & $-22.7\pm0.2$ &  [7]\\
REBELS-40  & Super8-1 & 23:50:34.66 & $-$43:32:32.5 & $7.49_{-0.08}^{+0.00}$ & $-21.9\pm0.1$ &  [5] 
\enddata
\tablenotetext{*}{Given that $>$15\% of the integrated redshift
  likelihood distribution $P(z)$ for the source is at $z<6$, this is
  one of two targets in the REBELS selection which could correspond to
  a lower-redshift interloper.}
\tablenotetext{$\dagger$}{[1] Smit et al.\ (2014), [2] Bowler et al.\ (2014), [3] Inoue et al. (2016), [4] Stefanon et al.\ (2017b, 2019a), [5] Bridge et al.\ (2019), [6] Bowler et al.\ (2020), [7] Endsley et al.\ (2021a), and [8] Schouws et al.\ (2022).}
\tablenotetext{$\ddagger$}{Identified here in selecting the base REBELS sample.}
\tablenotetext{b}{Including the F105W, F125W, and F160W-band data from GO 15931 (PI: Bowler) and GO 16879 (PI: Stefanon).}
\tablenotetext{c}{SXDF-NB1006-2 with $z_{[OIII]} = 7.2120 \pm 0.0003$ (Inoue et al.\ 2016).}
\end{deluxetable*}

\noindent \textit{Bowler et al.\ 2020}: Candidate $z\sim8$-11 galaxies
were identified using the very deep ground-based optical and near-IR
observations over $\sim$6 deg$^2$ the COSMOS/UltraVISTA and
VIDEO/XMM-LSS + UKIDSS/UDS fields.  Candidate $z\sim8$ galaxies were
required to be detected at 5$\sigma$ in either than $J$ or $H$ bands,
while candidate $z\sim9$ galaxies were required to show a $5\sigma$
detection in either the $H$ or $K_s$/$K$ bands.  No detection
($<$2$\sigma$) was allowed for sources in all of the deep optical
bands, as well as the $Y$ band for $z\sim9$ candidates.  The reality
of sources showing $5\sigma$ detections in the UDS $JHK$ data was
tested by looking for similar $2\sigma$ detections in the VISTA data.
The redshift likelihood distributions for candidate $z\sim8$-11
galaxies were then computed using a range of exponentially declining
star formation histories, metallicities of 0.2 $Z_{\odot}$ and
$Z_{\odot}$, a range of dust attenuation ($A_V < 6$), and the Madau
(1995) IGM absorption.  Bowler et al.\ (2020) identified 27
$z\sim8$-10 galaxies with $UV$ luminosities $M_{UV}$ ranging from
$-23.7$ to $-21.2$ mag.\\

\noindent \textit{Endsley et al.\ 2021a:} Candidate $z\sim7$ galaxies
were identified over the deep $\sim$2 deg$^2$ COSMOS/UltraVISTA and
$\sim$1 deg$^2$ UKIDSS UDS fields.  Sources were detected from
$\chi^2$ detection images (Szalay et al.\ 1999) constructed from the
deep $yYJHK_s$ HSC/UltraVISTA and $yYJHK$ HSC/VIDEO/UKIDSS imaging
observations.  Endsley et al.\ (2021a) then applied the following
color criteria: (1) $z-y>1.5$, (2) $z-Y>1.5$, (3) $NB_{921}-Y>1.0$,
and (4) $y-Y<0.4$.  Sources were required to be detected at $5\sigma$
in at least one of the $y$, $Y$, and $J$ bands and detected at
$3\sigma$ in all three and undetected at $<$2$\sigma$ in the HSC $g$
and $r$ bands.  Possible T dwarf contaminants were removed by requiring
either $Y-J<0.45$ or both $J-H > 0$ and $J-K_s > 0$.  Endsley et
al.\ (2021a) identified 50 sources over the COSMOS/UltraVISTA and UDS
fields.\\

\noindent \textit{Schouws et al.\ 2022:} Candidate $z\sim7$ galaxies
in Schouws et al.\ (2022) were selected from the $\sim$2 deg$^2$
UltraVISTA observations over the COSMOS field and required to show a
$>$6$\sigma$ detection in a stack of the $Y$, $J$, $H$, and $K_s$
images and a $\chi_{opt} ^2 < 4$ (Bouwens et al.\ 2011).  Best-fit
photometric redshifts and redshift likelihood distributions were then
derived with \textsc{eazy} using photometry of sources derived from
the available CFHT + Subaru optical imaging data, UltraVISTA $YJHK_s$
near-IR observations, and {\it Spitzer}/IRAC observations from
SCOSMOS, SPLASH, and SMUVS programs.  For inclusion in the Schouws et
al.\ (2022) $z>6.5$ sample, sources were required to have an
integrated $z>6$ probability of $>$50\% and to be better fit by a
galaxy SED template than one of the SED templates from the SpecX dwarf
star spectral library.  Schouws et al.\ (2022) identified some 30
bright $z\sim7$ as part of their study, with $M_{UV}$ ranging from
$-23.0$ mag to $-21.4$ mag.\\

\noindent \textit{This Paper (see also M. Stefanon et al.\ 2022, in
  prep):} Finally, we considered separate selections of $z\sim7-8$
galaxies drawn from the $\sim5$ deg$^2$ VIDEO/XMM-LSS + UKIDSS/UDS
fields generated using procedures similar to those described in
Stefanon et al. (2019a). The detection was performed on the combination
of the $J$, $H$, and $K_\mathrm{S}$ mosaics, either from the
UKIDSS/UDS or VIDEO programs. Candidates at $z\sim7$ were selected
through the $z-Y>0.7$\,mag and $\langle YJ\rangle - \langle JH \rangle
<0.5$\,mag color criteria, where $\langle .. \rangle$ denotes the
average flux density in the two indicated bands. To identify the
samples at $z\sim8$ we imposed $Y-J > 0.7$\,mag and $\langle JH
\rangle - \langle HK_\mathrm{S} \rangle <0.5$\,mag for $z\sim8$.
These criteria were coupled with the requirement of non-detection
($<2\sigma$) in all bands (CFHTLS, HSC, SUBARU) bluer than the nominal
Lyman break, together with a $5\sigma$ detection in the combined $J,
H$ and $K_\mathrm{S}$ bands, to increase the robustness of the
detection. The selection was refined computing the photometric
redshifts with \textsc{eazy} (Brammer et al.\ 2008), after adding the
flux densities in the $3.6\mu$m and $4.5\mu$m \textit{Spitzer}/IRAC
bands from the SWIRE, SpUDS, and SEDS programs.  Finally, the image
stamps of the candidates were visually inspected, to remove
contaminants and sources whose flux measurements are contaminated by
poorly subtracted neighbors.  Sources with better $\chi^2$ from the
templates of Burgasser (2014) were excluded as likely brown dwarf
contaminants. This led to the identification of $\sim40$ candidates at
$z\sim7$ and $\sim25$ at $z\sim8$ brighter than $J\sim25-26$\,mag.\\

We then created a master list of 60 $UV$-bright $z=6.5$-10 candidates
from the above analyses, as well as 5 other particularly bright
$z>6.5$ galaxies that had been identified over various HST legacy
fields (MACS0429Z-9372034910, BORG2229-0945-394, Super8-1, Super8-4,
Super8-5: Bradley et al.\ 2012; Bouwens et al.\ 2015, 2019; Smit et
al.\ 2014, 2015; Morashita et al.\ 2018; Bridge et al.\ 2019; Salmon
et al.\ 2020).  Combining the above analyses, the approximate area
probed in COSMOS/UltraVISTA field, UKIDSS/UDS + VIDEO/XMM-LSS fields,
and various HST archival fields (not included in the COSMOS or the UDS
fields) is 1.82 deg$^2$, 4.97 deg$^2$, and 0.2 deg$^2$, respectively,
or 6.99 deg$^2$ in total.  

For each of the 60 candidates identified over the UltraVISTA/COSMOS,
VIDEO/XMM-LSS, and UKIDSS/UDS fields, measurements of the optical,
near-IR, and {\it Spitzer}/IRAC fluxes were made by three members of
our team (Mauro Stefanon, Rebecca Bowler, and Ryan Endsley) applying
their own photometric procedures to the imaging data sets each had
compiled of the fields.  We refer to these flux measurements as PHOT1,
PHOT2, and PHOT3, respectively.  Very briefly, PHOT1 is based on
neighbor-subtracted aperture photometry of sources in 1.2$''$-diameter
and 1.8$''$-diameter apertures in the optical/near-IR and
\textit{Spitzer}/IRAC data using \textsc{mophongo} (e.g., see Stefanon
et al.\ 2019a), PHOT2 is based on aperture photometry in
1.8$''$-diameter apertures and neighbor-subtracted aperture photometry
in 2.8$''$-diameter apertures with \textsc{T-PHOT} (Merlin et
al.\ 2015) for the optical/near-IR and \textit{Spitzer}/IRAC data,
respectively (see e.g. Bowler et al.\ 2020), and PHOT3 is based on
1.2$''$-diameter aperture photometry for the optical/near-IR data, and
in 2.8$''$-diameter apertures after subtracting neighbors using a
\textsc{mophongo}/\textsc{T-PHOT}-like algorithm (see Endsley et
al.\ 2021a).  For each of these photometric catalogs, aperture
measurements are corrected to total based on the curve of growth.
More details on these photometric procedures will be provided in
M. Stefanon et al.\ (2022, in prep).  For sources found within various
HST legacy fields, use was made of published photometry, except in the
case of the Super8 sources where deeper IRAC imaging observations
(Holwerda et al.\ 2018) were utilized.

\begin{figure*}
\epsscale{1.1}
\plotone{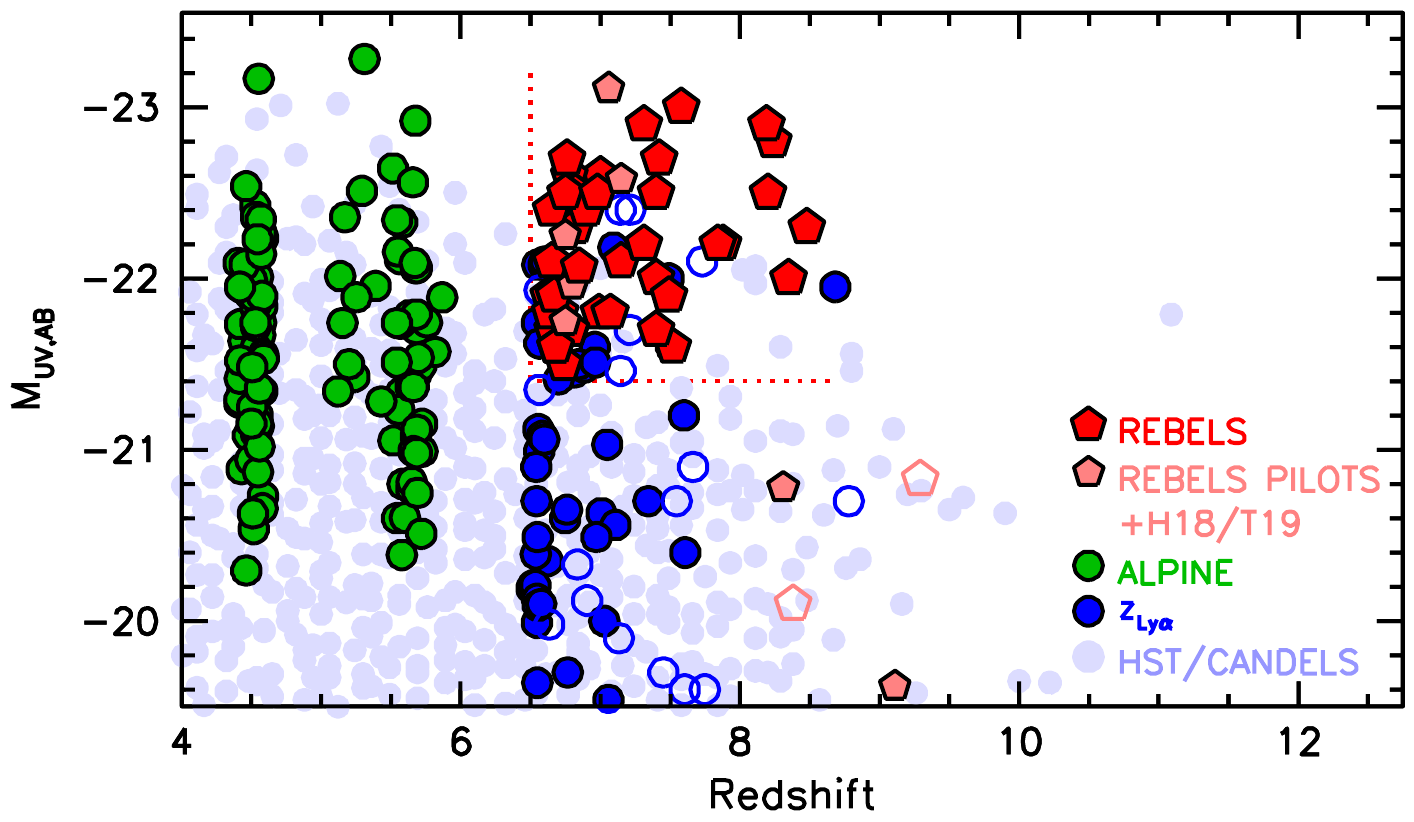}
\caption{$UV$ luminosities and photometric redshifts of the new sample
  of bright $z>6.5$ galaxies (\textit{red pentagons}) being targeted
  with the REBELS ALMA Large program relative to $z>6.5$ sources with
  spectroscopic redshifts from Ly$\alpha$ (\textit{dark blue
    circles}), $z>6.5$ sources from pilots to REBELS (\textit{light
    red pentagons}), sources over various {\it HST} legacy fields
  (Bouwens et al.\ 2021: \textit{light blue circles}), and sources
  with the ALPINE program (\textit{green circles}).  $z>6.5$ sources
  with redshift determinations based on Ly$\alpha$ and \oiii$\,$lines
  where the significance is $<$7$\sigma$ are shown with open blue
  circles and open light red pentagons, respectively.\label{fig:twod}}
\end{figure*}

Each of these members (MS, RB, RE) then made use of a separate
photometric redshift code (\textsc{eazy}, \textsc{lephare}, and
\textsc{beagle} [Chevallard \& Charlot 2016], respectively) to derive
a redshift likelihood distribution for each candidate on the basis of
their photometry PHOT1, PHOT2, and PHOT3, respectively.  A fourth
member of the REBELS team (Rychard Bouwens) then derived a fourth,
fifth, and sixth photometric redshift distribution for each source by
running the \textsc{eazy} photometric redshift code a second time on
the photometry derived by RB, RE, and MS, but using a different set of
SED templates SED4.  In particular, instead of using the EAZY\_v1.0
template set augmented by the Binary Population and Spectral Synthesis
code (BPASS: Eldridge et al.\ 2017) v1.1 for sub-solar metallicity (Z
= 0.2Z$_{\odot}$), which include nebular emission from \textsc{cloudy}
(Ferland et al. 2017), and 2Gyr-old passively evolving systems with
varying amounts ($A_V = 0$-8 mag) of dust extinction adopting the
Calzetti et al.\ (2000) law, i.e., SED template set SED1, the second
\textsc{eazy} runs use the EAZY\_v1.0 template set augmented by SED
templates from the Galaxy Evolutionary Synthesis Models (GALEV:
Kotulla et al.\ 2009), i.e., SED template set SED4.  Nebular continuum
and emission lines were included in the latter templates according to
the prescription provided in Anders \& Fritze-v. Alvensleben (2003), a
$0.2 Z_{\odot}$ metallicity, and scaled to a rest-frame EW for
H$\alpha$ of 1300\AA.

The mean of the six redshift likelihood distributions was then taken
and a single distribution derived (Figure~\ref{fig:pz}).  Care was
taken to explicitly verify that none of the candidates considered
resulted from the VISTA/VIRCAM electronic crosstalk artefact Bowler et
al.\ (2017) identified.  Bowler et al.\ (2017) found that such
artefacts occurred at multiples of 128 pixel separations from
saturated stars in the VISTA/VIRCAM data.  Similar crosstalk artefacts
are known to occur in the UKIDSS/UDS observations (Dye et al.\ 2006;
Warren et al.\ 2007), and as such, care was taken to confirm the
reality of sources based on multiple data sets, e.g., with
VISTA/VIRCAM, UKIDSS/UDS, and {\it Spitzer}/IRAC data.  Any source
suspected to correspond to an artefact was excluded from
consideration.

Sources were then ordered in terms of the likelihood of detecting an
ISM cooling line in each.  In computing these likelihoods, a
SFR$_{UV}$-to-\lcii$\,$conversion factor of 2$\times$10$^{7}$
$L_{\odot}$/(M$_{\odot}$/yr) was assumed (de Looze et al.\ 2014), with
a 0.3-dex scatter.  Additionally, $z=6.5$-7.2 candidate sources,
$z=7.2$-7.7 candidate sources, and $z=7.7$-9.5 candidate sources were
assumed to have an allocation of 2, 3, and 6 tunings, respectively.
To ensure a good sampling of luminous ISM reservoirs as a function of
redshift, 20 $z=6.5$-7.2 targets, 16 $z=7.2$-8.5 targets, and 4
$z=7.8$-9.4 targets were chosen.

\begin{deluxetable*}{ccccccc}
\tabletypesize{\footnotesize}
\tablewidth{17.5cm}
\tablecaption{$UV$-continuum Slopes, Stellar Population
    Parameters, and SFRs for Bright $z>6.5$ Candidate Galaxies
    Targeted by the REBELS Program\label{tab:targlist2}} \tablehead{
\colhead{REBELS ID} & \colhead{Redshift} &
  \colhead{$\beta$} & \colhead{$\log_{10}$ $M_{*}$ (M$_{\odot}$)\tablenotemark{*}}
  & \colhead{EW([OIII]+H$\beta$) ($\AA$)\tablenotemark{*}} &
  \colhead{SFR$_{UV}$ [$M_{\odot}$ yr$^{-1}$]\tablenotemark{$\dagger$}} &
  \colhead{SFR$_{IR}$ [$M_{\odot}$ yr$^{-1}$]\tablenotemark{$\ddagger$}}}
\startdata
REBELS-01  & 7.177 & $-2.04_{-0.20}^{+0.24}$ & 10.02$_{-0.53}^{+0.41}$ & $2.90_{-0.12}^{+0.28}$ & 45$\pm$5 & $<$35 \\
REBELS-02  & $6.65_{-0.13}^{+0.18}$ & $-2.24_{-0.36}^{+0.44}$ & 9.04$_{-0.49}^{+0.50}$ & $3.07_{-0.16}^{+0.14}$ & 22$\pm$4 & $<$30 \\
REBELS-03  & 6.969 & $-2.14_{-0.46}^{+0.63}$ & 9.13$_{-0.88}^{+0.64}$ & $3.05_{-0.22}^{+0.33}$ & 16$\pm$4 & $<$34 \\
REBELS-04  & $8.57_{-0.09}^{+0.10}$ & $-2.15_{-0.38}^{+0.20}$ & 8.72$_{-0.68}^{+1.03}$ & $3.25_{-0.31}^{+0.33}$ & 23$\pm$2 & 59$_{-36}^{+20}$\\
REBELS-05  & 6.496 & $-1.29_{-0.44}^{+0.36}$ & 9.16$_{-1.00}^{+0.85}$ & $3.12_{-0.32}^{+0.30}$ & 14$\pm$3 & 40$_{-16}^{+23}$\\
REBELS-06  & $6.79_{-0.11}^{+0.13}$ & $-1.24_{-0.35}^{+0.67}$ & 9.50$_{-0.79}^{+0.45}$ & $2.96_{-0.18}^{+0.27}$ & 15$\pm$4 & 49$_{-19}^{+28}$\\
REBELS-07  & $7.15_{-0.14}^{+0.20}$ & $-2.39_{-0.43}^{+0.37}$ & 8.69$_{-0.76}^{+0.74}$ & $3.18_{-0.17}^{+0.33}$ & 20$\pm$5 & $<$34 \\
REBELS-08  & 6.749 & $-2.17_{-0.58}^{+0.58}$ & 9.02$_{-0.68}^{+0.64}$ & $3.07_{-0.22}^{+0.23}$ & 16$\pm$6 & 64$_{-24}^{+48}$\\
REBELS-09  & $7.58_{-0.13}^{+0.27}$ & $-2.66_{-0.53}^{+0.93}$ & 8.65$_{-0.43}^{+0.43}$ & $3.77_{-0.02}^{+0.05}$ & 49$\pm$14 & $<$41 \\
REBELS-10  & $7.42_{-0.91}^{+0.23}$ & $-1.34_{-0.83}^{+0.48}$ & 10.16$_{-0.32}^{+0.31}$ & $2.95_{-0.09}^{+0.10}$ & 37$\pm$10 & $<$41 \\
REBELS-11  & $8.24_{-0.37}^{+0.65}$ & $-1.60_{-1.15}^{+0.17}$ & 9.36$_{-0.56}^{+0.52}$ & $3.08_{-0.19}^{+0.21}$ & 39$\pm$8 & $<$77 \\
REBELS-12\tablenotemark{a}  & 7.349 & $-1.99_{-0.76}^{+0.48}$ & 8.94$_{-0.70}^{+0.93}$ & $3.26_{-0.25}^{+0.29}$ & 30$\pm$8 & 62$_{-28}^{+38}$\\
REBELS-13  & $8.19_{-0.50}^{+0.84}$ & $-1.08_{-0.65}^{+0.59}$ & 9.80$_{-0.44}^{+0.43}$ & $2.98_{-0.09}^{+0.13}$ & 44$\pm$9 & $<$72 \\
REBELS-14  & 7.084 & $-2.21_{-0.47}^{+0.41}$ & 8.73$_{-0.70}^{+0.80}$ & $3.21_{-0.20}^{+0.35}$ & 35$\pm$13 & 41$_{-17}^{+24}$\\
REBELS-15  & $6.78_{-0.09}^{+0.11}$ & $-2.18_{-0.50}^{+0.52}$ & 8.81$_{-0.50}^{+0.50}$ & $3.73_{-0.54}^{+0.10}$ & 33$\pm$9 & $<$44 \\
REBELS-16  & $6.74_{-0.09}^{+0.09}$ & $-1.70_{-0.76}^{+0.48}$ & 9.47$_{-0.36}^{+0.34}$ & $3.02_{-0.10}^{+0.10}$ & 12$\pm$1 & $<$44 \\
REBELS-17  & $6.66_{-0.22}^{+0.16}$ & $-1.70_{-0.47}^{+0.33}$ & 9.07$_{-0.63}^{+0.58}$ & $3.04_{-0.19}^{+0.21}$ & 15$\pm$3 & $<$48 \\
REBELS-18  & 7.675 & $-1.34_{-0.32}^{+0.19}$ & 9.49$_{-0.73}^{+0.56}$ & $3.00_{-0.17}^{+0.22}$ & 31$\pm$4 & 41$_{-16}^{+23}$\\
REBELS-19  & 7.369 & $-2.33_{-0.64}^{+0.45}$ & 8.79$_{-0.69}^{+0.69}$ & $3.11_{-0.22}^{+0.21}$ & 14$\pm$3 & 52$_{-23}^{+31}$\\
REBELS-20  & $7.07_{-0.08}^{+0.10}$ & $-2.59_{-0.60}^{+0.57}$ & 8.59$_{-0.63}^{+0.63}$ & $3.17_{-0.11}^{+0.15}$ & 16$\pm$2 & $<$52 \\
REBELS-21  & $6.63_{-0.12}^{+0.09}$ & $-2.15_{-0.24}^{+0.42}$ & 10.38$_{-0.42}^{+0.25}$ & $2.84_{-0.12}^{+0.28}$ & 18$\pm$4 & $<$32 \\
REBELS-22  & $7.31_{-0.10}^{+0.11}$ & $-2.23_{-0.30}^{+0.21}$ & 9.65$_{-0.76}^{+0.42}$ & $2.91_{-0.14}^{+0.31}$ & 23$\pm$2 & $<$34 \\
REBELS-23  & $6.68_{-0.09}^{+0.12}$ & $-1.57_{-0.45}^{+0.28}$ & 9.11$_{-0.61}^{+0.54}$ & $3.03_{-0.16}^{+0.19}$ & 14$\pm$7 & $<$47 \\
REBELS-24  & $8.35_{-0.51}^{+0.66}$ & $-1.56_{-0.83}^{+0.56}$ & 8.97$_{-0.89}^{+0.89}$ & $3.13_{-0.21}^{+0.19}$ & 20$\pm$4 & $<$38 \\
REBELS-25  & 7.306 & $-1.85_{-0.46}^{+0.56}$ & 9.89$_{-0.18}^{+0.15}$ & $2.79_{-0.06}^{+0.21}$ & 15$\pm$3 & 185$_{-64}^{+101}$\\
REBELS-26  & $6.64_{-0.13}^{+0.22}$ & $-1.92_{-0.25}^{+0.19}$ & 9.54$_{-0.82}^{+0.52}$ & $2.98_{-0.21}^{+0.27}$ & 17$\pm$2 & $<$58 \\
REBELS-27  & 7.090 & $-1.79_{-0.45}^{+0.42}$ & 9.69$_{-0.34}^{+0.25}$ & $2.89_{-0.11}^{+0.27}$ & 20$\pm$4 & 35$_{-13}^{+19}$\\
REBELS-28  & $6.82_{-0.13}^{+0.13}$ & $-1.95_{-0.36}^{+0.29}$ & 8.61$_{-0.51}^{+0.70}$ & $3.26_{-0.20}^{+0.49}$ & 29$\pm$8 & $<$43 \\
REBELS-29\tablenotemark{a}  & 6.685 & $-1.61_{-0.19}^{+0.10}$ & 9.62$_{-0.19}^{+0.19}$ & $2.90_{-0.08}^{+0.12}$ & 25$\pm$3 & 35$_{-14}^{+20}$\\
REBELS-30  & 6.983 & $-1.95_{-0.22}^{+0.15}$ & 9.28$_{-0.61}^{+0.45}$ & $3.06_{-0.13}^{+0.20}$ & 27$\pm$2 & $<$36 \\
REBELS-31  & $6.65_{-0.06}^{+0.10}$ & $-2.27_{-0.33}^{+0.18}$ & 9.21$_{-0.37}^{+0.37}$ & $3.05_{-0.06}^{+0.06}$ & 27$\pm$4 & $<$52 \\
REBELS-32  & 6.729 & $-1.50_{-0.30}^{+0.28}$ & 9.55$_{-0.37}^{+0.35}$ & $3.01_{-0.11}^{+0.11}$ & 14$\pm$2 & 37$_{-17}^{+23}$\\
REBELS-33  & $6.68_{-0.12}^{+0.12}$ & $-2.04_{-0.71}^{+0.24}$ & 9.39$_{-0.51}^{+0.40}$ & $2.90_{-0.11}^{+0.27}$ & 13$\pm$2 & $<$49 \\
REBELS-34  & 6.633 & $-2.02_{-0.15}^{+0.07}$ & 9.33$_{-0.34}^{+0.33}$ & $3.03_{-0.06}^{+0.07}$ & 31$\pm$2 & $<$46 \\
REBELS-35  & $6.98_{-0.10}^{+0.10}$ & $-2.07_{-1.12}^{+0.27}$ & 8.91$_{-0.65}^{+0.66}$ & $3.18_{-0.18}^{+0.30}$ & 31$\pm$3 & $<$47 \\
REBELS-36  & 7.677 & $-2.57_{-0.47}^{+0.48}$ & 9.40$_{-0.95}^{+0.76}$ & $2.99_{-0.16}^{+0.25}$ & 24$\pm$4 & $<$32 \\
REBELS-37  & $7.75_{-0.17}^{+0.09}$ & $-1.24_{-0.27}^{+0.16}$ & 8.58$_{-0.71}^{+0.74}$ & $3.25_{-0.17}^{+0.32}$ & 28$\pm$3 & 74$_{-41}^{+18}$\\
REBELS-38  & 6.577 & $-2.18_{-0.42}^{+0.45}$ & 9.58$_{-1.27}^{+0.74}$ & $3.01_{-0.25}^{+0.35}$ & 18$\pm$4 & 98$_{-35}^{+54}$\\
REBELS-39  & 6.847 & $-1.96_{-0.28}^{+0.30}$ & 8.56$_{-0.57}^{+0.57}$ & $3.58_{-0.37}^{+0.17}$ & 37$\pm$6 & 52$_{-20}^{+30}$\\
REBELS-40  & 7.365 & $-1.44_{-0.36}^{+0.29}$ & 9.48$_{-0.99}^{+0.45}$ & $2.98_{-0.20}^{+0.32}$ & 17$\pm$1 & 35$_{-14}^{+20}$
\enddata
\tablenotetext{*}{Estimated using \textsc{beagle} (Chevallard \&
  Charlot 2016) assuming a constant star formation history
    (CSFH).  See M. Stefanon et al.\ (2022, in prep) for details.
  Note that alternate estimates of the EWs are provided assuming
  a delayed star formation history in Table~\ref{tab:ew} of Appendix
  B.}
\tablenotetext{$\dagger$}{Derived from the measured $UV$ luminosity
    using the prescription given in \S4.2}
\tablenotetext{$\ddagger$}{Derived from the measured $IR$
    luminosity $L_{IR}$ (Inami et al.\ 2022, in prep) using the
    prescription given in \S4.2.  Note that no use of this SFR is made
    for Figure~\ref{fig:mainseq}.  Instead for that figure, SFR$_{IR}$
    is taken to be ($10^{0.19}-1$)SFR$_{UV}$ as found for the average
    luminous $z\sim7$ source by Schouws et al.\ (2021).}
\tablenotetext{a}{Estimates of the stellar mass presented in
    Fudamoto et al.\ (2021) for REBELS-12 and REBELS-29 are higher by
    $\sim$0.7-1.0 dex than those quoted here using \textsc{BEAGLE} and
    assuming a constant star formation history and more in line with
    stellar mass estimates from \textsc{Prospector} relying on
    non-parameter star formation histories (Stefanon et al.\ 2022, in
    prep; Topping et al.\ 2022, in prep).  Estimates for the obscured
    SFR estimates from Fudamoto et al.\ (2021) are almost identical to
    what we present here, while the $UV$ SFRs are $\sim$0.1 dex
    higher.}

\end{deluxetable*}

\begin{figure}
\epsscale{1.17}
\plotone{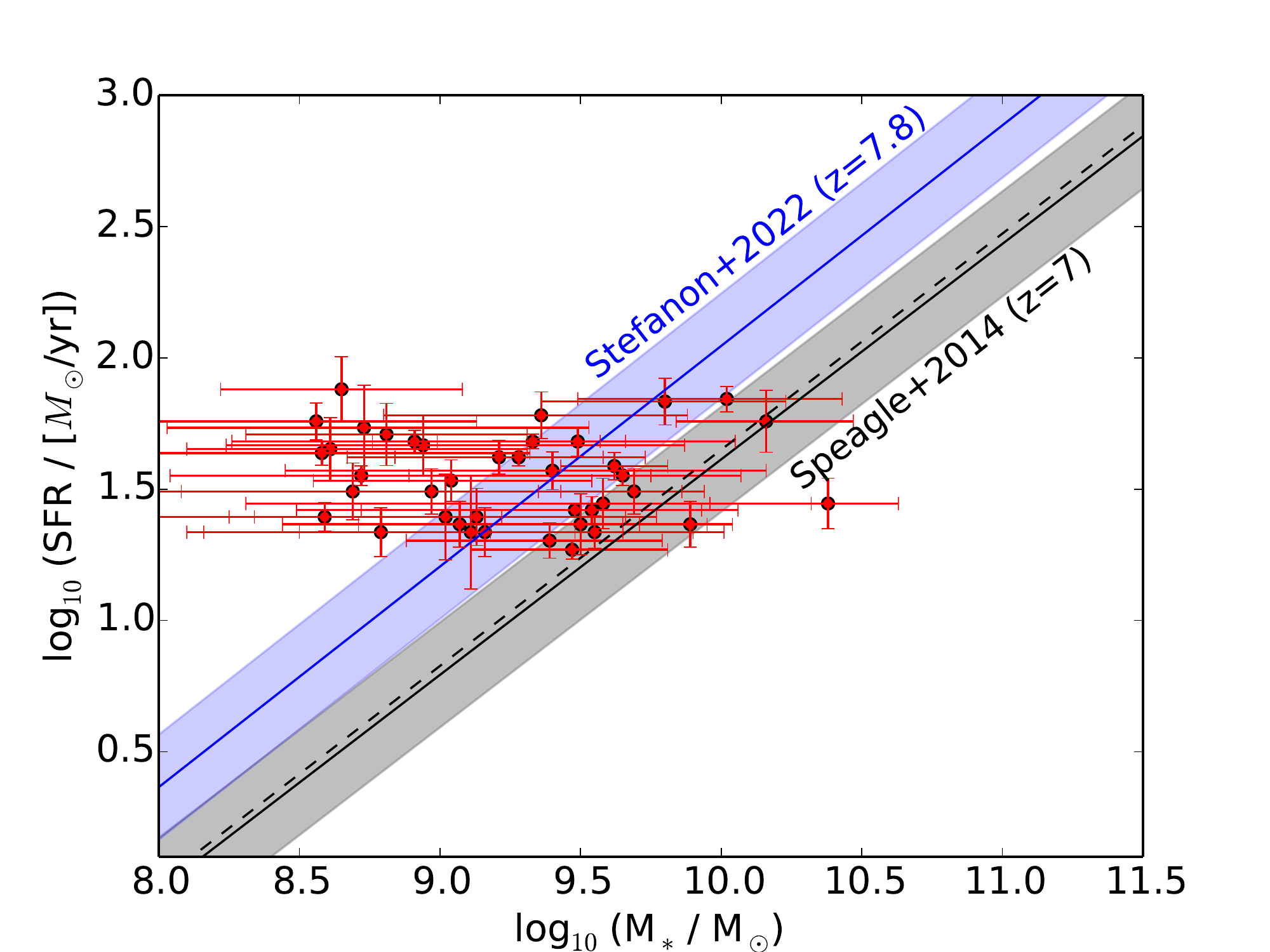}
\caption{Estimated star formation rate vs. stellar mass for the
    40 targets (\textit{solid red circles}) in the REBELS program.
    The shaded gray region shows the star-forming main sequence
    implied by the Spiegel et al.\ (2014) fitting formula results at
    $z\sim7$, while the dashed black line shows the implied main
    sequence at $z\sim7.8$.  The blue shaded region offsets this main
    sequence vertically to match the Stefanon et al.\ (2022) specific
    star formation rate results at $z\sim7.8$.  The stellar masses
    shown in this figure are derived from the REBELS photometry using
  \textsc{beagle} (Stefanon et al.\ 2022, in prep) and assuming a
    constant star formation history.  Meanwhile, the SFRs shown in
    this figure are the unobscured SFRs derived from the rest-$UV$
    data corrected upwards by $\sim$0.19 dex to account for the
    average contribution of the obscured SF (prior to the execution of
    the REBELS program) to the total (as found by Schouws et
    al.\ 2021).  Our assuming a constant star formation history may
    result in a significant underestimate of the stellar masses for
    sources with bursty star formation histories (Topping et
    al.\ 2022, in prep).  An improved version of the presented
    relation will be presented in Topping et al. (2022, in
    prep).\label{fig:mainseq}}
\end{figure}

\subsection{Characteristics of the REBELS Targets}

The final set of 40 targets for the REBELS program is presented in
Table~\ref{tab:targlist}.  Along with the right ascension and
declination for individual targets in our program, we also include the
original source names used in the ALMA observations, $UV$
luminosities, and photometric redshifts.  $UV$-continuum slopes
  $\beta$, stellar mass estimates, [OIII]$_{4959,5007}$+H$\beta$ EWs,
  and SFRs are presented in Table~\ref{tab:targlist2}.  $UV$
  luminosities for sources and photometry are based on flux
  measurements made within 2.4$''$-diameter apertures.  Fiducial
stellar masses and [OIII]$_{4959,5007}$+H$\beta$ EWs for sources in
REBELS are derived using the \textsc{beagle} photometric+spectroscopic
stellar population fitting code (Chevallard \& Charlot 2016), while
our $UV$-continuum slopes $\beta$ are derived based on power-law fits
to the best-fit SED (as performed in Stefanon et al.\ 2019a).  Our
procedures for quantifying the $UV$-continuum slopes $\beta$, stellar
masses, and [OIII]$_{4959,5007}$+H$\beta$ EWs for all 40 sources in
the REBELS sample will be detailed in M. Stefanon et al.\ (2022, in
prep).  Unobscured and obscured SFRs for targets are derived from
  the rest-frame $UV$ luminosities and $IR$ luminosities using
  prescriptions given in \S4.2.

In Figure~\ref{fig:twod}, we present the $UV$ luminosities $M_{UV}$ of
sources in the 40 source REBELS selection vs. their photometric
redshift.  For context, the light blue circles are included to show
the $UV$ luminosities and redshifts of sources from various legacy
fields covering an area of $\sim$0.3 deg$^2$ (Bouwens et al.\ 2021).
The light green hexagons indicate the $UV$ luminosities and redshifts
for other $z>6$ sources with spectroscopic redshift determinations
from ALMA.  The solid and open blue circles are sources with redshift
measurements from $\geq7\sigma$ and $<$7$\sigma$ Ly$\alpha$ lines,
respectively (Vanzella et al.\ 2011; Shibuya et al.\ 2012; Ono et
al.\ 2012; Schenker et al.\ 2012; Pentericci et al.\ 2012; Jiang et
al.\ 2013; Oesch et al.\ 2015; Sobral et al. 2015; Zitrin et
al.\ 2015; Song et al.\ 2016; Stark et al.\ 2017; Hoag et al.\ 2017;
Larson et al.\ 2017; Pentericci et al.\ 2018; Fuller et al.\ 2020;
Jung et al.\ 2020; Endsley et al.\ 2021; Pelliccia et al.\ 2021;
Laporte et al.\ 2021).

Figure~\ref{fig:mainseq} shows the approximate SFRs of targets in
  the REBELS program vs. their inferred stellar masses (\textit{solid
    red circles}).  These results are shown (\textit{shaded gray
    region}) relative to the main sequence at $z\sim7$ implied by the
  Speagle et al.\ (2014) star-forming main sequence fitting formula
  $\log_{10} SFR(M_{*}, t) = (0.84 \pm 0.02 - (0.026 \pm
  0.003)t)\log_{10} M_{*} - (6.51 \pm 0.24 - (0.11 \pm 0.03) t)$ where
  $t$ is the age of the universe in Gyr.  Also shown for comparison is
  the main sequence relation shifted to match the specific star
  formation results from Stefanon et al.\ (2022) at $z\sim7.8$
  (\textit{shaded blue region}).  Overall the REBELS targets are
  distributed both above and below the main sequence relation
  (particularly that derived by Stefanon et al.\ 2022).  Our assuming
  a constant star formation history may result in a significant
  underestimate of the stellar masses for sources with bursty star
  formation histories (Topping et al.\ 2022, in prep).  Topping et
  al.\ (2022, in prep) will provide an improved discussion of where
  the REBELS targets fall on the SFR vs. stellar mass relation.

Figure~\ref{fig:mass} from Appendix B shows the distribution of the
REBELS targets in stellar mass and compare this distribution with that
inferred (Faisst et al.\ 2020) for the $z=4$-6 ALPINE sample.  The
REBELS sample spans a very similar range in stellar mass as ALPINE.
The median stellar mass for the REBELS sample inferred from
  \textsc{beagle} and \textsc{Prospector} stellar population fitting
  codes (assuming constant and non-parametric star formation
  histories, respectively) is $10^{9.25}$ $M_{\odot}$ and $10^{9.79}$
  $M_{\odot}$, $\sim$0.4 dex lower and $\sim$0.1 dex higher than
  inferred for ALPINE.  Stellar mass estimates for sources in our
  sample from \textsc{Prospector} will be presented in Stefanon et
  al.\ (2022, in prep).  Thanks to the similarities between the
samples, ALPINE provides us with a convenient $z=4$-6 comparison
sample for assessing evolution in the galaxy population with redshift.

\begin{figure}
\epsscale{1.19}
\plotone{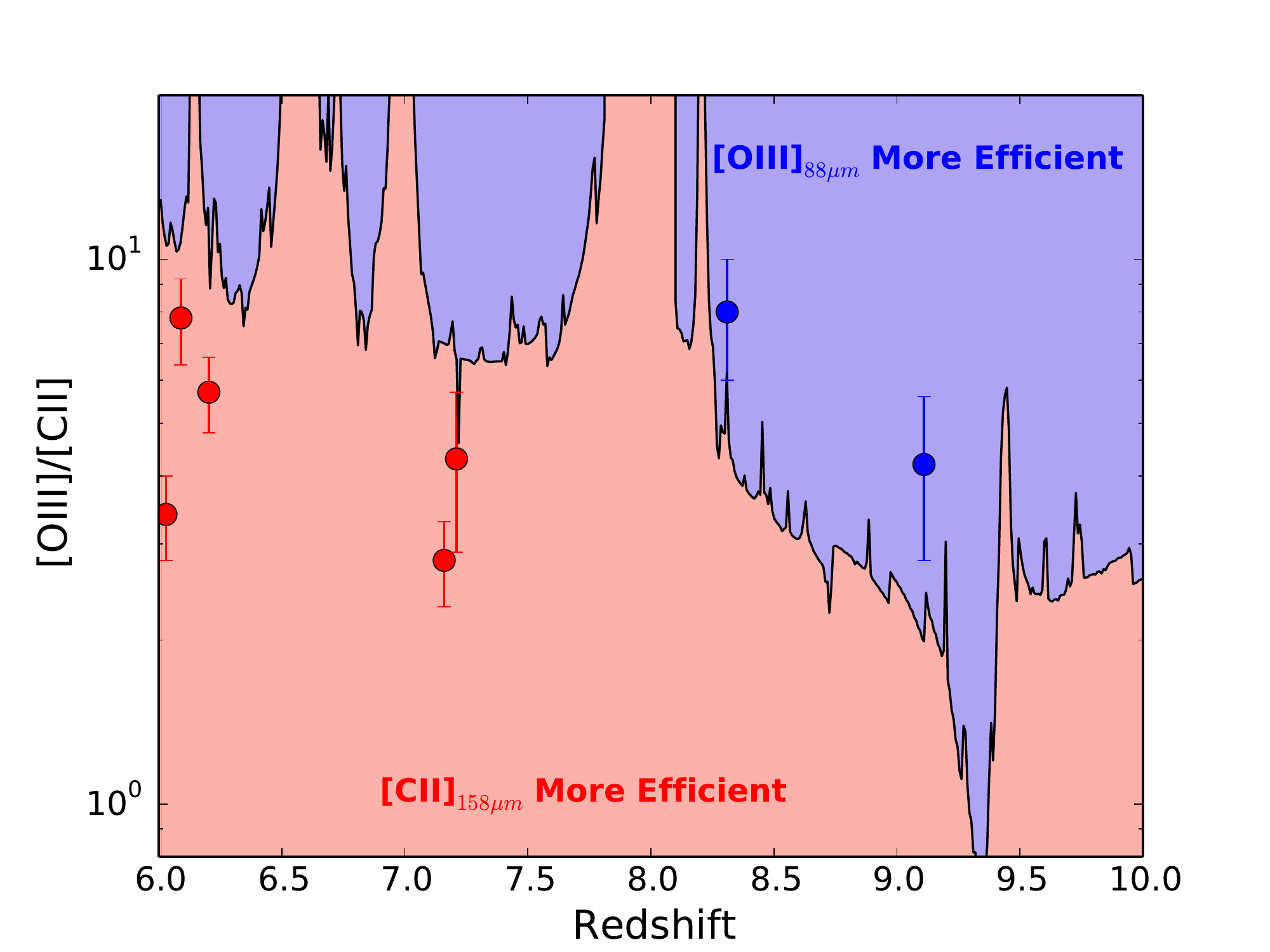}
\caption{The break-even \oiii/\cii$\,$luminosity ratio where
  \cii$\,$and \oiii$\,$scans are equally efficient to execute (\S3.1).
  The plotted red and blue circles show the measured
  \oiii/\cii$\,$luminosity ratios for specific high-redshift sources
  (Hashimoto et al.\ 2019; Bakx et al.\ 2020; Carniani et al.\ 2020;
  Harikane et al.\ 2020) featuring $\geq$5$\sigma$ line detections and
  where the emission is largely co-spatial (but see Carniani et
  al.\ 2017).  For luminosity ratios in the red regions, \cii$\,$scans
  are more efficient, while for luminosity ratios in the blue regions,
  \oiii$\,$scans are more efficient.  The breakeven ratios are
  computed based on the integration times required to detect
  \cii$\,$and \oiii$\,$lines at a given redshift and rely on the
  atmospheric transmission at a given frequency using the equations in
  the ALMA Technical Handbook (Cortes et al.\ 2020).  Use of the
  \cii$\,$line for spectral scans seems to be more efficient for
  galaxies at $z=6.5$-8.2, while use of the \oiii$\,$line for spectral
  scans at $z>8.5$ seems more efficient.\label{fig:oiiicii}}
\end{figure}

The distribution of REBELS targets in redshift, $UV$ luminosity,
$UV$-continuum slope $\beta$, and [OIII]$_{4959,5007}$+H$\beta$ EW is
illustrated in Figure~\ref{fig:paramdist} of Appendix B.  The latter
two distributions appear to be completely consistent to that found for
the $z\sim7$ galaxy population as a whole, suggesting that results
derived from REBELS should be representative of the general population
of massive star forming galaxies at $z>6.5$.  Despite the similar
  stellar mass characteristics of the two samples, REBELS does not
  include targets that extend as faint in $UV$ luminosities as ALPINE
  does (as Figure~\ref{fig:twod} makes clear).

\section{Observational Implementation of REBELS Survey}

\subsection{Choice of ISM Cooling Line for Spectral Scans}

Galaxies with the most luminous ISM reservoirs shine very brightly in
both the 157.74$\mu$m [CII] and 88.36$\mu$m [OIII] ISM cooling lines
and both lines are readily detectable with ALMA over a significant
fraction of the redshift range between $z\sim6$ and $z\sim10$.  As a
result, ALMA has already been successful in scanning for the
\cii$\,$line in five sources at $z\sim6$-7 (Smit et al.\ 2018;
Schouws et al.\ 2022) as well as uncovering serendipitous galaxies in
the immediate neighborhood of $z\sim6$-7 QSOs (Decarli et al.\ 2017;
Venemans et al.\ 2019, 2020).  Similarly, spectral scans for the
\oiii$\,$line have been shown to allow for efficient redshift
determinations for galaxies at $z>8$ (e.g., Hashimoto et al.\ 2018;
Tamura et al.\ 2019).

\begin{figure*}
\epsscale{1.17}
\plotone{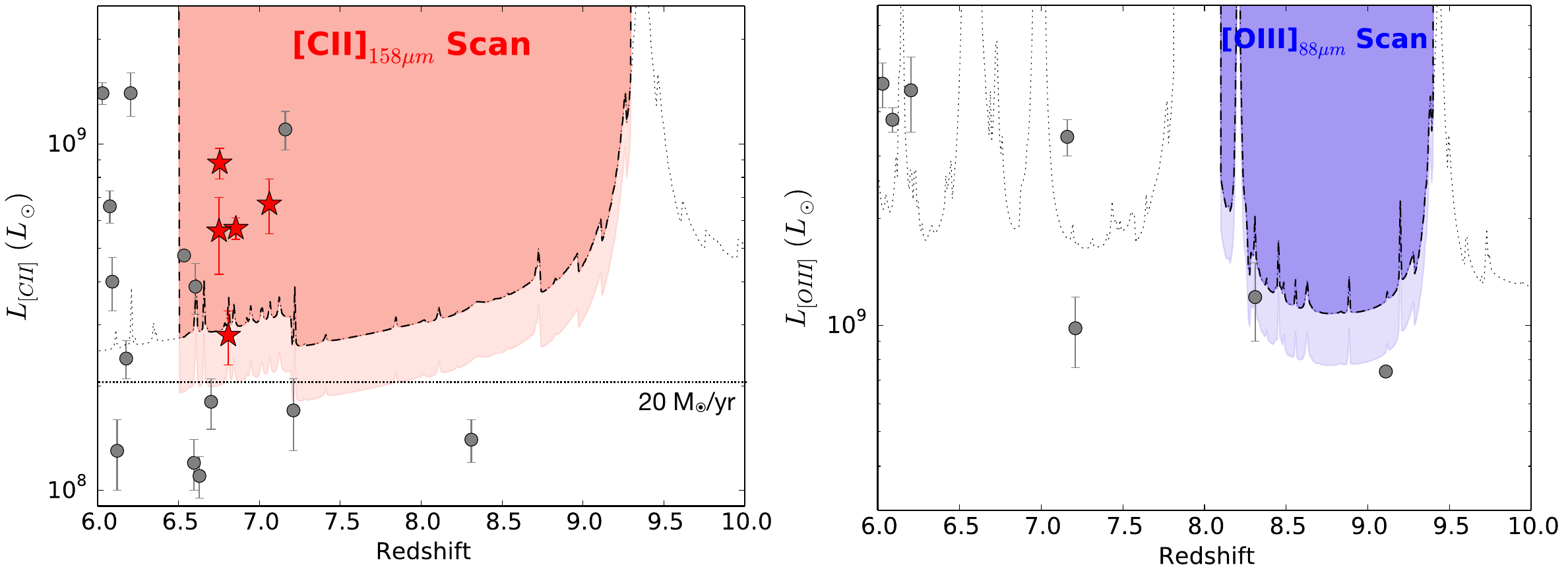}
\caption{The shaded regions indicate the parameter space we are
  scanning for \cii$\,$and \oiii$\,$in our 36 targeted $z\sim6.5$-8.5
  and 4 $z\sim 8.2$-9.4 galaxies (assuming a 5$\sigma$ detection
  threshold and the specified sensitivity).  The dotted line is
  computed using the equations in the ALMA Technical Handbook (Cortes
  et al.\ 2020), as in Figure~\ref{fig:oiiicii}.  The lighter
    shaded regions show the luminosities probed for the more typical
    sensitivity reached in the REBELS observations.  For context,
  \cii$\,$and \oiii$\,$detections (left and right panels,
  respectively) from the pilot program (\textit{red and blue stars}:
  Smit et al.\ 2018; Schouws et al.\ 2022) and the literature
  (\textit{gray circles}) are also included here.  The approximate
  limiting luminosity achieved in a typical \cii$\,$scan
  (for a $5\sigma$ detection threshold) is 20 M$_{\odot}$/yr
  (\textit{dotted horizontal line}) assuming the De Looze et
  al.\ (2014) $z\sim0$ \cii-SFR relation.\label{fig:scan}}
\end{figure*}

As such, we could potentially target either transition in searching
for bright ISM cooling lines in $z>6$ galaxies.  To help decide which
line would be more efficient, we calculated the approximate break-even
luminosity \loiii/\lcii$\,$ratio for \oiii$\,$vs. \cii$\,$spectral
scans as a function of redshift.  In calculating this ratio, we
expressly made use of the atmospheric transmission, using Eq. (9.8)
from the ALMA Technical
Handbook\footnote{https://almascience.eso.org/documents-and-tools/cycle8/alma-technical-handbook}
and the specified inputs to this equation from chapter 9 of the
handbook to derive sensitivities for a given integration time (Cortes
et al.\ 2020).  Observations were assumed to be conducted in
precipitible water vapor (PWV) octile consistent with that shown in
Figure 9.1 of the ALMA Technical Handbook, i.e., with band 6
observations being conducted in 5th octile weather and band 7
observations being conducted in generally better weather but depending
on the observational frequency. As a check on the results from our
sensitivity calculations, use was made of the ALMA sensitivity
calculator.  Also, account was made for the line scans with
\oiii$\,$covering a 1.8$\times$ smaller range in $\Delta z$ than
\cii$\,$ (due to the higher frequency of \oiii).  The FWHM of the
\cii$\,$and \oiii$\,$lines was assumed to be the same, i.e., 250 km/s,
for this calculation.  Such a FWHM is fairly typical for ISM-cooling
lines found for $UV$-bright $z>6$ galaxies (e.g., Matthee et
al.\ 2019; Harikane et al.\ 2020; Schouws et al.\ 2021) and is
consistent with theoretical expectations (Kohandel et al.\ 2019,
2020).  We emphasize that other FWHMs should yield essentially
identical results.

The results are shown in Figure~\ref{fig:oiiicii}.  For the typical
luminosity ratios observed for $z\sim 6$-8 galaxies, i.e., $\sim$4
(e.g., Inoue et al.\ 2016, Carniani et al.\ 2020; Bakx et al.\ 2020;
Harikane et al.\ 2020), not only is \cii$\,$clearly the most efficient
line to use for our spectral scans in the redshift range $z\sim 6.0$
to $z\sim 8.5$, but also is not subject to significant gaps in
redshift coverage.  For sources at $z>8.5$, by contrast, \oiii$\,$is
the most efficient line to use for spectral scans.  \oiii$\,$scans do
become significantly less efficient at a redshift $z\sim9.4$ due to
the H$_2$O$\, 5_{15}-4_{22}$ line in the earth's atmosphere at 325
GHz, but \cii$\,$is also not especially efficient at that redshift due
to the H$_2$O$\, 3_{13}-2_{20}$ line at 183 GHz.

\subsection{Set Up of the Spectral Scan Windows}

Here we describe our stategy for setting up the tunings for our
spectral scans.  For each target, we aim to optimally cover the
redshift likelihood distribution derived combining the results from
three independent sets of photometry (Figure~\ref{fig:pz}).

For bright sources in the redshift range $z=6.5$-7.2, the redshifts of
sources can be accurately estimated thanks to the Lyman-break and
high-EW [OIII]$_{4959,5007}$ doublet lying very close to edge of
multiple spectral elements ($z_{Subaru}$, $z_{HSC}$, $Y_{VISTA}$,
$y_{HSC}$, [3.6], and [4.5]) with deep coverage over the
COSMOS/UltraVISTA, VIDEO/XMM-LSS, and UKIDSS/UDS fields.  For
$z=6.5$-7.2 sources over the COSMOS/UltraVISTA field, the median
$\Delta z$ uncertainty is $\pm$0.11 while over the VIDEO/XMM-LSS +
UKIDSS/UDS fields, the median $\Delta z$ uncertainty is $\pm$0.15.
For most $z=6.5$-7.2 sources over the COSMOS/UltraVISTA field, we find
that we can use \cii$\,$scans to cover $\sim$89\% of the likelihood
distribution with two redshift tunings, each covering a contiguous
5.375 GHz frequency range (and a 10.75 GHz contiguous range in total).
REBELS-08 in Figure~\ref{fig:scan0} of Appendix A is one example of a
source where we used this tuning strategy.

From $z=7.2$ to $z=7.7$, the redshifts of sources are largely
constrained based on the position of the Lyman-break in the $Y$ band
filter.  Nevertheless, since there are fewer passbands with flux
measurements providing direct constraints on the redshifts of sources,
the median uncertainty on the redshifts of sources is larger, with a
value of $\pm$0.15.  For sources in this redshift range, we find that
we can use \cii$\,$scans to cover the 91\% of the redshift likelihood
distribution in 3 tunings covering a contiguous frequency range of
20.375 GHz (5.375 + 7.5 + 7.5 GHz).  REBELS-01 in
Figure~\ref{fig:scan0} of Appendix A is one example of a source where
we used this tuning strategy.

For sources with photometric redshifts in excess of 7.7, the redshift
uncertainties increase substantially due to the Lyman break falling
between the ground-based Y and J bands.  For these sources, the
redshift likelihood distribution typically extended from z$\sim$7.7 to
9.3, with a median uncertainty of $\pm$0.44.  To execute a spectral
scan for ISM cooling line over this range, we made use of
\cii$\,$scans in cases where such a scan had already begun as part of
a cycle-6 program (2018.1.00236.S, PI: Stefanon) probing the dust
continuum.  Otherwise, sources required six tunings in band 7 to
search for \oiii$\,$(probing the redshift range $z=8.10$ to
9.39) and approximately three tunings in band 6 to search
for \cii$\,$(probing the range $z=7.37$ to 8.00).\footnote{Because of some adjustments made to our program
after submission, there was insufficient time to scan the redshift
range $z=7.7$ to 8.1 for REBELS-11 and REBELS-13.}

We present the spectral scan windows we use for the 40 targets in the
REBELS LP in Figures~\ref{fig:scan0} and \ref{fig:scan1} of Appendix
A.  In total, 16 of the targets from the program required 2 tunings to
cover the redshift likelihood distribution, 17 targets required 3
tunings, 1 target required 4 tunings, 2 targets required 5 tunings, 1
target required 6 tunings, and 1 target required 8 tunings.  For 2
targets, only 1 tuning was allocated to extend scans which already
started in our second pilot program (Schouws et al.\ 2022).  In
total, the number of targets $\times$ tuning windows for the REBELS
program is equal to 91 for the \cii$\,$line and 22 for the
\oiii$\,$line.  Archival ALMA observations contributed an additional
12 tunings to our \cii$\,$scans (see Appendix A).

\begin{figure*}
\epsscale{1.19}
\plotone{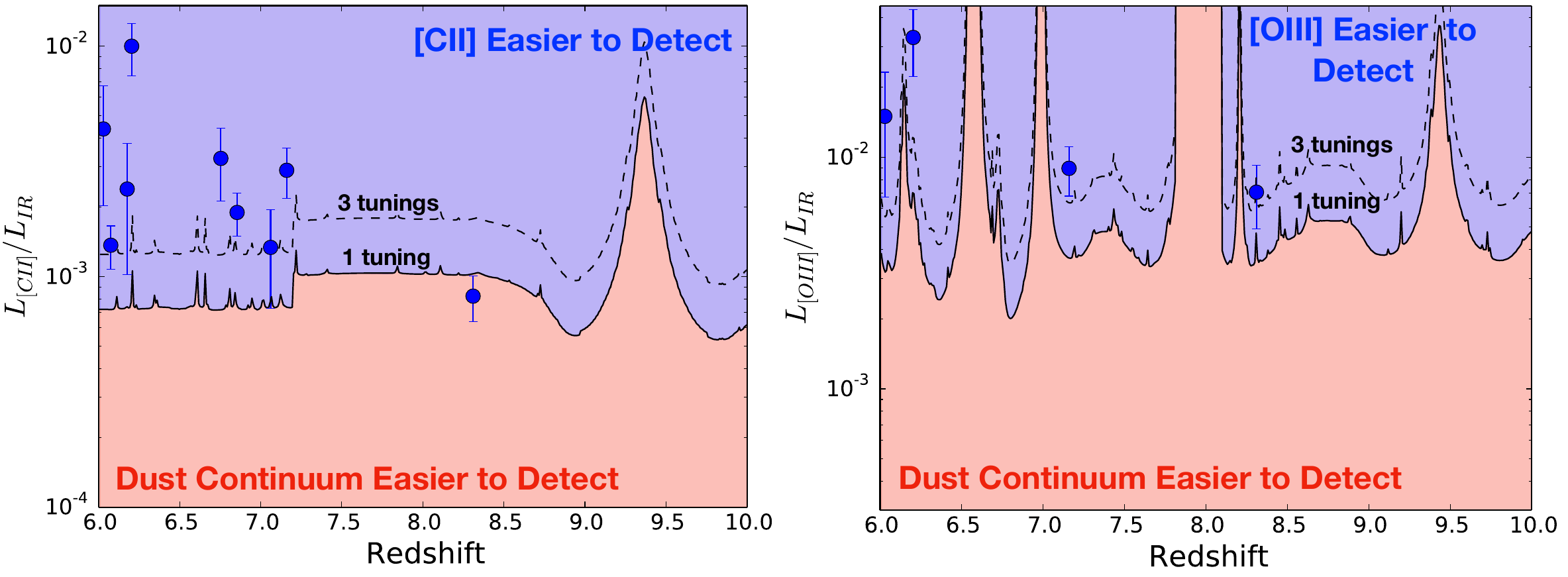}
\caption{The break-even \cii/IR and \oiii/IR luminosity ratio
  (\textit{left and right panels, respectively}) for both detection
  ($5\sigma$) of the \cii$\,$and \oiii$\,$lines and detection
  ($3.3\sigma$) of the dust continuum shown as a function of
  redshift.  Also shown on this diagram are $z\sim6$-10 sources from
  the REBELS pilot programs (Smit et al.\ 2018; Schouws et al.\ 2021,
  2022) and the literature (Hashimoto et al.\ 2019; Tamura et
  al.\ 2019; Bakx et al.\ 2020; Harikane et al.\ 2020).  The dashed
  lines show the break-even ratios assuming three separate tunings for
  a spectral scan.  The use of $>$3 tunings would shift the break-even
  ratio to even higher values.  For the typical source, one can
  clearly consider up to 6 tunings per source before the line
  sensitivity of ALMA becomes the limiting factor in probing both line
  and continuum emission.  This fits very well with the spectral scan
  strategy employed in REBELS where the aim is to probe both lines and
  the dust continuum.\label{fig:tdust_tline}}
\end{figure*}

\subsection{Sensitivity Requirements for Spectral Scans}

Our sensitivity requirements for REBELS relies on our experience with
searches in our pilot programs (Smit et al.\ 2018; Schouws et
al.\ 2022). There the detected \cii$\,$lines had peak fluxes of
1.5-4.0 mJy.  To guarantee the selection of similar sources at $z\sim
7$ with REBELS, we required that the peak flux sensitivity be
$\sim$340$\mu$Jy in a 66 km/s channel such that sources with a peak
flux of 1 mJy can be detected at 5$\sigma$ when combining multiple
channels.  This is equivalent to a $5\sigma$ limiting
point-source luminosity of $\sim3\times 10^{8} L_{\odot}$
at $z\sim 7$ (assuming a line width of 250 km/s), which requires
$\sim$20 minutes of integration time per tuning.  Reaching the same
limiting luminosity at $z\sim 8$ requires a 300$\mu$Jy sensitivity and
$\sim$30-40 minutes of integration time in band 5 or 6 (for $z > 8$
and $z< 8$ \cii$\,$searches, respectively). In scanning for the bright
\oiii$\,$lines in our $z\sim 9$ targets, the equivalent sensitivity
requirement is 450$\mu$Jy assuming an \oiii/\cii$\,$luminosity ratio
of 3.5.

For the purposes of illustration, the limiting \cii$\,$and \oiii
luminosities probed as a function of source redshift for the spectral
scans planned for sources in REBELS is provided in
Figure~\ref{fig:scan}.  This is for the requisite sensitivity
  specified for the REBELS program.  In practice, the sensitivity
  achieved is typically $\sim$1.4$\times$ better than specified
  (Schouws et al.\ 2022, in prep), allowing us to detect even lower
  luminosity ISM cooling lines (indicated with the lighter shading in
  Figure~\ref{fig:scan}).  For context, we have included the \cii
luminosities and redshifts of detected sources from the literature
(\textit{gray circles}) and our pilot programs (\textit{filled
  stars}).  As should be clear from the figure, REBELS will detect
line emission from sources if their \cii$\,$luminosities exceed
$\sim 3\times 10^8$ $L_{\odot}$ and $\sim2\times10^8$ $L_{\odot}$
  (requested and typical sensitivities, respectively) and, for $z>8.5$
targets, if their \oiii luminosities exceed $\sim 1.1\times 10^9
  L_{\odot}$ and $0.8\times 10^9 L_{\odot}$ (required and typical
  sensitivities, respectively).

To maximize the sensitivity of the REBELS spectral scan observations
and not overresolve the \cii$\,$line, observations were conducted in
the lowest spatial resolution configurations (C43-1 and C43-2), with
$\sim$1.2-1.6$''$ FWHM.  The REBELS observations were obtained in
frequency domain mode (FDM) at a spectral resolution of 488 MHz and
then spectrally averaged in bins of size 16, giving the output data a
spectral resolution of 7.813 MHz, equivalent to a velocity resolution
of $\sim$9 km/s for \cii$\,$line at $z\sim7$.  Given that \cii lines
in luminous $z\sim7$ galaxies have been found to have a minimum FWHM
of $\sim$80 km/s and more typically 250 km/s, this was expected to be
more than sufficient to study the kinematic structure of sources
revealed by the program.

63\% of the $z\sim 7$ galaxies targeted with observations from our
pilot programs yielded lines with luminosities in excess of
$2\times10^8$ $L_{\odot}$.  Given that similar selection criteria are
used for the REBELS large program, we would expect a similar detection
rate of \cii$\,$for the REBELS program, suggesting we will detect
lines in 25 out of 40 targets.  Combining the expected results with
previous \cii$\,$and \oiii$\,$line detections, we expected $\geq$35
$z>6.5$ galaxies with ALMA line detections once the program was
completed.  This is a sufficient number for a detailed physical
characterization of massive galaxies at $z\gtrsim 6.5$ and a study of
their evolution.

\subsection{Dust-Continuum Sensitivities}

The detection of dust-continuum emission from star-forming galaxies at
$z\sim5$-8 with ALMA has been found to be much more difficult than
detection of the \cii$\,$or \oiii$\,$ISM cooling lines.  This basic
difference in detectability was already evident in $z\sim5.5$ results
by Capak et al.\ (2015), who were able to detect all 10 luminous
$z=5.2$-5.7 galaxies they targeted in \cii, but only able to detect 4
of the 10 sources in the dust continuum.  Other results available on
$z>5$ galaxies have been similar, with \cii$\,$being detected in a
much larger fraction of sources than the dust continuum (e.g., Willott
et al.\ 2015; Maiolino et al.\ 2015; Inoue et al.\ 2016; Matthee et
al.\ 2017, 2019; Harikane et al.\ 2020; B{\'e}thermin et al.\ 2020).

The relative detectability of ISM cooling lines like \cii$\,$or
\oiii$\,$and the dust continuum can be quantified on the basis of
their measured \lcii-to-$L_{IR}$ ratios or \loiii-to-$L_{IR}$ ratios.
Compiling previous results from Capak et al.\ (2015), ALPINE (Le Fevre
et al.\ 2020; B{\'e}thermin et al.\ 2020; Faisst et al.\ 2020), Smit
et al.\ (2018), Schouws et al.\ (2021, 2022), we can calculate the
relative integration time required to detect sources in the dust
continuum and the integration time required to detect sources in \cii.
In performing this calculation, we assume a line width of 235 km/s
(FWHM) for both the \cii$\,$and \oiii$\,$lines, and we assume a
modified blackbody form for the dust-continuum SED with a dust
temperature of 50 K and dust emissivity index $\beta$ of 1.6, which
lies intermediate between the lowest and highest dust temperature
measurements at $z>7$ (e.g., Knudsen et al.\ 2017; Bakx et al.\ 2020).
We present results in Figure~\ref{fig:tdust_tline}.

Given the much longer exposure times required to detect sources in the
dust continuum than in either \cii$\,$or \oiii, our use of multiple
tuning windows for spectral scans really is an advantage in allowing
us to probe the dust continuum.  We have illustrated the approximate
IR luminosities we are able to probe with the REBELS LP in
Figure~\ref{fig:lir}.  Two spectral scan windows (10.75 GHz bandwidth)
are assumed at $z=6.5$-7.2, three spectral scan windows (20.375 GHz
bandwidth) at $z=7.2$-7.7, and six spectral scan windows (45 GHz
bandwidth) at $z>7.7$.

The REBELS dust-continuum probe also allows for a very valuable
assessment of incompleteness in our spectral scan results.  This is
because incompleteness can arise as a result of (1) scan range not
extending over a wide enough range in frequency to find the relevant
ISM cooling line or (2) the relevant ISM cooling line being fainter
than the $5\sigma$ detection limit for the spectral scan.  Given the
strong correlation between dust-continuum luminosities of galaxies and
the luminosities of \cii$\,$and \oiii, the detection of the dust
continuum in a source strongly suggests that the associated ISM
cooling is sufficiently bright to have been detected in our spectral
scans.  If the ISM line is not found but the dust continuum is, it
strongly suggests the spectral scan did not extend broadly enough in
frequency.  In cases where neither the line nor the dust continuum is
detected, it may mean that the line is fainter than the sensitivity
limits of the scans (or in the worst case at lower redshift, but our
careful selection suggests that the number of such targets in the
REBELS program is small).

\subsection{Summary and Execution of Program}

The REBELS LP observations began on November 15, 2019 when ALMA was in
the C43-2 configuration and continued until January 10, 2020 while
ALMA was in configurations C43-1 and C43-2.  60.6 hours of ALMA
observations have thus far been acquired, with 8 hours remaining to be
observed.

34 targets from the program have now been fully observed.
Observations are still incoming for REBELS-04, REBELS-06, REBELS-11,
REBELS-16, REBELS-24, and REBELS-37.  Two of the sources with
incoming observations are part of our 33-target $z=6.5$-7.7 sample.
The remaining four are part of our 7-target $z=7.7$-9.4 sample.  The
REBELS spectral scans for our $z=6.5$-7.7 and $z=7.7$-9.4 targets are
therefore 94\% and 43\% complete, respectively.

Following the first year of observations from the REBELS program and
successful detection of \cii$\,$in REBELS-18 and REBELS-36, several
tunings from REBELS-18 and REBELS-36 were shifted to REBELS-06,
REBELS-16, and REBELS-37 to extend the spectral scan range for
\cii$\,$in those sources.

The remaining observations appear likely to be executed in March 2022
when the configuration of ALMA is in C43-1 or C43-2 according to the
current JAO schedule.

\section{Spectral Scan Results}

The purpose of this section is to provide a quick overview of some of
the most prominent ISM cooling line detections obtained thus far with
the REBELS LP observations and to assess how effective the
observational strategy has been.

\begin{figure}
\epsscale{1.17}
\plotone{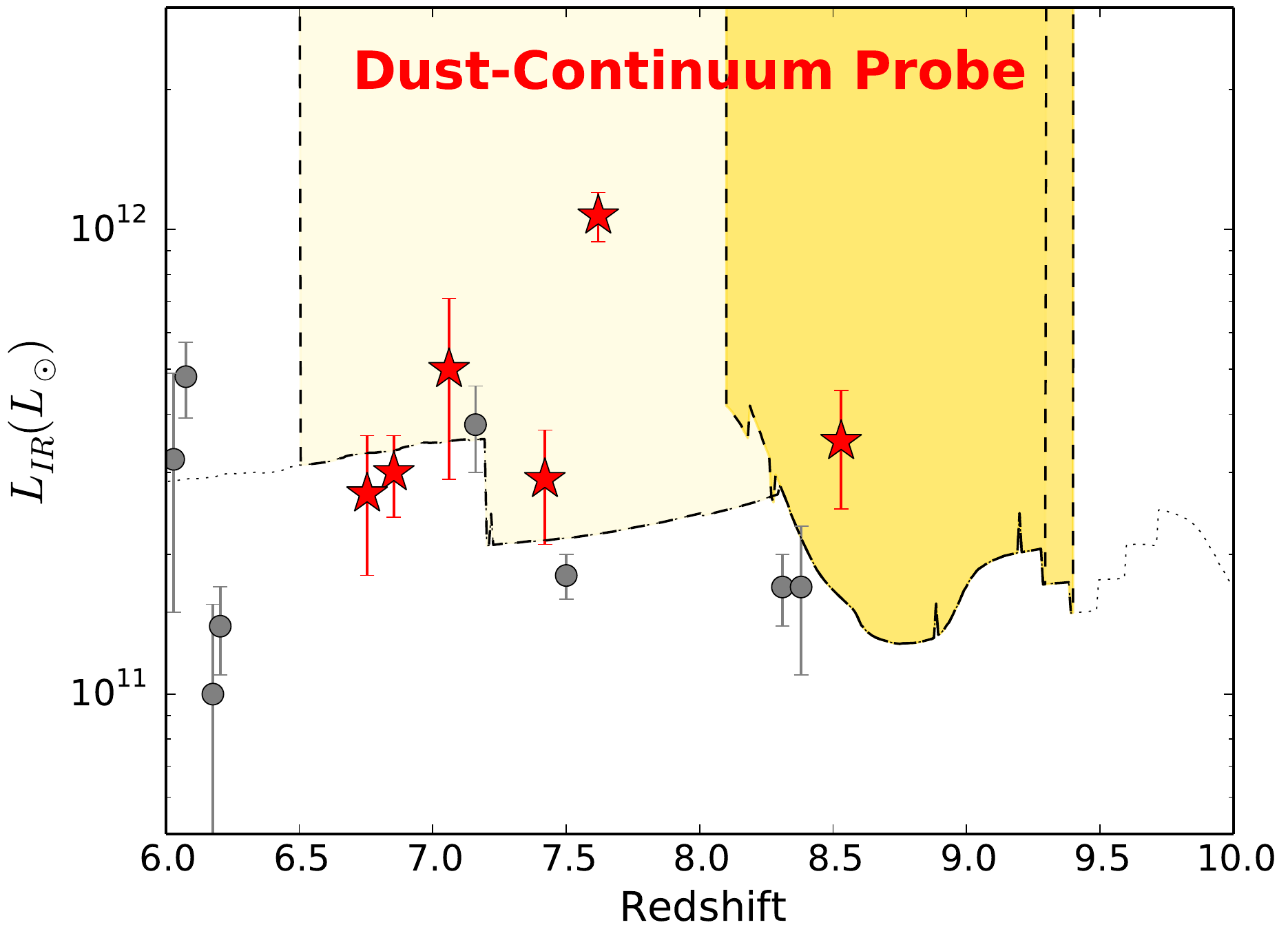}
\caption{The $3.3\sigma$ limiting IR luminosities probed by the REBELS
  scans for \cii$\,$and \oiii$\,$in 40 very bright reionization-era
  galaxies (\textit{light yellow} and \textit{gold} shaded regions,
  respectively).  The red stars show the $IR$ luminosities of six
  $z=7$-8 galaxies from the REBELS pilot programs (Schouws et
  al.\ 2021; Smit et al.\ 2018) while the grey circles show the $IR$
  luminosities inferred for various dust detected galaxies in the
  literature.  REBELS will detect sources in the dust continuum if
  their IR luminosities are greater than $3\times 10^{11} L_{\odot}$
  to $z\sim7.2$ and down to $2\times10^{11}$ $L_{\odot}$ at
  $z=7.2$-9.4 (equivalent to obscured SFRs of 36 M$_{\odot}$/yr and 24
  $M_{\odot}$/yr, respectively).  Extrapolating our pilot results, the
  REBELS program is projected to increase the number of dust
  detections by 3-4$\times$ at $z>6.5$ and will allow us to test
  whether large dust reservoirs are common (as suggested by our
  detections and also the Watson et al.\ (2015) and Laporte et
  al.\ (2017) sources at $z\sim 7.5$ and $z=8.38$, respectively) or
  whether there is a rapid decline in IR bright galaxies at
  $z>7.5$.\label{fig:lir}}
\end{figure}

\subsection{Processing of the ALMA Data and Initial Results}

Here we provide a very brief summary of the procedures used to reduce
and calibrate ALMA observations from the REBELS program.  More details
will be given in S. Schouws et al.\ (2022, in prep) and H. Inami et
al.\ (2022, in prep).

\begin{figure*}
\epsscale{1.18}
\plotone{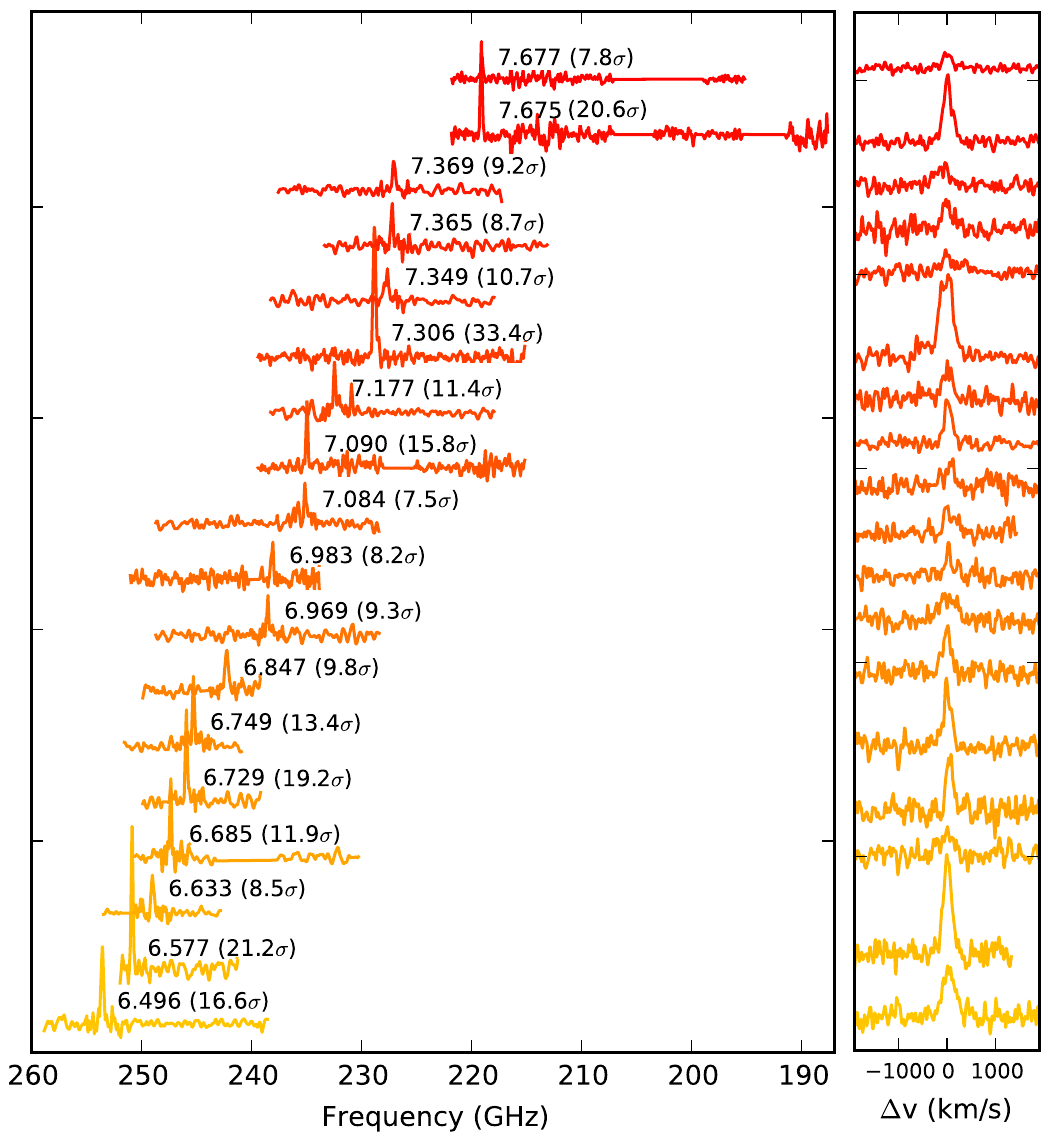}
\caption{(\textit{left}) Illustration of the highest significance
  ($\geq$7$\sigma$) ISM-cooling lines identified to-date in the REBELS
  program and the band-5/6 spectral scans used to locate these lines
  (\S4.1).  Shown on the figure next to the \cii$\,$line detections
  are the systemic redshifts of the sources as well as an estimate of
  the significance of the line detection.  (\textit{right}) Velocity
  structure of the same line detections as shown in the left panel.
  More details on these line detections and on the characteristics of
  even fainter, lower S/N line detections in the REBELS first year
  data will be presented in S. Schouws et al.\ (2022, in
  prep).\label{fig:lines}}
\end{figure*}

\begin{deluxetable}{cccccc}
\tablecaption{Discovered $\geq 7\sigma$ \cii$\,$lines in the first-year data from the REBELS program\label{tab:lines}}
\tablehead{\colhead{REBELS} & \colhead{$\nu_{[CII]}$} & \colhead{} & \colhead{} & \colhead{}\\
\colhead{ID} & \colhead{(GHz)} & \colhead{$z_{[CII]}$} & \colhead{S/N$_{[CII]}$\tablenotemark{$\dagger$}} & \colhead{S/N$_{continuum}$\tablenotemark{$\dagger$}}}
\startdata
REBELS-01 & 232.42 & 7.177 & 11.4 & $<$3.3\\
REBELS-03 & 238.49 & 6.969 & 9.3 & $<$3.3\\
REBELS-05 & 253.53 & 6.496 & 16.6 & 5.5\\
REBELS-08 & 245.25 & 6.749 & 13.4 & 5.0\\
REBELS-12 & 227.65 & 7.349 & 10.7 & 3.4 \\
REBELS-14 & 235.09 & 7.084 & 7.5 & 5.1 \\
REBELS-18 & 219.08 & 7.675 & 20.6 & 4.9 \\
REBELS-19 & 227.09 & 7.369 & 9.2 & 3.4 \\
REBELS-25 & 228.80 & 7.306 & 33.4 & 17.4 \\
REBELS-27 & 234.94 & 7.090 & 15.8 & 5.3 \\
REBELS-29 & 247.31 & 6.685 & 11.9 & 4.0 \\
REBELS-30 & 238.08 & 6.983 & 8.2 & $<$3.3\\
REBELS-32 & 245.89 & 6.729 & 19.2 & 4.0 \\
REBELS-34 & 248.98 & 6.633 & 8.5 & $<$3.3\\
REBELS-36 & 219.02 & 7.677 & 7.8 & $<$3.3\\
REBELS-38 & 250.83 & 6.577 & 21.2 & 7.1 \\
REBELS-39 & 242.19 & 6.847 & 9.8 & 5.2\\
REBELS-40 & 227.20 & 7.365 & 8.7 & 3.6
\enddata
\tablenotetext{$\dagger$}{Based on the peak flux.}
\end{deluxetable}

\begin{figure*}
\epsscale{1.05}
\plotone{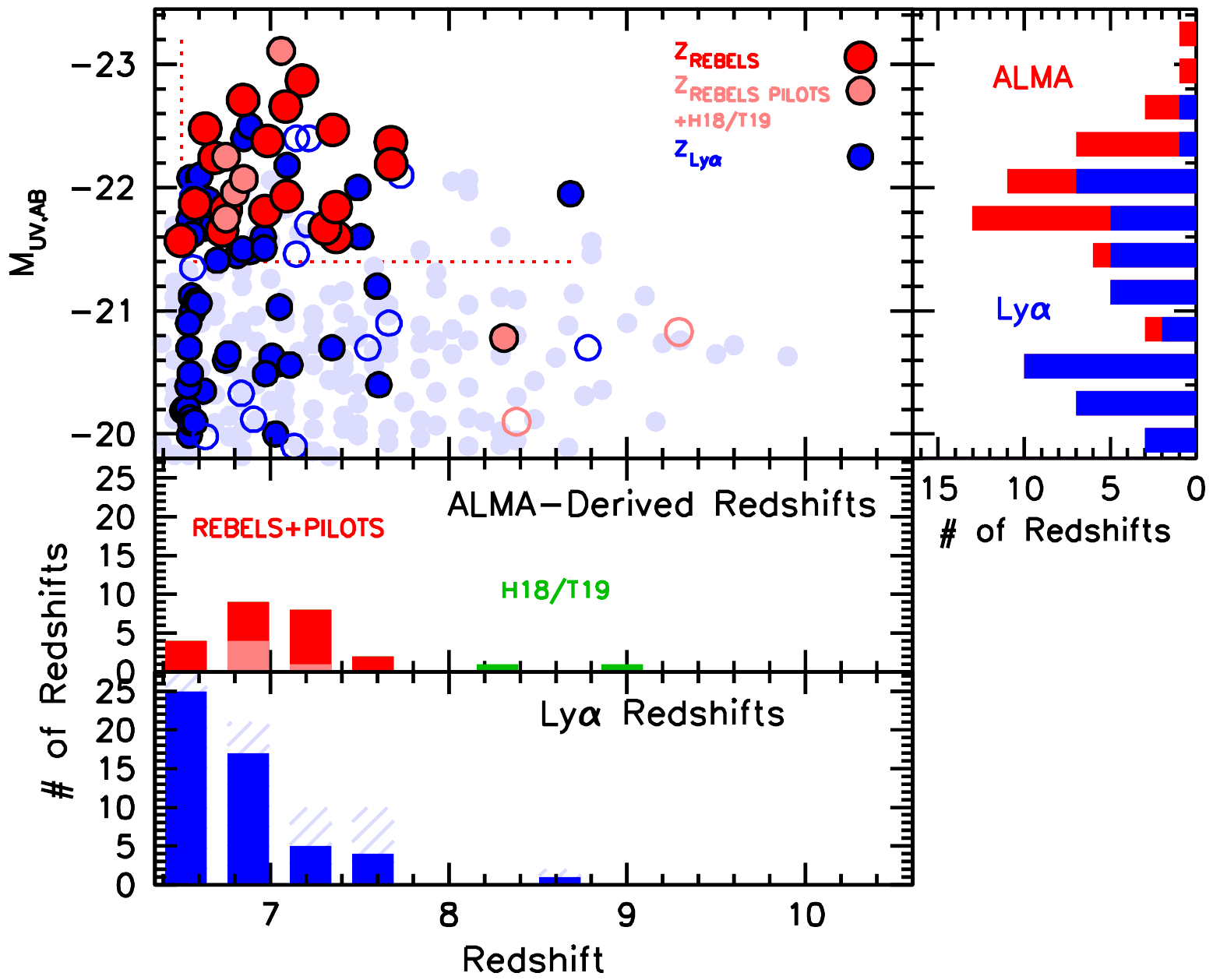}
\caption{(\textit{upper panel}) $UV$ luminosity vs. spectroscopic
  redshift measurement from \cii$\,$based on observations from the
  REBELS ALMA large program (\textit{red circles}).  Also shown
  (\textit{light red circles}) are the $UV$ luminosities and
  spectroscopic redshift for sources from the REBELS pilot programs
  (Smit et al.\ 2018; Schouws et al.\ 2021, 2022) and other spectral
  scan programs in the literature (Hashimoto et al.\ 2018; Tamura et
  al.\ 2019).  For context, the $UV$ luminosities and existing
  spectroscopic redshift measurements from the Ly$\alpha$ line are
  shown as the blue circles.  $z>6.5$ sources with redshift
  determinations based on Ly$\alpha$ and \oiii$\,$lines where the
  signficance is $<$7$\sigma$ are shown with open blue circles and
  open light red circles, respectively.  The light blue circles are
  also shown for context and are based on the $UV$ luminosities and
  photometric redshifts of sources over {\it HST} legacy fields
  (Bouwens et al.\ 2021).  (\textit{middle panel}) The number of
  spectroscopic redshifts available as a function of redshift relying
  on only the most prominent ($>$7$\sigma$) \cii-detected galaxies in
  the REBELS large program and pilot programs (\textit{red and light
    red histogram, respectively}).  The green histogram shows the
  numbers based on other spectral scan programs in the literature
  (Hashimoto et al.\ 2018; Tamura et al.\ 2019).  (\textit{lower
    panel}) The number of spectroscopic redshifts available as a
  function of redshift relying on $\geq7\sigma$ and 5-7$\sigma$
  Ly$\alpha$-detected galaxies at $z>6.5$ (\textit{blue and light blue
    hatched histogram, respectively}).  (\textit{right panel}) Number
  of spectroscopic redshifts from ALMA (\textit{red histogram}) and
  Ly$\alpha$ (\textit{blue histogram}) on the basis of $\geq$7$\sigma$
  lines.  Notice how there are already $\approx$2$\times$ as many
  spectroscopic redshifts available for the brightest
  ($M_{UV,AB}<-22$) sources from ALMA as available from
  Ly$\alpha$.\label{fig:specz}}
\end{figure*}

\begin{figure*}
\epsscale{1.08}
\plotone{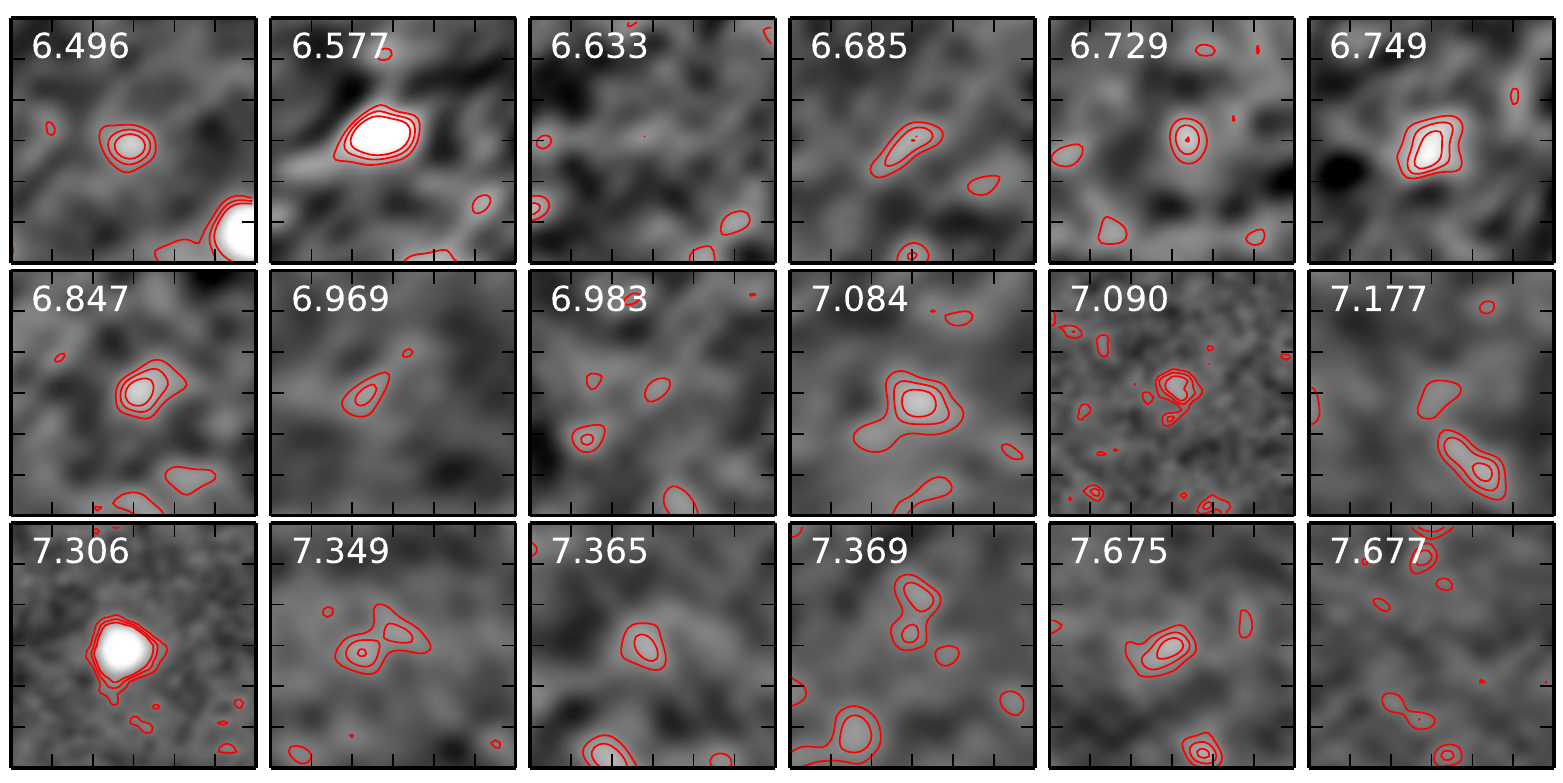}
\caption{Illustration of the dust-continuum observations available for
  sources showing $\geq$7$\sigma$ ISM cooling line detections in
  Figure~\ref{fig:lines}.  Each stamp is 7.2$''$$\times$7.2$''$ on a
  side.  The redshifts of the sources are indicated in the upper left
  corner of each panel.  The red contours correspond to 2, 3, and
  4$\sigma$.  The majority of the sources with highly significant
  $7\sigma$ ISM cooling line detections (13 out of 18) also show
  $\gtrsim$3.3$\sigma$ detections in the dust continuum.  A more
  detailed characterization of the REBELS targets in the dust
  continuum will be presented in H. Inami et al.\ (2022, in
  prep).\label{fig:dust}}
\end{figure*}

Reduction and calibration of data from the program were performed
using the standard ALMA calibration pipeline as implemented in the
Common Astronomy Software Applications package (CASA) version 5.6.1.
Data cubes were reimaged with the \texttt{tclean} task using a natural
weighting to maximize our sensitivity for detecting ISM cooling lines
in ALMA band-5/6/7 observations obtained from the program.  Cleaning
was done down to 2$\sigma$ in producing the data cubes.

Line searches were performed using three different line-search
algorithms on the reduced data cubes.  Searches for $>$$5\sigma$ lines
were perfomed within 1'' of the target center in the rest-$UV$.  Line
widths from 80 km/s to 600 km/s were considered in searching for lines
throughout our data cubes.  S. Schouws et al.\ (2022, in prep) will
provide a detailed description of our line search procedures and
catalogs, while carefully quantifying both the completeness and purity
of the line searches.

As an illustration of the effectiveness of the spectral scans employed
in the REBELS program to-date, we include in Figure~\ref{fig:lines}
the 18 \cii$\,$lines detected with a significance of $>$7$\sigma$.
Redshifts of the detected \cii$\,$lines range from $z=6.496$ to 7.677.
$7\sigma$ was adopted as the detection threshold in this paper to keep
the focus on the brightest and most significant lines found in the
survey.  Details on the purity of our ISM line searches,
characteristics of the \cii$\,$line detections as well as several
additional lower S/N \cii$\,$lines will be presented in S. Schouws et
al.\ (2022, in prep).

Even if we consider only the $>$7$\sigma$ line detections in the
REBELS data obtained thus far, the program is already having a
substantial impact on the number of spectroscopic redshifts derived
with ALMA at $z>6.5$.  Figure~\ref{fig:specz} (\textit{top panel})
shows how the sources with $\geq7\sigma$ ISM-cooling lines are
distributed as a function of redshift and $UV$ luminosity.  Also shown
in the top panel of Figure~\ref{fig:specz} are sources from the two
pilot programs to REBELS (Smit et al.\ 2018; Schouws et al.\ 2021,
2022).

As the lower two panels of Figure~\ref{fig:specz} demonstrate, the
number of spectroscopic redshift measurements from ALMA is already
fairly similar to that available from similarly significant Ly$\alpha$
lines at $z>7$, if one makes use of the most significant ISM-cooling
line detections from the REBELS program and from the two pilot
programs.  This is especially the case for the brightest and most
massive sources at $z>6.5$ where the number of redshifts from ALMA
already exceeds that available from Ly$\alpha$ by a factor of 2.

Simultaneous with the REBELS spectral scans for bright ISM cooling
lines, we are able to probe dust continuum emission from sources in
REBELS.  Figure~\ref{fig:dust} shows the continuum observations
available for the 18 targets from REBELS with $\geq 7\sigma$
ISM-cooling line detections, and it is clear that the majority of
these \cii$\,$-detected sources (13/18) show nominal $3.3\sigma$
detections in the dust continuum.  These dust-continuum images were
generated using the \texttt{tclean} task in CASA.  Any channels
containing emission from the \cii$\,$line were excluded when producing
the continuum maps.  More details on our current dust-continuum
results from REBELS will be presented in H. Inami et al.\ (2022, in
prep).

\subsection{Effectiveness of the REBELS Spectral Scan Strategy\label{sec:results}}

We can already use the results from the REBELS first-year data to
assess the effectiveness of the spectral scan strategy.  Perhaps the
most relevant variable to utilize in gauging the effectiveness of the
scans is in terms of the SFRs of individual targets in our program.

For our SFR estimates for individual sources in the REBELS program, we
make use of the luminosities of sources in both the rest-$UV$ and
far-IR continuum.  The unobscured SFR$_{UV}$s for sources are
estimated based on source $UV$ luminosities using the relation
SFR$_{UV} = 7.1\times 10^{-29} L_{\nu} [\textrm{ergs/s/Hz}]$, while
the obscured SFR$_{IR}$s of sources are estimated based on source $IR$
luminosities using the relation SFR$_{IR} = 1.2\times 10^{-10}
L_{IR}/L_{\odot}$. The inferred SFRs are $\sim$14\% lower for a given
$UV$/$IR$ luminosity than the calibrations adopted for ALPINE
(e.g. Schaerer et al.\ 2020).  A detailed motivation and discussion of
these relations will be provided in M. Stefanon et al.\ (2022, in
prep), M. Topping et al.\ (2022, in prep), and H. Inami et al.\ (2022,
in prep).

In computing the obscured SFR$_{IR}$ of sources, the IR luminosities
of sources $L_{IR}$ are taken to equal 14.2$_{-4.7}^{+7.6}$
$\nu$$L_{\nu}$ where $\nu$ is the frequency of the \cii$\,$line
(H. Inami et al.\ 2022, in prep; Sommovigo et al.\ 2022).  Using this
scaling, sources in REBELS have $L_{IR}$ luminosities ranging from
3$\times$10$^{11}$ $L_{\odot}$ to 1$\times$10$^{12}$ $L_{\odot}$
(equivalent to obscured SFRs of 36 M$_{\odot}$/yr to 120
M$_{\odot}$/yr, respectively).  The precise scaling we use here is
equivalent to that found from a modified blackbody SED with a dust
temperature of 50 K and a dust opacity index $\beta$ of 1.6.  This is
a slightly higher dust temperature than ALPINE use in the analysis of
their $z=4$-6 sample (B{\'e}thermin et al.\ 2020), but lower than is
found by Bouwens et al.\ (2020) in modeling the dust temperature
vs. redshift measurements available at the time.  It is also
consistent with the general range of dust temperatures found by
semi-analytical models (Sommovigo et al.\ 2022) and numerical
simulations (R. Schneider et al.\ 2022, in prep) for REBELS-like
sources.  A more extensive motivation for these conversion factors
will be provided in H. Inami et al.\ (2022, in prep).

\begin{figure}
\epsscale{1.17}
\plotone{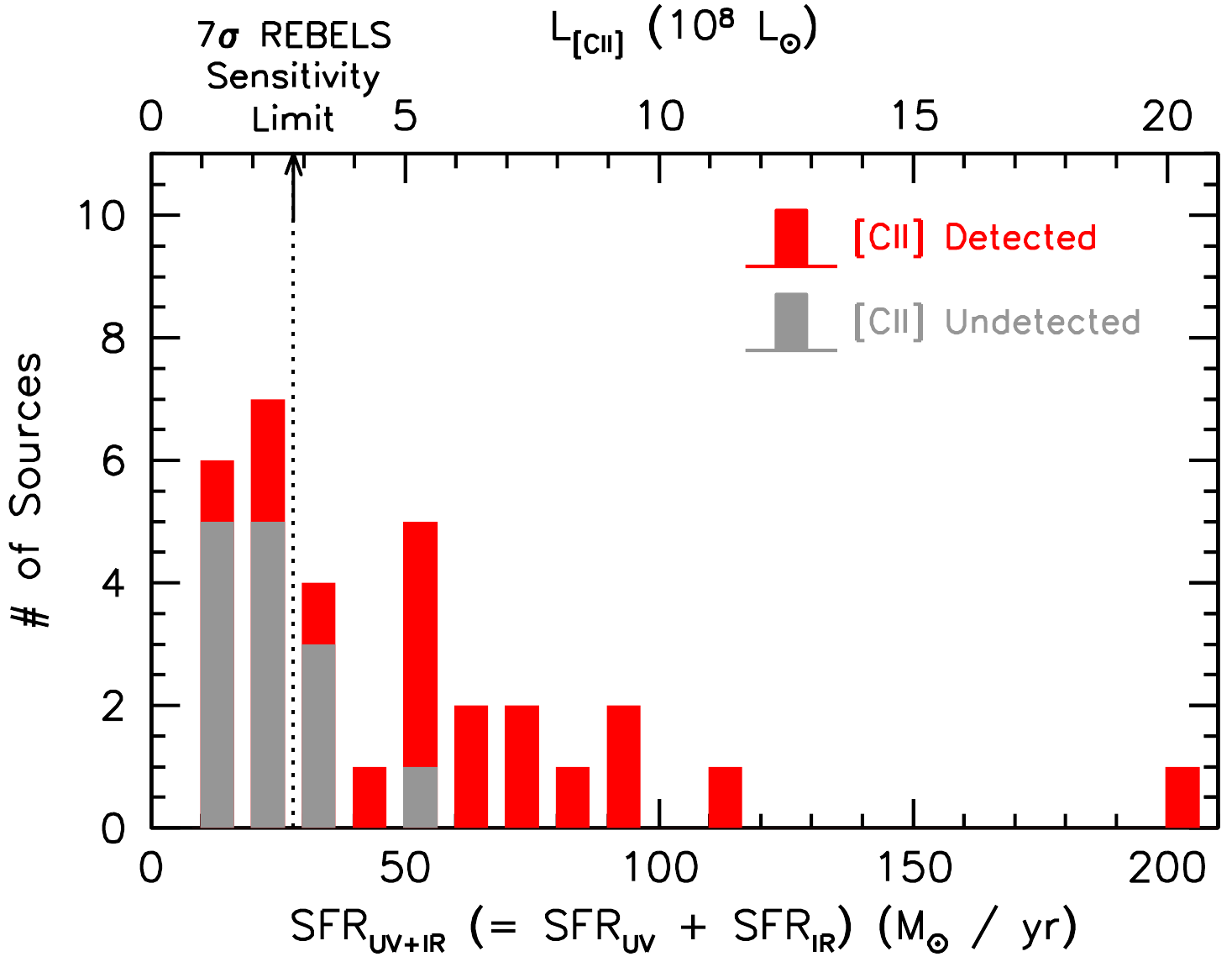}
\caption{Number of sources with \cii$\,$detected (\textit{red
    histograms}) at 7$\sigma$ vs. SFR$_{UV+IR}$ ($=$ SFR$_{UV}$ +
  SFR$_{IR}$).  The gray histogram indicate the number of sources
  where \cii$\,$is yet to be detected at $7\sigma$ in the REBELS
  program.  Only sources where the spectral scans are complete -- and
  where sources are considered to be securely at $z>6$ -- are included
  in the numbers.  The upper horizontal axis shows the
  \cii$\,$luminosity that is equivalent to a given SFR$_{UV+IR}$s
  using the $z\sim0$ \lcii-SFR relation from de Looze et al.\ (2014).
  Our prescriptions for computing SFR$_{UV}$ and SFR$_{IR}$ are
    given in \S4.2.  The detection rate of \cii$\,$shows a dramatic
  increase at SFRs higher than 28 M$_{\odot}$/yr (equivalent to
  the approximate luminosity limit for searches for \cii$\,$in REBELS
  adopting a 7$\sigma$ detection threshold, i.e., $2.8
  \times10^{8} L_{\odot}$ [Figure~\ref{fig:scan}]).\label{fig:sfrtot}}
\end{figure}

\begin{figure}
\epsscale{1.17}
\plotone{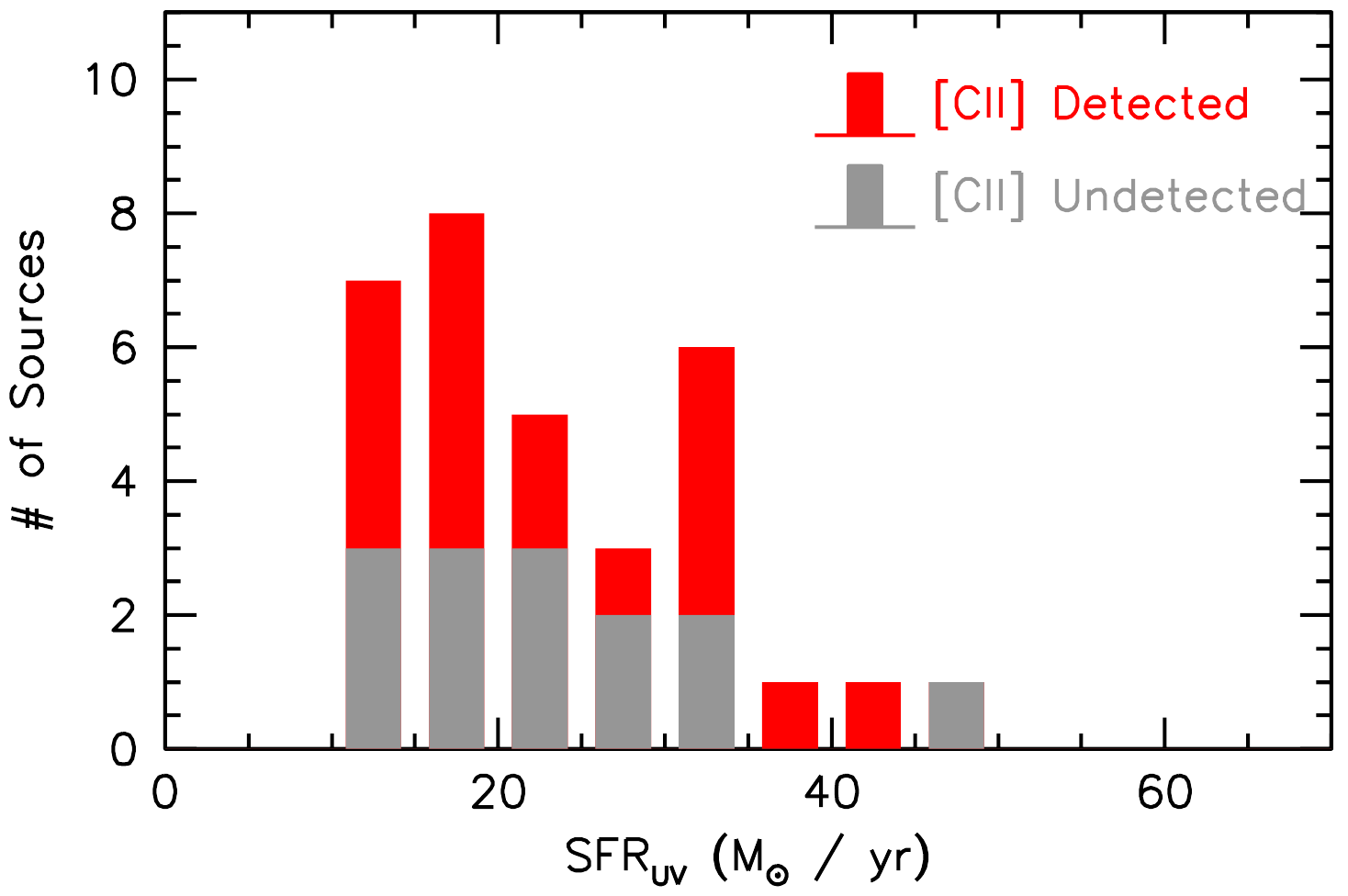}
\caption{Similar to Figure~\ref{fig:sfrtot} but as a function of the
  unobscured SFR$_{UV}$.  While the \cii-detected fraction of sources
  in REBELS does show some dependence on SFR$_{UV}$, the dependence is
  significantly weaker than on SFR$_{UV+IR}$
  (Figure~\ref{fig:sfrtot}).  This demonstrates the essential value
  that ALMA observations have in helping us to characterize the ISM of
  star-forming galaxies at $z>6.5$.\label{fig:sfruv}}
\end{figure}

Using the above scalings to estimate the SFRs of individual sources,
we present the number of sources showing prominent ISM-cooling line
detections as a function of the total SFR of sources in
Figure~\ref{fig:sfrtot}.  We only include in the analysis the 32
sources for which our ISM cooling line scans are complete and which
appear to be securely at $z>6$.\footnote{Significantly deeper
\textit{Spitzer}/IRAC observations became available for REBELS-10
following the selection of targets for the REBELS program (Stefanon et
al.\ 2019b).  Using the new photometry, REBELS-10 now appears to be
more likely at $z<6$ than at $z>6$.  In addition to REBELS-10, there
are also concerns about the robustness of REBELS-13, given the formal
integrated likelihood of its being at $z<6$ and its lacking a red
[3.6]$-$[4.5] color, as is typical for star-forming galaxies in the
redshift range $z=7.0$-9.1 due to the [OIII]$_{4959,5007}$+H$\beta$
emission lines (Roberts-Borsani et al.\ 2016).  M. Stefanon et
al.\ (2022, in prep) will discuss each of these cases in more detail.}
Since only one of our \oiii$\,$spectral scans is complete to present
in the first year data, it makes sense to frame these search results
in terms of our spectral scans for \cii.

Looking over the results, the efficiency of our spectral scans for
\cii$\,$show a dramatic increase in efficiency above 28
M$_{\odot}$/yr.  Fifteen of the 19 sources with SFRs in
excess of 28 M$_{\odot}$/yr, i.e., 79\%, show prominent
7$\sigma$ ISM-cooling lines in the observations taken to-date.
Meanwhile, below a SFR of 28 M$_{\odot}$/yr, only three of
thirteen sources show a prominent $7\sigma$ \cii$\,$detection.
We note that the $7\sigma$ limiting luminosity for \cii$\,$scans
in the REBELS program (Figure~\ref{fig:scan}),
2.8$\times$10$^{8}$ $L_{\odot}$/yr, is equivalent to a SFR
threshold of 28 M$_{\odot}$/yr, using the $z\sim0$ relation from
de Looze et al.\ (2014).  This result is suggestive of only minimal
evolution in the \cii-SFR relation to $z\sim0$ (see also Matthee et
al.\ 2017, 2019; Carniani et al.\ 2018, 2020; Harikane et al.\ 2020;
Schaerer et al.\ 2020).

We also note that some sources with SFRs higher than 28
  M$_{\odot}$/yr remain undetected in \cii$\,$, while some sources
  below 28 M$_{\odot}$/yr are detected.  This is suggestive of there
  being some intrinsic scatter in the \lcii vs. SFR relation at
  $z\sim7$, similar to the 0.27 dex scatter observed at $z\sim0$ by de
  Looze et al.\ (2014).  We will further characterize both the
  evolution and the scatter in the \lcii vs. SFR relation in Schouws
  et al.\ (2022, in prep).

It is also interesting to present the efficiency of our \cii$\,$scans
in terms of the unobscured SFRs for sources.  Figure~\ref{fig:sfruv}
shows the number of sources where \cii$\,$is prominently detected
($>$7$\sigma$: \textit{red histogram}) as a function of the unobscured
SFRs and also the number of sources where no prominent
\cii$\,$emission is found (\textit{gray histogram}).  While the
\cii$\,$ detection fraction does show some dependence on the
unobscured SFR$_{UV}$s, the dependence is significantly less steep
than on the total SFR$_{UV+IR}$s (Figure~\ref{fig:sfrtot}).  This
demonstrates that the essential value ALMA observations have for
characterizing the ISM of star-forming galaxies at $z>6.5$.

We emphasize that only the most prominent \cii$\,$detections from the
REBELS program are included here in evaluating the efficiency of the
spectral scans.  There are indeed a larger number of
\cii$\,$detections in REBELS, but these detections require a more
careful demonstration of their robustness.  These line detections will
be presented in detail in S. Schouws et al.\ (2022, in prep).

\section{Science Objectives}

In facilitating the construction of a significant sample of especially
luminous ISM reservoirs in the $z>6.5$ universe, the REBELS Large
Program enables us to pursue a wide variety of different scientific
objectives.

These include the following:\\

\noindent \textbf{Significantly Expand the Overall Number of Sources
  in Current Spectroscopic Samples at z$\gtrsim$7:} Based on the
success rate of observations from various pilot programs (Smit et
al.\ 2018; Hashimoto et al.\ 2018; Schouws et al.\ 2022), REBELS is
providing new redshift measurements for 25-30 galaxies at $z\sim 7$-9
in the reionization epoch, almost tripling the number of such
measurements available now with ALMA at $z>7$.  See histograms in the
lower two panels of Figure~\ref{fig:specz}.  In addition to the
increase in numbers, the new spectroscopic redshifts we are obtaining
with ALMA sample much more massive sources than has generally been
identified thus far with rest-$UV$ lines like Ly$\alpha$.\\

\noindent \textbf{Probe the Emergence of Dust in Bright, Massive
  $\mathbf{z>6.5}$ Galaxies:} REBELS is providing a sensitive probe of
dust-continuum emission in $z > 6$ galaxies due to our scan strategy
for finding \cii+\oiii$\,$line emission. As Figure~\ref{fig:lir}
shows, our scan strategy allows us to observe faint enough in the
far-IR continuum to probe dust growth over most of the reionization
era, effectively providing the community with a $z = 6$-10 extension
to $z=4$-6 results from ALPINE (Le Fevre et al.\ 2020; B{\'e}thermin
et al.\ 2020; Faisst et al.\ 2020).  Probing dust emission in $z>7$
galaxies is very interesting, given the huge uncertainties that exist
regarding the origin and production of early dust in the universe. In
particular, both the relative contribution of dust creation from
high-z supernovae (SNe) or Asymptotic Giant-branch (AGB) stars, as
well as grain growth and dust destruction, are poorly constrained and
continue to be actively debated (e.g., Mancini et al.\ 2015;
Michalowski 2015; Popping et al.\ 2017; Graziani et al.\ 2020). Of
considerable interest to such discussions has been the detection and
IR luminosities of $z > 6$ galaxies like A1689-zD1 ($z\sim 7.5$:
Watson et al.\ 2015; Knudsen et al.\ 2017) and A2744 YD4 ($z=8.38$:
Laporte et al.\ 2017), which have been argued to show too much dust
mass/emission to match most models (e.g. Mancini et al.\ 2015;
Michalowski 2015).  One exception is Behrens et al.\ (2018), who
reproduced the observed SED from their simulations (Pallottini et
al.\ 2017) finding a SFR $\approx$4 higher than deduced from SED
fitting by Laporte et al. (2017), as well as a low dust-to-metal ratio
(implying inefficient early dust formation), and higher dust
temperatures (as high as Td $\sim$ 91 K: e.g., Ferrara et al.\ 2016,
2021; Sommovigo et al.\ 2020, 2021).  Whatever the resolution of this
issue, REBELS is allowing for a characterization of the build-up of
dust in galaxies, based on comparisons it allows from galaxies at
$z\sim 8.5$-9.5 to $z\sim6$, 7, and 8 (Dayal et al.\ 2022; H. Inami et
al.\ 2022, in prep; R. Schneider et al.\ 2022, in prep).

Follow-up of bright sources from our sample should allow for a
definitive measurement of the dust temperature, the emissivity, as
well as evolution in the IRX-$\beta$ relation, as well as the
evolution of these quantities with cosmic time.  This will be valuable
for determining whether the dust temperatures of galaxies evolve
strongly in the first 2 billion years of the universe (B{\'e}thermin
et al.\ 2015; Schreiber et al.\ 2018; Bouwens et al.\ 2020) or whether
the evolution is relatively mild (Faisst et al.\ 2020).  The REBELS
probe of the dust content in galaxies is also allowing for a
definitive determination of the obscured SFR density in $UV$ bright
galaxies in the epoch of reionization as well as a search for highly
dust obscured galaxies in the immediate neighborhood of the
$UV$-bright population as was recently reported in Fudamoto et
al.\ (2021) using the REBELS data set.  This will allow for an
investigation into whether the cosmic star formation rate density
flattens, as found in FIR and radio observations dusty UV-obscured
systems (Novak et al.\ 2017; Gruppioni et al.\ 2020; Talia et
al.\ 2021).\\

\noindent \textbf{Provide First Significant Probe of the
  High-Luminosity End of the IR and \cii$\,$luminosity functions in
  the Reionization Epoch:} Using the luminosities of the \cii$\,$lines
we detect in REBELS and volume densities of the UV samples from which
our targets are drawn, we are deriving the first significant \cii and
IR luminosity functions at $z > 6$ at the high luminosity end, which
provide a valuable constraint for theoretical models that look at
\cii$\,$emission from galaxies (e.g., Vallini et al.\ 2015, 2017,
2020; Pallottini et al.\ 2017, 2019; Olsen et al.\ 2017; Katz et
al.\ 2017; Popping et al.\ 2017; Lagache et al.\ 2018; Kohandel et
al.\ 2019; Arata et al.\ 2020). Having constraints on the IR
luminosity function provides us with a direct pathway to compute the
SFR function itself ($= SFR_{UV} + SFR_{IR}$) at $z>6.5$.\\

\noindent \textbf{Rest-optical Emission Line EWs, stellar masses,
  specific star formation rates, and the galaxy stellar-mass function
  at $z\sim 7$:} Our estimates on the stellar masses of $z > 6$
galaxies are poor due to the impact of very strong optical nebular
line emission which have a huge (factor of $\sim$1.7) impact on
observed 3-5$\mu$m fluxes of galaxies at $z > 6$ (e.g., Schaerer \& de
Barros 2009; Raiter et al.\ 2010; Stark et al.\ 2013; Smit et
al.\ 2014; Roberts-Borsani et al.\ 2016). Our ALMA scans give us the
precise redshift of sources, enabling us to model the impact of line
emission on the {\it Spitzer}/IRAC fluxes much more precisely. These
data are thus allowing us to dramatically improve the robustness of
the high-mass end of the galaxy stellar mass function at $z\sim 7$.\\

\noindent \textbf{Dynamical mass estimates:} To reveal the physics
behind early massive galaxy build-up, we are deriving dynamical mass
estimates from the emission-line widths of our sample (following,
e.g., Capak et al.  2015).  These estimates can then be related to the
stellar mass estimates that we can uniquely derive for this sample,
allowing us to investigate if the stellar mass in these early galaxies
builds up smoothly as their gas reservoirs grow or if feedback
processes plays a significant role in the conversion from gas to
stars, based on the scatter in the stellar-to-dynamical mass ratio.\\

\noindent \textbf{Major mergers and rotation-dominated systems:} Our
observations are also providing robust detections of the \cii$\,$or
\oiii$\,$lines in our galaxies, which reveal if major mergers are
ongoing in our sources by fitting for double-peaked line profiles
(such as observed by Willott et al.\ 2015, and Matthee et al.\ 2017).
We are comparing the major merger rate found in our program with
recent {\it HST} imaging analysis of major merger rates between
redshift $1 < z < 7$ (e.g., Duncan et al. 2019), which suggest a
steady increase in the major merger rate with redshift, such that at
$z > 6$ roughly $\sim$50\% of galaxy growth is contributed by merger
activity (but see however Dayal et al.\ 2013).  We are testing these
claims in a statistically significant sample at z$\gtrsim$7, using the
first such analysis from spectroscopy.

Furthermore, we are performing a Briggs-weighted analysis of the data
to investigate the low-resolution kinematics of our sources and derive
velocity gradients (following e.g. Smit et al.\ 2018).
Forster-Schreiber et al.\ (2009) estimate the level of rotational
support in a galaxy based on the observed velocity gradient and the
integrated velocity dispersion of the source, defining
$v_{obs}/2\sigma_{int} > 0.4$ as rotation-dominated systems.  Using
sources with single peaked line profiles and monotonically rising
velocity gradients (i.e. sources without major mergers), we aim to
obtain a census of the rotation-dominated fraction of the z$\gtrsim$7
galaxy population.\\

\noindent \textbf{Outflows and \cii$\,$halos:} We are performing a
stacking analysis of \cii$\,$(in the $uv$ plane) to reveal the low
surface brightness components of \cii$\,$and search for broad line
emission in the stacked spectra. By comparing the outflow velocities
to SFRs, dynamical, and stellar masses, we are investigating if
outflow velocities increase with redshift to $z\sim 7$, as has been
found at $z\sim 0$-2 (e.g. Sugahara et al.\ 2019). Recent observations
furthermore suggest that diffuse and extended ‘\cii$\,$halos’ might be
present in $z\sim 3$-7 star-forming galaxies (Fujimoto et al.\ 2019;
Rybak et al.\ 2019; Herrera-Camus et al.\ 2021). Our stacking analysis
is revealing if such diffuse components are ubiquitous in the earliest
galaxies, providing the first insight into the physical properties of
the circumgalactic medium at $z\gtrsim 7$.\\

\noindent \textbf{Legacy value:} REBELS is explicitly designed to
deliver a large sample of very high luminosity \cii$\,$and
\oiii-emitting galaxies at z$\gtrsim$7, at a modest cost. The REBELS
line emitters are prime targets for deep spectroscopic follow-up with
JWST, given the clear value of these galaxies for understanding the
early build-up of luminous ISM reservoirs, and in fact one such
program has been approved for execution in cycle 1 (Stefanon et
al.\ 2021).  Additionally, the identification of such line emitters is
absolutely essential for future work probing the kinematics of
z$\gtrsim$7 galaxies at high resolution.  This will be essential for
robustly discriminating between rotation-dominated galaxies and those
undergoing mergers -- which can be degenerate for some choices of
parameters at the REBELS spatial resolution.

Furthermore, the characterisation of the physical conditions of the
gas in high redshift galaxies through observations of [NII]$_{205\mu
  m}$, [NII]$_{122\mu m}$, [CI]/CO, [OIII]$_{52\mu m}$ or [OI]$_{63\mu
  m}$/[OI]$_{146\mu m}$ is only feasible for the brightest galaxies in
the Epoch of Reionization.  New systemic redshift measurements from
REBELS would have huge value for studies of low-S/N rest-UV lines in
these bright galaxies (including constraints on the neutral Hydrogen
fraction of the IGM, based on the declining prevalence of Ly$\alpha$
as a function of redshift), as is being presented e.g. in Endsley et
al.\ (2022).  After shifting to a common rest-frame wavelength,
archival/future spectra of these galaxies can be immediately stacked.

\section{Summary}

The purpose of this paper is to summarize the scientific motivation,
observational strategy, and sample selection of the 70-hour
Reionization Era Bright Emission Line Survey (REBELS) ALMA Large
Program (2019.1.01634.L), while showcasing some of the most exciting
initial results from the program.

The motivation for the REBELS program has been to construct and to
perform a first characterization of a significant sample of especially
luminous ISM reservoirs in the $z>6.5$ Universe.  To achieve this,
REBELS has been systematically scanning 40 of the brightest $z>6.5$
galaxies identified over 7 deg$^2$ for bright ISM cooling lines, while
probing dust-continuum emission from the same sources.

The utility of targeting $UV$-bright galaxies with ALMA has become
increasingly clear from work over the past few years, with many of
brightest ISM-cooling lines at $z>4$ being present among the bright
population (Capak et al.\ 2015; Willott et al.\ 2015; Smit et
al.\ 2018; Matthee et al.\ 2019; Harikane et al.\ 2020; B{\'e}thermin
et al.\ 2020; Schouws et al.\ 2022).  In fact, 79\% of the sources
with SFRs of $>$28 M$_{\odot}$/yr show $\geq7\sigma$ \cii$\,$lines,
which given the typical sensitivites of the REBELS spectral scans
would give the sources \cii$\,$luminosities in excess of $2.8\times
10^{8}$ $L_{\odot}$ (Figure~\ref{fig:cii_vs_muv}).

To maximize the impact of the REBELS program, considerable effort was
thus devoted towards targeting the brightest and most robust selection
of $z>6.5$ galaxies visible to ALMA (\S2).  The use of targets from at
least eight different high-redshift selections (e.g., Bowler et
al.\ 2014, 2017, 2020; Stefanon et al.\ 2017b, 2019a; Endsley et
al.\ 2021, 2022; M. Stefanon et al.\ 2022, in prep) was considered.
For each potential target, photometry was done independently by at
least three different members of the REBELS team, and redshift
likelihood distributions were derived from three independent codes.
Targets were then ranked according to their brightness and robustness,
and ultimately the likelihood of detecting an ISM cooling line in a
spectral scan.  Only the highest ranked sources were included in our
final set of 40 targets.

Consideration was given both to spectral scans targeting the
157.74$\mu$m [CII] line and 88.36$\mu$m [OIII] line for the REBELS
program.  We computed the limiting luminosity to which we could probe
each line as a function of redshift to assess the relative efficiency.
After considering the line ratios observed for $z>6$ galaxies in the
literature (Figure~\ref{fig:oiiicii}), we decided that \cii$\,$line
would be best choice for executing line scans out to $z\sim8.5$ and
that the \oiii$\,$line would be the best choice for sources with
likely redshifts above $z\sim8.5$ (\S3.1).

For each target in our sample, we set up the tunings for our spectral
scans to cover the bulk of the redshift likelihood distribution for
each source (typically $>$80-90\%).  Results obtained from three
independent sets of photometry were used in deriving this likelihood
distribution (\S3.2).  With 1, 2, and 3 tunings, we can cover a
contiguous frequency range of 5.375 GHz, 10.75 GHz, and 20.375 GHz,
respectively, if the spectral scan is in band 6.  We have arranged the
tunings for sources to optimize the number of ISM cooling lines we can
detect for a $\sim$70-hour ALMA allocation.  Appendix A illustrates in
detail the layout of the tunings that make up the spectral scans for
the 40 targets in our program.

The integration time for each tuning is set by the requirement that we
detect a $3\times10^{8}$ $L_{\odot}$ \cii$\,$line at $5\sigma$ when
executing our band-6/5 scans and a $1.1 \times10^9$ $L_{\odot}$
\oiii$\,$line at $5\sigma$ when executing our band-7 scans.  Those
sensitivity requirements translated into $\sim$20 minute integration
times for REBELS \cii$\,$searches to $z\sim7.2$, $\sim$30-40 minute
integration times for \cii$\,$searches to $z\sim8.5$, and $\sim$20-40
minute integration times in searches for the \oiii$\,$line at $z>8.2$.

Simultaneous with the REBELS scans for \cii$\,$and \oiii, REBELS
probes the $IR$ luminosities of our targets based on their
dust-continuum emission.  Given the poorer sensitivity of REBELS for
dust continuum (Figure~\ref{fig:tdust_tline}), our use of multiple
spectral scan tunings to search for ISM cooling lines really is an
advantage in allowing us to probe the dust continuum.  As shown in
Figure~\ref{fig:lir}, REBELS probes to $3\times10^{11}$ $L_{\odot}$ in
galaxies out to $z\sim7.2$.  At $z>7.2$, it probes even deeper to
$2\times10^{11}$ $L_{\odot}$.  This is equivalent to probing to
obscured SFRs of 36 M$_{\odot}$/yr and 24 M$_{\odot}$/yr,
respectively.

During the first year of observations from the REBELS program
(November 2019 to January 2020), 60.6 out of the total 69.6 hours
allocated to REBELS have been executed.  A search for prominent
cooling lines in the data revealed 18 prominent $\geq 7\sigma$
\cii$\,$lines (Figure~\ref{fig:lines} and Table~2).  No
especially significant $\geq7\sigma$ \oiii$\,$lines have been
identified in the existing data from the REBELS program, but
observations are complete for only one of four sources using
\oiii$\,$scans.  The majority of the sources showing $\geq$7$\sigma$
detections of \cii$\,$(13/18) also show $\geq$3.3$\sigma$ dust-continuum
emission (Figure~\ref{fig:dust}).  Remarkably, adding the newly
identified \cii$\,$lines to those from the literature and pilot
programs to REBELS (Smit et al.\ 2018; Schouws et al.\ 2022), the
number of redshift determinations from ALMA is already starting to be
comparable to the number of Ly$\alpha$-derived redshifts at $z>7$
(Figure~\ref{fig:specz}).

It is interesting to already make use of the prominent $7\sigma$
\cii$\,$line detections to quantify the efficiency of spectral scans
with ALMA.  Looking specifically at the fraction of \cii-detected
galaxies as a function of their total SFR, we find a dramatic increase
in the fraction above a SFR of 28 M$_{\odot}$/yr
(Figure~\ref{fig:sfrtot}).  Fifteen of the 19 sources with
SFRs in excess of 28 M$_{\odot}$/yr, i.e., 79\%, show
prominent 7$\sigma$ \cii-cooling lines in the observations taken to
date.  Meanwhile, below a SFR of 28 M$_{\odot}$/yr, only
three of thirteen sources shows a prominent
\cii$\,$detection.  We note that a SFR threshold of 28
M$_{\odot}$/yr corresponds to a \cii$\,$luminosity of
2.8$\times$10$^{8}$ $L_{\odot}$ using the $z\sim0$ De Looze et
al.\ (2014) relation.  Since 2.8$\times$10$^{8}$ $L_{\odot}$ is
also the approximate $7\sigma$ limit for our \cii$\,$scans in
the REBELS program (Figure~\ref{fig:scan}), this is suggestive of
minimal evolution in the \cii-SFR relation from $z\sim8$ to $z\sim0$,
as Schaerer et al.\ (2020) also conclude on the basis of the ALPINE
program (see Matthee et al.\ 2017, 2019; Carniani et al.\ 2018, 2020;
Harikane et al.\ 2020).

In this paper, we have presented the motivation, observational
strategy, and some initial observational results from a cycle-7 ALMA
large program known as REBELS.  In the future, we look forward to the
completion of spectral scans for the final seven targets in the
program, the bulk of which probe galaxies at $z\geq 8$.  Also of
considerable importance will be an exciting array of follow-up
observations being acquired on targets from our program, including
from JWST, JVLA, ALMA, Keck, and the VLT.

\acknowledgements

We would like to acknowledge our collaborators on the BoRG project,
Joanna Bridge, Benne Holwerda, and Michele Trenti for the collective
work done on the identification and characterization of sources from
pure parallel {\it HST} programs.  This allowed for the inclusion of
Super8-1 in the present selection.  We would like to thank Jorryt
Matthee for useful conversations related to the redshift scan range
for 1-2 sources from the REBELS selection.  This paper is based on
data obtained with the ALMA Observatory, under the Large Program
2019.1.01634.L. ALMA is a partnership of ESO (representing its member
states), NSF(USA) and NINS (Japan), together with NRC (Canada), MOST
and ASIAA (Taiwan), and KASI (Republic of Korea), in cooperation with
the Republic of Chile. The Joint ALMA Observatory is operated by ESO,
AUI/NRAO and NAOJ.  RJB and MS acknowledge support from TOP grant
TOP1.16.057.  SS acknowledges support from the Nederlandse
Onderzoekschool voor Astronomie (NOVA).  RS and RAB acknowledge
support from STFC Ernest Rutherford Fellowships [grant numbers
  ST/S004831/1 and ST/T003596/1].  RE acknowledges funding from
JWST/NIRCam contract to the University of Arizona, NAS5-02015.  PAO,
LB, and YF acknowledge support from the Swiss National Science
Foundation through the SNSF Professorship grant 190079 ‘Galaxy
Build-up at Cosmic Dawn’.  HI and HSBA acknowledge support from
  the NAOJ ALMA Scientific Research Grant Code 2021-19A.  HI
acknowledges support from the JSPS KAKENHI Grant Number JP19K23462.
JH gratefully acknowledges support of the VIDI research program with
project number 639.042.611, which is (partly) financed by the
Netherlands Organisation for Scientific Research (NWO).  MA
acknowledges support from FONDECYT grant 1211951, ``CONICYT + PCI +
INSTITUTO MAX PLANCK DE ASTRONOMIA MPG190030'' and ``CONICYT+PCI+REDES
190194''.  PD acknowledges support from the European Research
Council's starting grant ERC StG-717001 (``DELPHI''), from the NWO
grant 016.VIDI.189.162 (``ODIN'') and the European Commission's and
University of Groningen's CO-FUND Rosalind Franklin program.  LG and
RS acknowledge support from the Amaldi Research Center funded by the
MIUR program ``Dipartimento di Eccellenza'' (CUP:B81I18001170001). YF
further acknowledges support from NAOJ ALMA Scientific Research Grant
number 2020-16B ‘ALMA HzFINEST: High-z Far-Infrared Nebular Emission
STudies.’ AF acknowledges support from the ERC Advanced Grant
INTERSTELLAR H2020/740120. Any dissemination of results must indicate
that it reflects only the author’s view and that the Commission is not
responsible for any use that may be made of the information it
contains. Partial support from the Carl Friedrich von
Siemens-Forschungspreis der Alexander von Humboldt-Stiftung Research
Award is kindly acknowledged (AF).  IDL acknowledges support from ERC
starting grant 851622 DustOrigin.  JW acknowledges support from the
ERC Advanced Grant 695671, ``QUENCH’’, and from the Fondation MERAC.
This paper utilizes observations obtained with the NASA/ESA {\it
  Hubble} Space Telescope, retrieved from the Mikulski Archive for
Space Telescopes (MAST) at the Space Telescope Science Institute
(STScI). STScI is operated by the Association of Universities for
Research in Astronomy, Inc. under NASA contract NAS 5-26555.  This
work is based [in part] on observations made with the {\it Spitzer}
Space Telescope, which was operated by the Jet Propulsion Laboratory,
California Institute of Technology under a contract with NASA. Support
for this work was provided by NASA through an award issued by
JPL/Caltech.

\newpage

\begin{figure*}
\epsscale{1.17}
\plotone{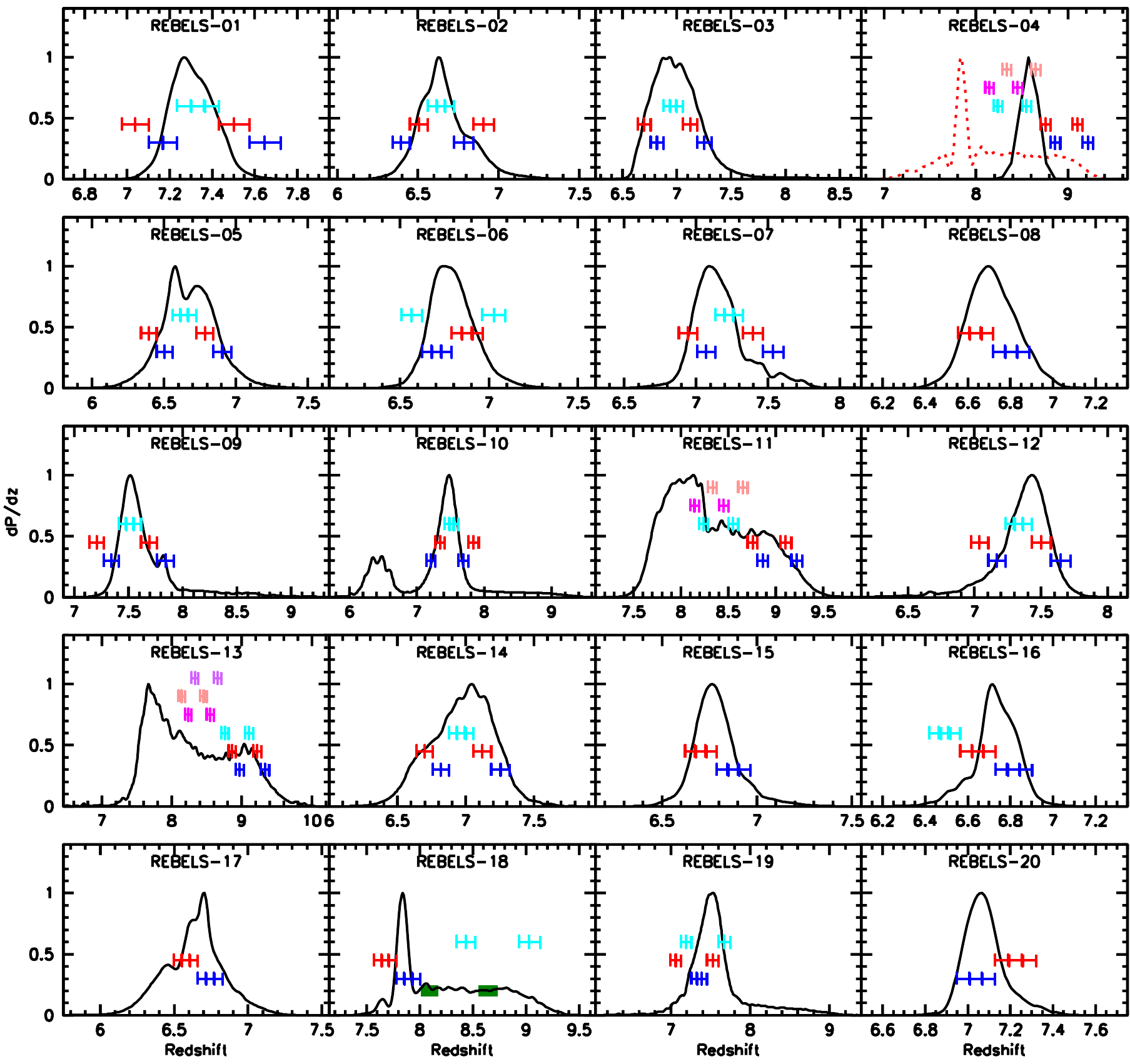}
\caption{Spectral scan windows used to search \cii$\,$and \oiii$\,$in
  REBELS sources 1 through 20.  The black lines show the redshift
  likelihood distributions we derive combining the six separate
  redshift likelihood distributions we derived for each source (e.g.,
  see Figure~\ref{fig:pz}).  The dotted red lines show the redshift
  likelihood distributions derived before including the {\it HST}
  F105W, F125W, and F160W observations from GO 15931 (PI:
  Bowler) and GO 16879 (PI: Stefanon).  Scans are for \cii
  except in the case of REBELS-04, REBELS-11, and REBELS-13 where the
  scans are for \oiii.  The green horizontal bars indicate the
  redshift range probed by previous ALMA observations (Bowler et
  al.\ 2018; Schouws et al.\ 2021, 2022).  \label{fig:scan0}}
\end{figure*}

\begin{figure*}
\epsscale{1.17}
\plotone{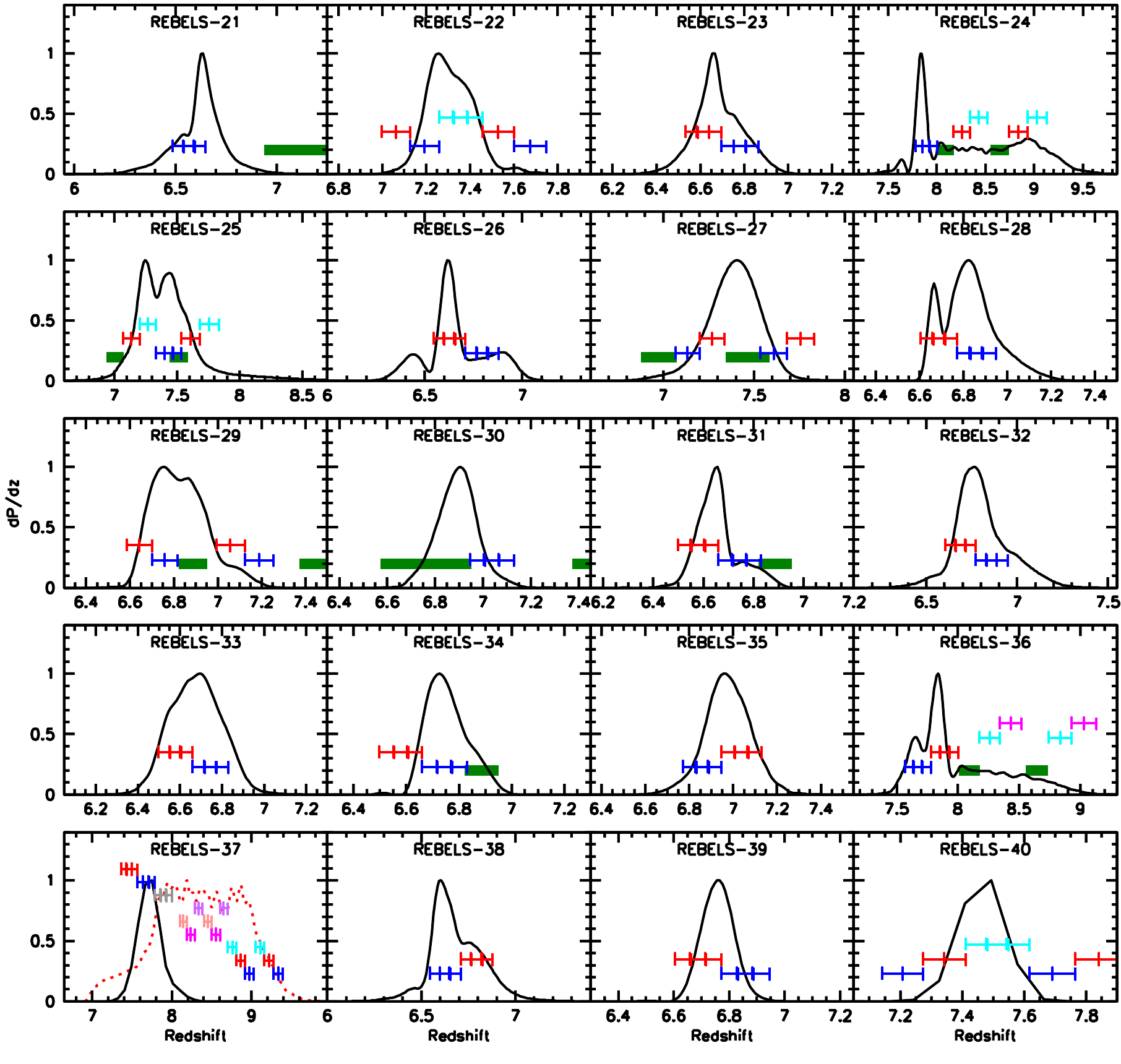}
\caption{As in Figure~\ref{fig:scan0} but for REBELS sources 21
  through 40.  Scans are for \cii$\,$except in the case of REBELS-37
  where the scan is for \oiii.\label{fig:scan1}}
\end{figure*}

\begin{figure}
\epsscale{0.7}
\plotone{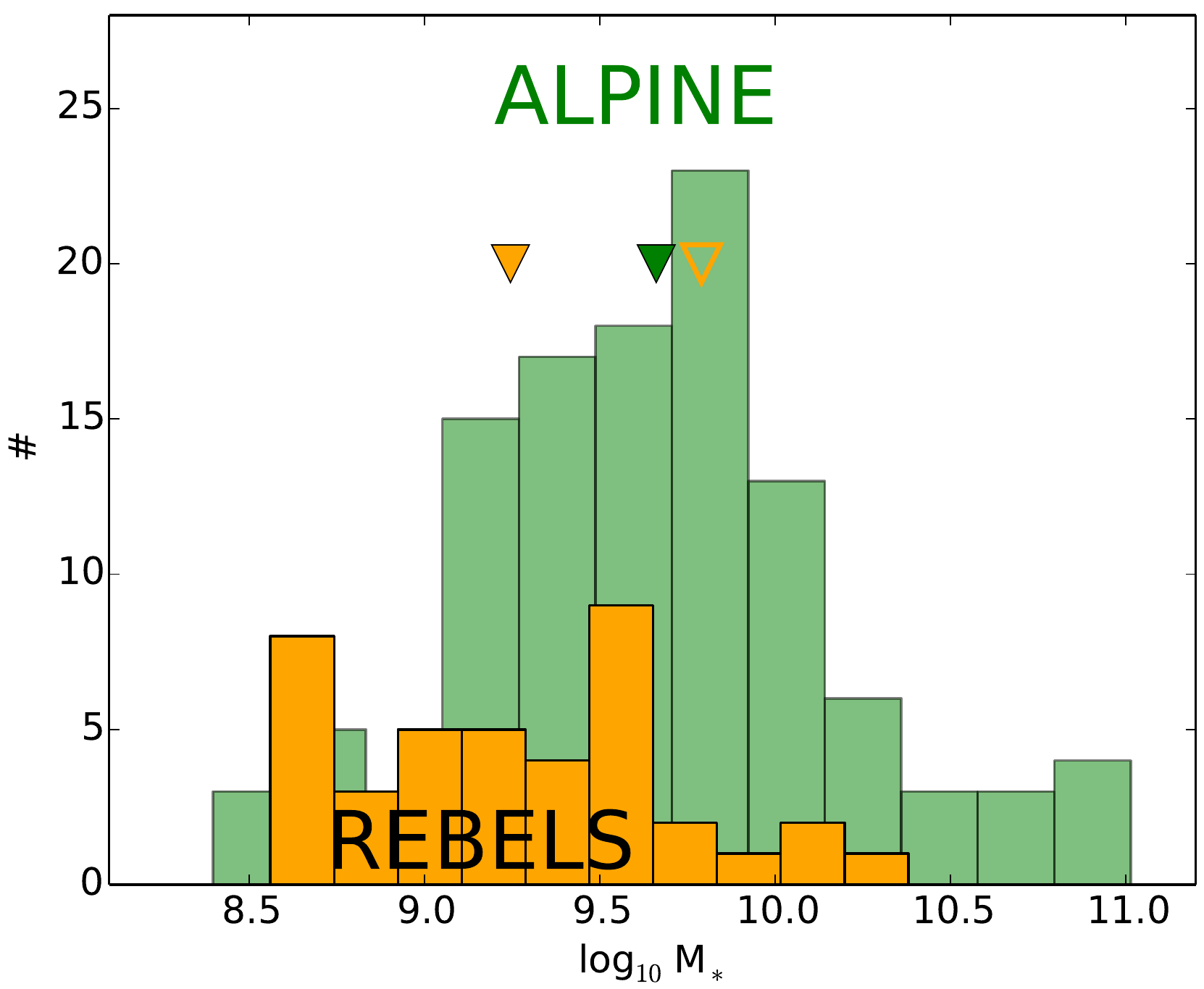}
\caption{Distribution of stellar masses (\textit{orange histogram})
  inferred using \textsc{beagle} for the 40 targets in the REBELS LP
assuming a constant star formation history.  For context, the green
histogram shows the stellar mass distribution for $z=4$-6 targets in
the ALPINE program (Faisst et al.\ 2020).  The median stellar masses
inferred using \textsc{beagle} and using \textsc{Prospector} (Leja et
al.\ 2017: assuming constant and non-parametric star formation
histories, respectively) are shown as downward pointing filled and
open orangle triangles, respectively.  Given the overall similarity of
the two distributions and similar medians, the ALPINE program provides
us with a convenient lower redshift sample to characterize evolution
in galaxies from $z>6.5$ to $z=4$-6.  A similar presentation of the
REBELS sample in redshift, $UV$ luminosity, $UV$-continuum slope
$\beta$, and the [OIII]$_{4959,5007}$+H$\beta$ EW is provided in
Figure~\ref{fig:paramdist}.\label{fig:mass}}
\end{figure}

\begin{figure}
\epsscale{1.18}
\plotone{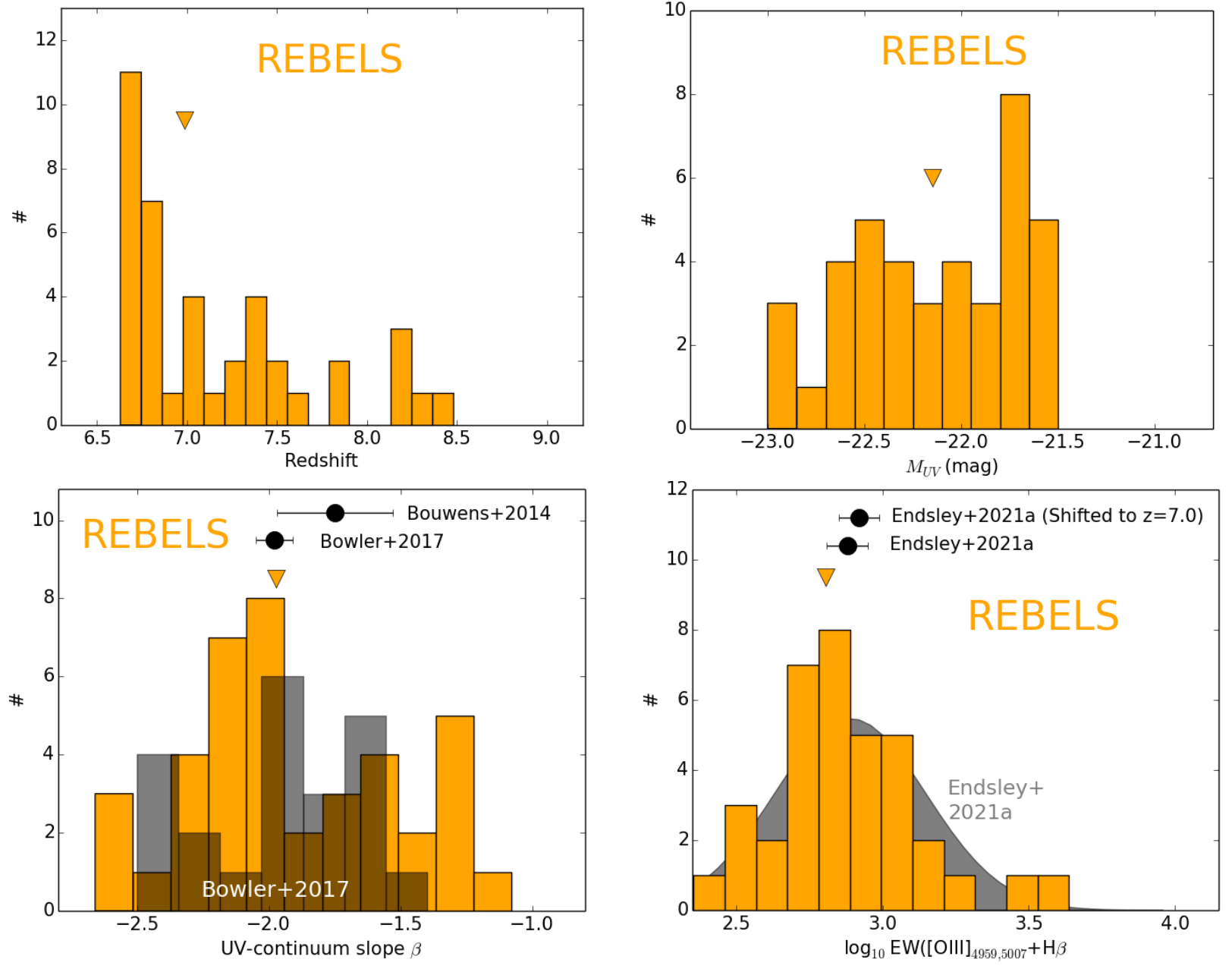}
\caption{Distribution of the 40 REBELS targets in redshift
  (\textit{upper left}), $UV$ luminosity (\textit{upper right}),
  $UV$-continuum slope $\beta$ (\textit{lower left}), and
  [OIII]$_{4959,5007}$+H$\beta$ EW (\textit{lower right}).  The shaded
  histogram shown in the lower left panel is the $UV$-continuum slope
  $\beta$ distribution derived by Bowler et al.\ (2017) for a similar
  luminosity sample of $z\sim7$ galaxies identified over the COSMOS
  and UKIDSS/UDS regions, while the shaded region shown in the lower
  right region is the [OIII]$_{4959,5007}$+H$\beta$ EW distribution
  derived by Endsley et al.\ (2021a) at $z\sim6.75$.  The median
  photometric redshift and $UV$ luminosity are 6.99 and $-22.2$ mag,
  respectively and indicated by the orange downward-pointing
  triangles.  The median $UV$-continuum slope $\beta$ is $-1.98$
  (\textit{indicated by the orange downward-pointing triangle}) and is
  consistent with the mean $\beta$, i.e., $-1.98\pm0.07$, for the
  Bowler et al.\ (2017) $z\sim7$ sample as well as the biweight mean
  $\beta$, i.e., $-$1.75$\pm$0.18$\pm$0.13, derived by Bouwens et
  al.\ (2014) in their highest luminosity $z\sim7$ bin (\textit{black
    circle} with $1\sigma$ uncertainties).  The median
  [OIII]$_{4959,5007}$+H$\beta$ EW for targets in the REBELS program
  using a delayed star formation history is 638$\AA$
  (\textit{indicated by the orange downward-pointing triangle}) and is
  consistent with the median EW 759$_{-113}^{+112}$$\AA$
  (\textit{lower black circle} with $1\sigma$ uncertainties) derived
  by Endsley et al.\ (2021a) at $z\sim6.75$ using a similar fitting
  procedure, as well as an sSFR-evolution-corrected EW of
  820$_{-120}^{+119}$$\AA$ (\textit{upper black circle} with $1\sigma$
  uncertainties) at $z\sim 6.99$.  The similarity of the
  $UV$-continuum slopes $\beta$ and the [OIII]$_{4959,5007}$+$H\beta$
  distribution to the population averages at $z\sim7$ suggest that the
  results derived from REBELS should be reasonably representative of
  the general galaxy population at $z\sim7$.\label{fig:paramdist}}
\end{figure}

\appendix

\section{A.  Spectral scan windows}

The purpose of this appendix is to summarize the tunings utilized by
REBELS in scanning for \cii$\,$and \oiii$\,$in the 40 $z>6.5$ galaxies
the LP targeted.

Figures~\ref{fig:scan0} and \ref{fig:scan1} show the complete set of
tunings used in scanning for the \cii$\,$and \oiii$\,$ISM cooling
lines.  Also shown are the redshift likelihood distributions derived
for each target, which are identical to those presented earlier in
Figure~\ref{fig:pz} with the thick black lines.

Also shown in Figure~\ref{fig:scan0} are the redshift likelihood
distributions for two sources REBELS-04 and REBELS-37 derived
including 1-orbit {\it HST} F105W, 3/4 orbit F125W, and 5/4 orbit
F160W observations from a 2-orbit program led by Rebecca Bowler (GO
15931) and 4-orbit program led by Mauro Stefanon (GO 16879).

With the exception of four targets in our program (REBELS-04,
REBELS-11, REBELS-13, and REBELS-37) where the bulk of the redshift
likelihood distribution is $z>8$, the \cii$\,$line is targeted with
the REBELS spectral scans.  For REBELS-04, REBELS-11, and REBELS-13,
the \oiii$\,$line is targeted.  For REBELS-37, both the \cii$\,$line
(at $z<8$) and \oiii$\,$line (at $z>8$) are targeted as part of the
spectral scans.

Following the first year of observations from the REBELS program and
successful detection of the \cii$\,$ISM cooling line in REBELS-18 and
REBELS-36, several tunings from REBELS-18 and REBELS-36 were shifted
to REBELS-06, REBELS-16, and REBELS-37 to extend the redshift range of
spectral scans for \cii$\,$in these sources.

\begin{deluxetable*}{ccc}
\tablewidth{0cm} \tabletypesize{\footnotesize}
\tablecaption{[OIII]$_{4959,5007}$+H$\beta$ EWs inferred for targets from the REBELS sample using different assumptions about the star formation history\label{tab:ew}}
\tablehead{\colhead{REBELS} & \multicolumn{2}{c}{$\log_{10}$ EW([OIII]+H$\beta$)($\AA$)\tablenotemark{a}}\\
  \colhead{IDs} & \colhead{CSFH} & \colhead{Delayed SFH}}
\startdata
REBELS-01 & $2.90_{-0.12}^{+0.28}$ & 2.97$_{-0.24}^{+0.20}$\\
REBELS-02 & $3.07_{-0.16}^{+0.14}$ & 2.73$_{-0.36}^{+0.29}$\\
REBELS-03 & $3.05_{-0.22}^{+0.33}$ & 2.78$_{-0.47}^{+0.36}$\\
REBELS-04 & $3.25_{-0.31}^{+0.33}$ & 3.15$_{-0.29}^{+0.26}$\\
REBELS-05 & $3.12_{-0.32}^{+0.30}$ & 3.02$_{-0.34}^{+0.30}$\\
REBELS-06 & $2.96_{-0.18}^{+0.27}$ & 2.80$_{-0.27}^{+0.31}$\\
REBELS-07 & $3.18_{-0.17}^{+0.33}$ & 2.82$_{-0.50}^{+0.42}$\\
REBELS-08 & $3.07_{-0.22}^{+0.23}$ & 2.90$_{-0.35}^{+0.26}$\\
REBELS-09 & $3.77_{-0.02}^{+0.05}$ & 1.39$_{-0.59}^{+1.03}$\\
REBELS-10 & $2.95_{-0.09}^{+0.10}$ & 2.16$_{-0.64}^{+0.50}$\\
REBELS-11 & $3.08_{-0.19}^{+0.21}$ & 2.74$_{-0.39}^{+0.36}$\\
REBELS-12 & $3.26_{-0.25}^{+0.29}$ & 3.27$_{-0.25}^{+0.20}$\\
REBELS-13 & $2.98_{-0.09}^{+0.13}$ & 2.35$_{-0.54}^{+0.37}$\\
REBELS-14 & $3.21_{-0.20}^{+0.35}$ & 3.10$_{-0.44}^{+0.31}$\\
REBELS-15 & $3.73_{-0.54}^{+0.10}$ & 3.64$_{-0.59}^{+0.22}$\\
REBELS-16 & $3.02_{-0.10}^{+0.10}$ & 2.61$_{-0.55}^{+0.25}$\\
REBELS-17 & $3.04_{-0.19}^{+0.21}$ & 2.74$_{-0.59}^{+0.41}$\\
REBELS-18 & $3.00_{-0.17}^{+0.22}$ & 2.79$_{-0.25}^{+0.26}$\\
REBELS-19 & $3.11_{-0.22}^{+0.21}$ & 2.82$_{-0.34}^{+0.33}$\\
REBELS-20 & $3.17_{-0.11}^{+0.15}$ & 3.18$_{-0.47}^{+0.22}$\\
REBELS-21 & $2.84_{-0.12}^{+0.28}$ & 2.39$_{-0.61}^{+0.58}$\\
REBELS-22 & $2.91_{-0.14}^{+0.31}$ & 2.75$_{-0.47}^{+0.36}$\\
REBELS-23 & $3.03_{-0.16}^{+0.19}$ & 2.97$_{-0.28}^{+0.21}$\\
REBELS-24 & $3.13_{-0.21}^{+0.19}$ & 2.86$_{-0.36}^{+0.30}$\\
REBELS-25 & $2.79_{-0.06}^{+0.21}$ & 2.53$_{-0.35}^{+0.29}$\\
REBELS-26 & $2.98_{-0.21}^{+0.27}$ & 2.97$_{-0.29}^{+0.26}$\\
REBELS-27 & $2.89_{-0.11}^{+0.27}$ & 2.57$_{-0.44}^{+0.34}$\\
REBELS-28 & $3.26_{-0.20}^{+0.49}$ & 2.98$_{-0.27}^{+0.21}$\\
REBELS-29 & $2.90_{-0.08}^{+0.12}$ & 2.20$_{-0.43}^{+0.40}$\\
REBELS-30 & $3.06_{-0.13}^{+0.20}$ & 2.80$_{-0.33}^{+0.30}$\\
REBELS-31 & $3.05_{-0.06}^{+0.06}$ & 2.55$_{-0.38}^{+0.26}$\\
REBELS-32 & $3.01_{-0.11}^{+0.11}$ & 2.81$_{-0.21}^{+0.20}$\\
REBELS-33 & $2.90_{-0.11}^{+0.27}$ & 2.72$_{-0.28}^{+0.23}$\\
REBELS-34 & $3.03_{-0.06}^{+0.07}$ & 2.56$_{-0.35}^{+0.35}$\\
REBELS-35 & $3.18_{-0.18}^{+0.30}$ & 2.89$_{-0.33}^{+0.35}$\\
REBELS-36 & $2.99_{-0.16}^{+0.25}$ & 3.04$_{-0.28}^{+0.22}$\\
REBELS-37 & $3.25_{-0.17}^{+0.32}$ & 3.05$_{-0.24}^{+0.18}$\\
REBELS-38 & $3.01_{-0.25}^{+0.35}$ & 3.05$_{-0.33}^{+0.23}$\\
REBELS-39 & $3.58_{-0.37}^{+0.17}$ & 3.52$_{-0.35}^{+0.15}$\\
REBELS-40 & $2.98_{-0.20}^{+0.32}$ & 2.73$_{-0.40}^{+0.36}$
\enddata
\tablenotetext{a}{Derived using \textsc{beagle} (M. Stefanon et al.\ 2022, in prep).}
\end{deluxetable*}

\section{B.  Distribution of the REBELS sample in Parameter Space}

The purpose of Appendix B is to illustrate the distribution of REBELS
targets in parameter space.  Figure~\ref{fig:mass} shows the stellar
mass distribution of sources in the REBELS sample inferred using
  \textsc{beagle} and compares it against the stellar mass
distribution inferred for ALPINE (Faisst et al.\ 2020).  Overall
  the two distributions span a fairly similar mass range.

The median stellar mass we estimate for the REBELS sample from
\textsc{beagle} is $10^{9.25}$ $M_{\odot}$ (\textit{shown in
  Figure~\ref{fig:mass} as a filled downward-pointing orange
  triangle}) assuming a constant star formation rate, while using
\textsc{prospector} (Stefanon et al.\ 2022, in prep), the median
stellar mass we estimate is $10^{9.8}$ $M_{\odot}$ (\textit{shown in
  Figure~\ref{fig:mass} as an open downward-pointing orange
  triangle}).  These masses are $\sim$0.4 dex lower and $\sim$0.1 dex
higher, respectively, than inferred by Faisst et al.\ (2020) for
ALPINE.

Figure~\ref{fig:paramdist} shows the distribution of galaxies in
redshift, $UV$ luminosity, $UV$-continuum slope $\beta$, and
[OIII]$_{4959,5007}$+H$\beta$ EW.  The EW distribution shown in the
lower right panel is derived using a similar procedure to what Endsley
et al.\ (2021a) use to derive [OIII]$_{4959,5007}$+H$\beta$ EWs in
their paper, i.e., assuming a delayed star formation history
(SFR$\propto t e^{-t/\tau}$) and a SMC extinction curve (Pei 1992).
This is to ensure the comparison with Endsley et al.\ (2021a) is done
in the most fair way possible.  For completeness, we include the EWs
derived both assuming a constant star formation history, as shown in
Table~\ref{tab:targlist2}, and assuming a delayed star formation
history in Table~\ref{tab:ew}.  The median photometric redshift, $UV$
luminosity, $\beta$, and [OIII]$_{4959,5007}$+H$\beta$ EW of the
REBELS sources is 6.99, $-22.2$, $-1.98$, and 638$\AA$,
respectively.

The median $UV$-continuum slope $\beta$ is consistent with the
  mean $\beta$, i.e., $-1.98\pm0.07$, for the Bowler et al.\ (2017)
  $z\sim7$ sample.  It is also consistent with the biweight mean
$-$1.75$\pm$0.18$\pm$0.13 derived by Bouwens et al.\ (2014) in their
highest luminosity $z\sim7$ bin.

Meanwhile, the median [OIII]$_{4959,5007}$+H$\beta$ EW is consistent
with the median EW 759$_{-113}^{+112}$$\AA$ derived by Endsley et
al.\ (2021a) at $z\sim6.75$ derived using an analogous
  procedure.  The similarity of the $UV$-continuum slope $\beta$ and
[OIII]$_{4959,5007}$+H$\beta$ EW distributions seen in REBELS to
  the $z\sim7$ population suggest that the results derived from
REBELS should be reasonably representative of the overall
galaxy population at $z\sim7$.

\end{document}